\documentclass[prl,twocolumn,superscriptaddress,showpacs,nofootinbib, amsmath, amssymb]{revtex4-1}

\usepackage{graphicx}
\usepackage{hyperref}
\usepackage{color}
\usepackage[utf8]{inputenc}

\begin{document}

\title{Impact of cosmic-ray physics on dark matter indirect searches}

\author{Daniele Gaggero}
\affiliation{GRAPPA Institute, Institute of Physics, University of Amsterdam, 1098 XH Amsterdam, The Netherlands}
\email{d.gaggero@uva.nl}
\author{Mauro Valli}
\affiliation{INFN, Sezione di Roma, P.le A. Moro 2, I-00185 Roma, Italy}
\email{mauro.valli@roma1.infn.it}

\begin{abstract}
The quest for the elusive dark matter (DM) that permeates the Universe (and in general the search for signatures of Physics beyond the Standard Model at astronomical scales) provides a unique opportunity and a tough challenge to the high-energy astrophysics community. 

In particular, the so-called DM  {\it indirect searches} -- mostly focused on a class of theoretically well-motivated DM candidates such as the weakly-interacting massive particles -- are affected by a complex astrophysical background of cosmic radiation. The understanding and modeling of such background requires a deep comprehension of an intricate classical plasma physics problem, i.e. the interaction between high-energy charged particles, accelerated in peculiar astrophysical environments,  and magneto-hydrodynamic turbulence in the interstellar medium of our Galaxy. 

In this review we highlight several aspects of this exciting interplay between  the most recent claims of DM annihilation/decay signatures from the sky and the Galactic cosmic-ray research field. 
Our purpose is to further stimulate the debate about viable astrophysical explanations, discussing possible directions that would help breaking degeneracy patterns in the interpretation of current data. We eventually aim to emphasize how a deep knowledge on the physics of CR transport is therefore required to tackle the DM indirect search program at present and in the forthcoming years.
\end{abstract}

\maketitle

\section{Exotica: Where to find them?}

The particle dark matter (DM) \cite{Bertone:2010zza} discovery may potentially undertake a different path than the beaten track of collider searches \cite{Abdallah:2015ter} and direct detection experiments \cite{Undagoitia:2015gya}.  
Early Universe thermal relics, in particular, may be well-motivated DM candidates \cite{ellis1984,jungman1996,Olive:2003iq} expected to annihilate (or decay) even in today galactic halos, producing Standard Model (SM) particle yields. Therefore, the measurement of charged particle and gamma-ray fluxes of cosmic origin in a wide energy range -- say from few MeV all the way up to the multi-TeV domain -- may be a quite unique tool at our disposal in order to probe the putative particle nature of DM \cite{Maurin:2002ua,Strigari:2013iaa}. 

It is widely recognized that the indirect extraction of a DM signal in this context is an extremely challenging task. The DM problem in general and the indirect searches in particular have been already presented in comprehensive review papers (see e.g. \cite{Bergstrom:2000pn,bertone2005}). Here, we wish to focus our attention on the important interplay between particle DM signatures and the background signals expected from astrophysics, discussing in particular the phenomenological relevance of cosmic-ray physics.

As far as charged particles are concerned, the first consideration in order is that high-energy protons, nuclei and electrons are injected in copious amounts as cosmic rays (CRs) by different classes of astrophysical sources (such as shocks associated to supernova explosions or super-bubbles, or possibly accretion-powered mildly relativistic jets), which provide a huge and irreducible background. 
On the other hand, the paucity of anti-particles, mostly produced by secondary interactions of CRs, could in principle drastically improve the signal-to-noise ratio in favor of a putative DM detection, within a rather low expected astrophysical background; this possibility of detecting early Universe relics by studying Galactic antiparticles was first outlined in the early 1980s in several pioneering papers (e.g. \cite{silk1984,stecker1985}) and has been studied in much larger detail in particular during the last decade, mainly thanks to the dramatic improvement in the quality of the data provided by PAMELA \cite{pamela2006} and AMS-02 \cite{ams2012} experiments. For a recent discussion on cosmic antimatter opportunities, see for example \cite{Reinert:2017aga,Blum:2017iwq}.

Despite the low background, even in the case of antimatter searches the large uncertainties involved in the modeling of both conventional astrophysical production and Galactic transport play a major role and have hindered a firm DM detection so far, although some recent tentative claims (in particular, among others, \cite{PAMELApositron, AMSpositron, AMSantiproton,Cuoco:2016eej,Cuoco:2017okh}) have triggered an important debate in the community. 
In sections~4,5,6 we will describe in detail the current status for positron, anti-proton and antinuclei indirect searches: Our purpose is to provide a case-by-case discussion mostly focused on the relevance of CR transport physics in antimatter channels for DM indirect detection.

While charged antiparticles may be promising indirect messengers of the particle DM nature, they do not retain the directionality from their emission point and, hence, cannot provide the morphological characterization of a DM signal: This possibility is accomplished instead with the analysis of gamma-ray data (for early studies, see e.g. \cite{bergstrom1988, rudaz1989, giudice1989}; more recently, \cite{bergstrom1998,Bringmann:2012ez}).
If DM particles eventually decay or annihilate into gamma radiation, it is crucial to identify the most promising regions of the sky where either the expected signal is large, or the astrophysical background is low: Among the most important targets, we can certainly mention the inner Galaxy (which satisfies the first criterion), and the DM-dominated satellite galaxies orbiting around our Galaxy. The most tentative claims and interesting bounds from the study of the gamma-ray sky will be reviewed in detail in section~7.

Before going through an extensive discussion of all these channels for DM indirect searches, in the next two sections we will set up the stages of this review by briefly presenting some relevant aspects of DM models connected with the scope of the present paper, and then highlighting the several key aspects of the Galactic CR transport problem. Those concepts will be recalled all across the paper when the most relevant interpretations of CR and gamma-ray anomalies are discussed.


\section{Targeting DM indirect searches on High Energy Physics}

The quest for the fundamental origin of DM may require to consider a priori an impressive range of energy scales (see \cite{Bertone:2016nfn, deSwart:2017heh} for interesting historical retrospectives). For instance, sitting on the extremes of the viable mass window for DM searches, today we may be looking for imprints on the cosmological matter power spectrum of super-light candidates such as axion-like particles \cite{Hu:2000ke,Marsh:2015xka,Hui:2016ltb} from the string landscape \cite{Svrcek:2006yi,Arvanitaki:2009fg}, as well as aiming to detect the gravitational echoes of massive black hole merging \cite{Abbott:2016blz,Bird:2016dcv}, possibly originated from primordial density fluctuations in the early Universe \cite{1975Natur.253..251C,Carr:2016drx}. 

However, the phenomenology of DM candidates may be intriguingly correlated to the long-standing puzzles pertaining to the realm of the SM (see, e.g., \cite{bertone2005,Feng:2010gw} for broad reviews on the topic). Of particular significance, one of the main driving forces of research on High Energy Physics has been the quadratic UV sensitivity of the Higgs boson mass to any New Physics energy scale above the electroweak one \cite{Giudice:2008bi}. On general grounds, the attention of this review is mostly paid to DM candidates motivated by New Physics at the electroweak scale. Many extensions of the SM theory addressing the electroweak hierarchy problem can indeed accommodate such DM candidate in their spectrum; see, e.g., \cite{jungman1996} for a review on the widely mentioned case of supersymmetry; \cite{Regis:2006hc,Panico:2008bx} for higher dimensional theories aimed at stabilizing the electroweak scale; \cite{Ryttov:2008xe,Belyaev:2010kp,Frigerio:2012uc,Bruggisser:2016ixa} for examples of strongly coupled UV completions. Moreover, DM candidates sitting around the electroweak scale may also be well-motivated in the context of non-natural theories addressing other possible SM issues, such as proton stability and/or gauge coupling unification; see e.g. \cite{Giudice:2004tc, ArkaniHamed:2004yi,Masiero:2004ft} for the case of supersymmetry, \cite{Servant:2002aq,Agashe:2004ci,Agashe:2004bm,Hooper:2007qk} for interesting universal/warped extra-dimensional proposals. 

In all these scenarios, the DM particle is usually stable due to its charge under a (discrete) symmetry of the new theory, while a benchmark range of masses and couplings can be eventually individuated on the basis of the UV guiding principles. Most importantly, the emerging DM phenomenology from these studies typically falls in the experimental window of sensitivities for antimatter and gamma-ray searches discussed in the next sections. 

Marginalizing over the specific details of UV models, the DM reference framework we mainly refer to, in this review, corresponds to early Universe cold thermal relics. Assumed to be in thermal equilibrium with the primordial plasma at the very early stages, DM decoupling as non-relativistic species eventually leads to \cite{Gondolo:1990dk} (for more details, see also \cite{Profumo:2017hqp}):
\begin{equation}
\frac{\Omega}{0.25} \sim \frac{3 \times 10^{-26} \, \textrm{cm}^{3} \, \textrm{s}^{-1}}{\langle \sigma v \rangle} \simeq \frac{10^{-8} \, \textrm{GeV}^{-2}}{\sigma} \ .
\label{eq:thermal_relic}
\end{equation}
$\Omega$ is the DM cosmological relative abundance observed today \cite{Ade:2015xua}, $\langle \sigma v \rangle$ is the DM thermally-averaged particle annihilation cross section and in the last step we make use of the approximation $\langle \sigma v \rangle \simeq \sigma \, c/3$.
By means of dimensional analysis, we can naively estimate the DM annihilation cross section in terms of its mass $m$ and dimensionless coupling constant $g$, obtaining:  
\begin{equation}
m \sim \left(\frac{g^{2}}{10^{-1}}\right)  \left(\frac{10^{-4} \, \textrm{GeV}^{-1}}{\sqrt{\sigma}}\right) \ \textrm{TeV} \ ;
\label{eq:WIMP_mass}
\end{equation}
Eq.~(\ref{eq:thermal_relic}) in conjunction with the estimate in Eq.~(\ref{eq:WIMP_mass}) characterizes the so-called \textit{Weakly Interacting Massive Particle (WIMP) miracle}: A cold thermal relic charged under weak interactions, and with mass close by the electroweak scale, naturally accounts for the present DM abundance.

The WIMP mass range may be bracketed from below according to the seminal works in \cite{Vysotsky:1977pe,Lee:1977ua} as $m \gtrsim 10$~GeV, while from the above unitarity arguments on the DM cross section pinpoint to $m \lesssim 10^{2}$~TeV \cite{Griest:1989wd} (but caveats exist \cite{Harigaya:2016nlg}). Typical expectations from the WIMP paradigm may be notably disregarded by the presence of resonant regimes, near mass thresholds, co-annihilation with other particles in the thermal bath \cite{Griest:1990kh}. Many other examples of variants to the standard WIMP freeze-out scenario have been investigated in literature \cite{Carlson:1992fn,Pospelov:2007mp,Hochberg:2014dra,DEramo:2010keq,DAgnolo:2015ujb,Pappadopulo:2016pkp,Dror:2016rxc,DAgnolo:2017dbv,DEramo:2017gpl}. Non-perturbative effects such as Sommerfeld enhancement \cite{ArkaniHamed:2008qn,Feng:2009hw,Feng:2010zp,Hryczuk:2010zi} and bound-state formation \cite{Shepherd:2009sa,vonHarling:2014kha,An:2016gad,Cirelli:2016rnw,Mitridate:2017izz} have also more recently acknowledged to be of possible dramatic impact in the broad context of WIMP phenomenology.

Importantly, Eq.~(\ref{eq:thermal_relic}) shows that the DM relic abundance is mostly sensitive to the annihilation cross section: It follows that a WIMP-less miracle is perfectly conceivable \cite{Feng:2008ya}, opening on general grounds a broader range of viable mass scales and couplings for the phenomenology of DM thermal relics. Therefore, indirect signals from WIMP-like scenarios -- whose signatures have been comprehensively inspected in \cite{Cirelli:2010xx,Buch:2015iya} and are of particular importance for this review -- are after all: \textbf{1)} intimately connected to the possibility of being effectively visible today (in particular, non exhibiting important velocity suppression); \textbf{2)} possibly connected to the tantalizing discovery of New Physics near the electroweak scale \cite{Arcadi:2017kky}.

\section{Mini-guide to Galactic CR physics}

\subsection{Origin of CRs}

A clear identification of the classes of astrophysical sources able to accelerate particles from GeV all the way up to PeV energies (and, for extragalactic accelerators, up to $\sim 10^{20}$ eV) is crucial for DM indirect detection. 

We will not digress here on the long debate about the origin of CRs and their acceleration mechanisms (see e.g. \cite{bell2013} for an excellent review).
For the purpose of this review, let us start by mentioning the {\it supernova paradigm} as the main guideline. Supernova remnants (SNRs) were proposed as potential sources out of energy budget arguments in \cite{baade1934,baade1934PR}; the picture was better defined later in terms of SNRs located in our own Galaxy \cite{morrison1957,ginzburg1956}; however, a physical process capable of such a powerful CR acceleration had not been proposed yet at that time.  
Later, the theory of {\it diffusive shock acceleration} was presented in four famous seminal papers \cite{blandford1978,bell1978,axford1977,krymskii1977}, and is currently considered the main reference framework in the field. 

However, it is important to keep in mind that other classes of sources (e.g. pulsar wind nebulae for leptonic CRs \cite{aharonian1995}, OB associations \cite{murphy2016}, X-ray binaries \cite{heinz2002}), and several other acceleration mechanisms, have been proposed as well. As we will see below, many of these potential accelerators are expected to play a role in the explanation of some tentative claims of DM detection.

\subsection{CR transport: Preliminary considerations}

Let us now turn our attention to a crucial aspect of CR physics, which has an extremely relevant impact on DM indirect searches, i.e. CR propagation in our Galaxy.

The usual starting point is a collection of several key observations that characterize the cosmic-ray flux.

\begin{itemize}
\item The isotropy of the arrival direction (at the level of $\sim 10^{-3}$ in the TeV - PeV range, recently measured with high precision by many experiments \cite{anisoKASCADE,anisoMILAGRO,anisoEASTOP,anisoTIBET,anisoIceTop,anisoIceCube}); 
\item The much larger abundance of Lithium, Beryllium and Boron compared to the solar system abundances, which is naturally interpreted as the  signature of the interaction of {\it primary} species such as protons and heavy nuclei with a column density of interstellar gas as large as few $g/cm^2$: Such a {\it grammage} implies that the primary species have crossed the Galactic disk many times; 
\item The presence of a diffuse gamma-ray emission across the whole Galactic disk, already predicted in the early 1960s \cite{pollack1963}, and first measured by pioneering satellite experiments such as OSO-3 (1967) and SAS-2 (1972); nowadays, Fermi-LAT has provided a state-of-the art description of this emission in the $300$ MeV - $300$ GeV range, as detailed below.
\end{itemize}

These pieces of information, combined together, suggest a ``conventional scenario'' for CR transport that was shaped by the pioneering work of Ginzburg and colleagues (see \cite{ginzburg1964} and references therein, and \cite{Berezinskii1990}), based on a random walk through the Galaxy governed by the quasi-linear theory of pitch-angle scattering on Alfvénic turbulence (QLT), first presented in the 1960s \cite{Jokipii1966,Jokipii1968}.

\subsection{The CR transport equation}

Magneto-hydrodynamic turbulence -- which is ubiquitous in the interstellar medium (ISM) and covers a very wide range of scales from astronomical units (AUs) to kpc \cite{armstrong1995} -- is widely considered as the main responsible for this diffusive regime. 
In more detail, the relativistic motion of charged particles in our Galactic environment is affected by the presence of both a coherent large-scale magnetic field component, $\vec{B}$ \cite{farrar2012,ferriere2017}, on top of which magnetic inhomogeneities, $\delta \vec{B}$ are propagating. 
These fluctuations in the magnetized interstellar medium (ISM) are associated to a turbulent cascade that is believed to be either initiated at large scales $\sim 10^2$ pc (by supernova explosions, differential rotation of the Galactic disk, or other mechanisms \cite{scalo2004}), or (especially at small scales) triggered by CR themselves via {\it streaming instability} \cite{cesarsky1980,blasi2012}. 
This cascade has been usually considered, in the basic scenario of QLT, as isotropic and mainly composed of \textit{Alfv\'en waves}, i.e. transverse magnetic perturbations moving at the Alfv\'en speed\footnote{The Alfv\'en speed is supersonic in most typical ISM environments: hot HII regions, warm intercloud gas, molecular gas, with $\beta \equiv (v_s/v_A)^2 \simeq 0.1 \div 0.3$ everywhere \cite{Ptuskin2006}.}:

\begin{equation}
v_{A}  \simeq \, 2 \times 10^{6} \ \frac{|\vec{B}|}{\mu {\rm G}} \,\sqrt{\frac{ {\rm cm}^{-3} }  
{\rho_{\textrm{\tiny ISM}}}} \ \rm{cm} \, \rm{s}^{-1}  \ . 
\label{eq:vA}
\end{equation}

Following in part the approach of \cite{Blasi2013Rev}, we recap here the main features of QLT (see also~\cite{ginzburg1964,Berezinskii1990}).

The rationale of QLT is to consider the interaction of a charged particle of momentum $\vec{p} = m \vec{v}$ with magnetic inhomogeneities $\delta \vec{B}$ that are sufficiently small (with respect to the regular field $\vec{B}$) at the scale of interest.
The process is well described by a stochastic equation for the {\it pitch angle}, defined as $\mu = \cos(\hat{p} \wedge \vec{B})$. On average the variance of the pitch angle can be shown to feature a {\it resonance condition} \cite{Berezinskii1990}. According to it, the particle only interacts with the inhomogeneities of wavelength $\sim 2\pi / k$ matching the particle Larmor gyroradius $r_{L}$\footnote{As a reference, the reader can keep in mind that the energy scales from GeV to PeV, characteristic of Galactic CRs, resonate with scales from $\mathcal{O}$(AU) to $\mathcal{O}$(pc).}:

\begin{equation}
\langle \frac{\Delta \mu \Delta \mu}{\Delta t} \rangle \,=\, \frac{\pi \, v}{\mu r_{L}^{2}} \, \frac{|\vec{\delta B}|^2}{|\vec{B}|^2} \, (1-\mu^{2}) \, \delta\left( k - \frac{1}{ \mu  r_{L}} \right) \ .
\end{equation}

Let us now consider an ensemble of particles described by a phase-space distribution ${f}(\vec{x},\vec{p},t)$, with probability density $\Psi(\vec{p},\Delta \vec{p}\,)$ for transitions $\vec{p} \to \vec{p} + \Delta \vec{p}$ in momentum space, due to interactions with stochastic fluctuations in the magnetized environment. We can state that, after a lapse $\Delta t$, {\it in the Alfv\'en wave rest frame} (primed), the evolved phase-space distribution must be equal to:

\begin{eqnarray}
\label{eq:phaseSpaceEvolution}
&{f}&(\vec{x}\, '+ \vec{v}\, ' \cdot \Delta t,\vec{p}\, ',t+ \Delta t) = \nonumber\\
&\,=\,& \int d \Delta \vec{p}\, ' \, \Psi(\vec{p}\, ' -\Delta \vec{p}\, ',\Delta \vec{p}\, ' \,) \, {f}(\vec{x}\, ',\vec{p}\, ' -\Delta \vec{p}\, ',t) \ 
\end{eqnarray}

being $\vec{v}\ '$ the CR particle velocity in the wave frame. 

We assume that detailed balance holds, i.e. a transition $\vec{p}\, ' \rightarrow \vec{p}\, '-\Delta \vec{p}\, '$ described by a probability $\Psi(\vec{p}\, ',-\Delta\vec{p}\, ')$, is equivalent to the one described by $\Psi(\vec{p}\, '-\Delta\vec{p}\, ',\Delta\vec{p}\, '\,)$.
Applying this principle in the limit of $\Delta p\, ' / p\, ' \ll 1$ (originating from $|\delta \vec{B}\,| / |\vec{B}\,| \ll 1$) that characterizes QLT, we can write:

\begin{equation}
\label{eq:Delta_p_variance}
\langle \Delta p_{i} ' \, \rangle_{\Delta \vec{p}\, '}  = \frac{1}{2} \sum_{j} \frac{\partial}{\partial p_{j} '} \langle \Delta p_{i} ' \Delta p_{j} ' \rangle_{\Delta\vec{p}\, '} \ ,
\end{equation}

where $\langle  \dots \rangle_{\Delta \vec{p}\, '} \equiv \int d \Delta \vec{p}\, ' \ \Psi(\vec{p}\, ',\Delta \vec{p}\, '\,) $, and a Taylor expansion of $\Psi(\vec{p}\, ',\Delta \vec{p}\, '\,)$ has been performed.

Assuming the static limit, i.e. $\Delta t /t \ll 1$, starting from a Taylor expansion of Eq.~(\ref{eq:phaseSpaceEvolution}), with the help of Eq.~(\ref{eq:Delta_p_variance}) we finally get, in the same wave frame:

\begin{equation}
\frac{\partial {f}}{\partial t} + \vec{v}\, ' \, \frac{\partial {f}}{\partial \vec{x}\, ' } =  \frac{\partial}{\partial \vec{p}\, '}   \left( D_{\vec{p}\, ' \vec{p}\, '} \, \frac{\partial {f}}{\partial \vec{p}\, ' } \right) .
\label{eq:diff_in_p}
\end{equation}

This is a Boltzmann equation where in the right-hand side the ``collision operator'' qualifies Brownian motion in momentum space with 
\textit{diffusion coefficient}

\begin{equation}
D_{p_{i} 'p_{j} '} \equiv \frac{1}{2} \langle \frac{\Delta p_{i} ' \Delta p_{j} '}{\Delta t} \rangle_{\Delta \vec{p}\, '}  \ \ \ \textrm{for} \ i, j = 1,2,3 \, , 
\label{eq:diff_in_momentum}
\end{equation}

describing indeed the momentum isotropization due to CR stochastic scattering with Alfv\'en waves. \footnote{Equivalently, in term of pitch angle, in the direction along the regular field we can write:

\begin{equation}
\frac{\partial {f}}{\partial t} + v \mu \, \frac{\partial {f}}{\partial x} =  \frac{\partial}{\partial \mu}  \left( D_{\mu \mu} \, \frac{\partial {f}}{\partial \mu} \right) .
\label{eq:diff_in_mu}
\end{equation}
}

Let us now perform a transformation to the Galactic rest frame. 
In this frame of reference, Eq.~(\ref{eq:diff_in_p}) features a spatial diffusion operator {\it along the direction of the regular field} as well: This is the most important term that governs CR transport in the Galaxy. 
The spatial diffusion coefficient $D_{z z}$ is related to $D_{pp}$ by \cite{Skilling1975}:

\begin{equation}
D_{z z} D_{p p} \propto v_{A}^{2} \, p^{2} \ ,
\label{DxxDpp}
\end{equation}
for $\vec{B} = B \, \hat{z}\,$. Inspired and guided by the results of QLT, a general transport equation is usually considered, mainly based on the aforementioned process of diffusion in both position and momentum space, but featuring a wider set of terms associated to other physical phenomena.
The full equation reads:

\begin{eqnarray}
&\,&\frac{\partial N}{\partial t} + \vec{\nabla} \left( \vec{u} \, N \right) - \frac{1}{3} \frac{\partial}{\partial p} \left[ p \left( \vec{\nabla} \cdot \vec{u}  \right) \, N \right] - \vec{\nabla}  \left( D_{\vec{x} \vec{x}} \vec{\nabla} N \right)  - \nonumber \\
&\,& - \frac{\partial}{\partial p} \left[ p^{2} D_{pp} \frac{\partial}{\partial p} \left( \frac{N}{p^{2}}\right)\right] + \frac{\partial}{\partial p} \left( \frac{d p }{d t } \, N \right) = \nonumber \\
&\,&= Q_0 + Q_{\rm sec} - \frac{N}{\tau_{N}} \, , \ \ \ \ \ 
\label{eq:general_CR_transport}
\end{eqnarray}
with $N(\vec{x}, p, t)$ and $Q(\vec{x}, p, t)$ being respectively the CR density species and CR injecting density source per unit of momentum.
In the left-hand side, the diffusion term is usually {\it isotropic} and described by a scalar, position-independent coefficient, despite the fact that QLT predicts a highly anisotropic transport along the regular field direction (see e.g. the discussion in \cite{Cerri:2017joy}). The scalar spatial diffusion coefficient is generally taken as

\begin{equation}
D = \frac{c\, r_L}{{\cal F}(k)} \ ,
\end{equation}
where $\mathcal{F}(k)$ is defined as the (normalized) power associated to the turbulent modes with wave number $k \propto 1/p$ resonating with the particles carrying momentum $p$.
Since the turbulent power scales as a power law, the rigidity dependence of the diffusion coefficient is usually parametrized as:
\begin{equation}
D = D_0 \, \left( \frac{p}{p_0} \right)^{\delta} \ ,
\end{equation}
with $D_0$ and $\delta$ as free parameters to be fixed by comparison with CR data. The spatial dependence of such normalization, that stems from the spatial variations of the turbulent power, is usually neglected with some relevant exceptions \cite{Evoli2012,Gaggero:2014xla,Cerri:2017joy,gaggero2017prl,Tomassetti2015twoHalo,feng2016}.

The momentum diffusion is also called {\it stochastic reacceleration}, and the relation \ref{DxxDpp} is assumed to hold.

The left-hand side also involves an advection term originally present in Eq.~(\ref{eq:diff_in_p}) as well, now characterized by the bulk velocity of the plasma in the lab frame: 
Galactic winds affecting CR motion may be described by such a term, together with adiabatic energy losses, involving velocity gradients even of $\mathcal{O}(10^{2})$ km s$^{-1}$ kpc$^{-1}$ perpendicularly to the Galactic disk. 

The physics of advective-diffusive transport is enriched by two more relevant phenomena included in Eq.~(\ref{eq:general_CR_transport}): net energy losses and spallation.  In fact, we need to consider CR energy loss processes, characterized by the continuous loss rate $d p / d t$, particularly important e.g. at high energies for light charged species such as leptons (see e.g. \cite{StrongMoskalenko2007} and \cite{Evoli2017I}). 

Eventually, on the right-hand side of Eq.~(\ref{eq:general_CR_transport}):
\begin{itemize}
\item The primary source term $Q_0$ captures the primary accelerators of CRs: As mentioned above, while the {\it supernova paradigm} is still the most accredited one, other classes of sources can certainly be at work; 
\item The secondary source term $Q_{\rm sec}$ describes the production of a given species from spallation of the heavier ones onto interstellar gas;
\item A loss term due to inelastic collisions characterized by an interaction time $\tau_{N}$ is also introduced.
\end{itemize}

An important remark is needed at this point. Our picture of MHD turbulence has dramatically improved during the latest decades: According to the current scenarios \cite{GS1994,GS1995}, MHD turbulence is composed of an {\it anisotropic} cascade of both Alfv\'en waves, and isotropic fast magnetosonic modes, as theoretically demonstrated and numerically confirmed by several simulations. As a consequence of the anisotropy of the Alfv\'enic cascade, the scattering efficiency on Alfvén waves turns out to be very low \cite{Chandran2000}. posing a tough challenge to the whole scenario discussed above. Among others, a possible solution \cite{Yan2002,Yan2008} is that magnetosonic modes dominate gyroresonance interaction for most of the pitch-angle range. 

However, although the actual microphysics underlying the CR random walk is still far from being exhaustively addressed, the QLT can still be considered a useful guideline to be taken as a reference, and Eq.~(\ref{eq:general_CR_transport}) should be understood as a phenomenological tool to tame the complexity of the plasma physics problem, allowing us to make predictions against a plethora of data.

\subsection{Modeling CR transport: A glimpse}

In order to solve the complicated CR transport equation (for each CR species), today we have at our disposal several public numerical codes, equipped with different numerical and astrophysical ingredients, aimed at solving Eq.~(\ref{eq:general_CR_transport}), most notably (in chronological order): \texttt{GALPROP}\cite{Galprop1,Galprop2,Galprop3,2001ICRC....5.1942S}, \texttt{DRAGON} \cite{Evoli:2008dv,Gaggero:2013rya,Evoli2017I,Evoli2017II}, \texttt{PICARD} \cite{Picard1,Picard2}. A semi-analytical approach is instead followed by the \texttt{USINE} project \cite{Usine}. 


While a detailed and realistic study of Galactic CR propagation requires the extensive use of those numerical or complex semi-analytical methods, we can extract some physical insight useful for the next sections looking at a simplified version of Eq.~(\ref{eq:general_CR_transport}). 

Indeed, at the basis of CR transport may be conceived an important hierarchic game of  scales: convective and re-acceleration effects are typically related to low-energy regimes, while energy-loss rates are negligible for high-energy hadrons such as protons or heavier nuclei. So, for energies $E \gtrsim$ few GeV, the Galactic motion of heavy species can be approximately described to be in a purely diffusive regime: Then, we may trade the spatial diffusion operator for an effective time of confinement, $\tau_{D}$, i.e. 
\begin{equation}
\vec{\nabla}  \left( D_{\vec{x} \vec{x}} \vec{\nabla} N \right) \, \longrightarrow \frac{N}{\tau_D} \ ,
\end{equation}
and treat the Galaxy as a box where CRs perform a random walk up to the box boundaries, beyond which they leak out.  In the steady-state limit of this simplified framework, CR secondary species, produced by the interaction of source-injected (namely, primary) CR particles,  can be estimated as:

\begin{equation}
N_{s} = Q_{s}  \, \tau_{D} \propto  N_{p} \, \tau_{D}  .
\label{eq:leaky_box_sol_sec}
\end{equation}

Hence, the timescale for confinement of CRs in the Galaxy is intimately linked to relative abundances of secondary and primary species. As anticipated in the above, anti-matter species are typically produced as secondaries, and therefore Eq.~(\ref{eq:leaky_box_sol_sec}) is of direct relevance for DM indirect searches in anti-matter channels.  

Note that by dimensional analysis, $\tau_{D} \sim H^{2} / D$, where $H$ captures the typical size of the box. Therefore, secondary fluxes are sensitive to both the spatial diffusion coefficient and the height of the CR propagation halo. As we will discuss below, our current poor knowledge of the value of $H$ generally translates into an important source of uncertainty in DM indirect detection studies.

As stated above, the spatial diffusion coefficient scales as a power law with rigidity. Supplementing Eq.(\ref{eq:leaky_box_sol_sec}) with such expectation, we have a theoretical prediction that nicely fits the trend of available experimental data \cite{PhysRevLett.117.231102}: at energies above few GeV, the measurements of local CR secondary-over-primary observables like B/C are indeed compatible with a power-law behavior (with index $\delta$) of the diffusion coefficient.  
Typical estimates of this parameter are in the range $0.3 < \delta < 0.6$ \cite{trotta2011,trotta2016,Evoli:2015vaa,yuan2017ams,niu2017}, with a normalization at GeV corresponding to $D_{0} \simeq 10^{28}$ cm$^{2}$ s$^{-1}$, but pertain only to a local measurement (see \cite{Gaggero:2014xla} for a possible indirect inference of $\delta$ across the Galaxy).


\section{The antiproton channel}\label{sec:pbar}

\begin{figure}
\centering{
\includegraphics[scale = 0.32]{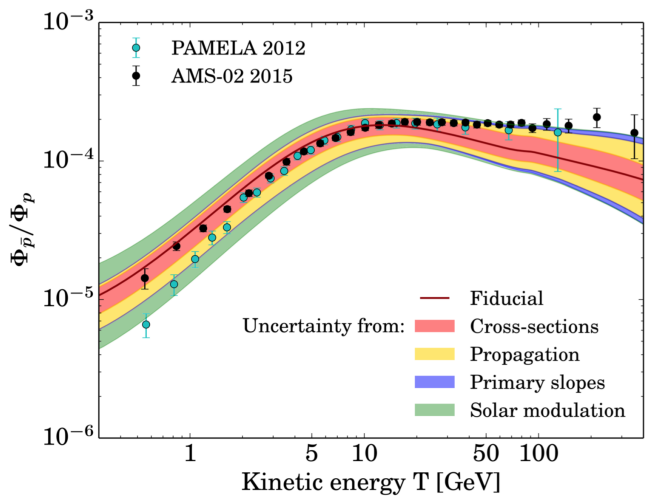}
}
\caption{Impact of cross-section, CR transport, and modulation uncertainties on the conventional predictions for the antiproton compared to the related dataset collected by the AMS-02 collaboration. Credit to \cite{giesen2015}, Fig.~2.}
\label{fig:antiproton}
\end{figure}

In the conventional scenario sketched above, antiprotons are produced in the Galactic environment by {\it spallation} of heavy nuclei and protons onto interstellar gas. 

The early measurements of the antiproton flux date back to the 1970s and early 1980s; a first tentative claim of anomaly with respect to the conventional expectations based on the picture of nuclear spallation goes back to \cite{buffington1981}.
However, DM connections were not outlined at that time: Cosmic antiprotons were considered a promising channel for DM searches only some years later, in several seminal papers (see, for instance, \cite{silk1984,ellis1988, rudaz1988} and, more recently, \cite{bergstrom1999}).

A dramatic improvement in the accuracy of the data was provided by the PAMELA collaboration in 2009 \cite{pamela2009pbar} (with further refinement in \cite{pamela2010pbar}): The measurement showed a reasonably good agreement with conventional models based on purely secondary origin, as confirmed by  \cite{dibernardo2010pbar} and \cite{trotta2011pbar}.
A note here is in order about the meaning of {\it conventional model}: In what follows, we will use this expression for a model based on the simplest version of Eq.~(\ref{eq:general_CR_transport}), taking a single class of sources (SNRs) at work, with antiparticles produced only as secondary products from primary spallation, and featuring constant and homogeneous diffusion, tuned on local CR data.

Given the absence of significant unexpected spectral features (such as bumps), the potential constraining power of PAMELA dataset for DM searches was soon demonstrated in a series of papers (from the early ones as \cite{bringmann2007,donato2009} to the more recent \cite{cirelli2014}), which provided a comprehensive discussion on the upper bound on the WIMP annihilation cross section from the detection of cosmic antiprotons. The most relevant point made in those papers is the crucial role of CR transport. 
A major source of uncertainty, in particular, is the size of the diffusive halo $H$ introduced in the previous section, i.e. the volume where Galactic CRs are effectively confined by the presence of a turbulent magnetic field. Models of CR transport based on larger diffusion halos usually feature larger average values of the diffusion coefficient in order to correctly reproduced the secondary/primary ratio data: Therefore, in these scenarios, the antiproton flux probe a larger region of the Galaxy, and exhibits a larger constraining power. On the other hand, cases where one assumes a very thin halo (smaller than $\simeq 2$ kpc) turn out to be much less restricting on the particle DM properties indirectly probed.

Another source of uncertainty certainly lies on the properties of CR transport in a much smaller environment: the Heliosphere. We refer to \cite{cirelli2014} and references therein for a comprehensive discussion on this aspect.

As a consequence of this complicated puzzle, it was not possible to firmly exclude some relevant tentative DM claims made in other channels (see the gamma-ray section for more details), and the most severe limitation came from the poor constraints we actually have on the size of the Galactic CR diffusion halo.
A possible improvement in this direction may come from more accurate measurements of the Beryllium isotopes, and from a careful analysis of the current and forthcoming data on the vertical profile (with respect to the Galactic plane) of the synchrotron emission from the Galaxy (following \cite{strong2011,DiBernardo:2012zu}, we point out that the current data seem to favor large values for the size of the diffusion halo).

In 2015 a much more accurate dataset was published by the AMS-02 collaboration \cite{AMSantiproton}. The debate on the antiproton channel has then included tentative claims of anomalies with respect to the conventional  scenario, possibly explained in terms of DM indirect detection.

First of all, the AMS-02 collaboration itself initially claimed the presence of an excess at high energies over 100 GeV. Right after, the significance of this anomaly was better characterized \cite{giesen2015,evoli2015pbar}, pointing only to a mild overshooting of the expected background. The relevance of this putative discrepancy an the estimated impact of the different sources of uncertainty on the model predictions is well depicted in Fig. \ref{fig:antiproton} (taken from \cite{giesen2015}).\footnote{We outline in particular the relevant role of the cross-section uncertainties (see also \cite{Evoli2017II} and references therein). In this regard, there has been a remarkable activity in the latest years both concerning semi-empirical parametrizations (tuned on experimental datasets) and Monte Carlo event generators: As far as the former category is concerned, new models have been proposed based on the data provided by the NA49 and BRAHMS collaborations \cite{Duperray:2003bd,diMauro:2014zea,Kappl:2014hha,Winkler:2017xor}; concerning the latter, several codes (e.g. EPOS 1.99 \cite{PhysRevC.92.034906}, SIBYLL \cite{Engel:1999db}, and QGSJET-II-04 \cite{PhysRevD.83.014018}) have been recently tuned to LHC data (see e.g. \cite{PhysRevD.83.014018,PhysRevC.92.034906}).}

Taking at face-value the original tentative claim from AMS, the interpretation of the excess requires scenarios beyond the conventional on  of CR transport pictured in the mini-guide of section~3. For instance, a mechanism that may be at work and explain the discrepancy is the {\it secondary production at the accelerator}: The idea, proposed before AMS data in \cite{blasi2009} as a possible explanation to the positron ratio anomaly (see section~6), is that secondary products of hadronic interactions inside the sources can participate in the acceleration process and subsequently escape into the interstellar medium as an extra component featuring a very flat spectrum. 
In \cite{mertsch2014} this scenario was applied to (pre-AMS) antiproton data as well, and the authors demonstrated that the boron-over-carbon ratio has much constraining power for this interpretation. 

More recently, refined Bayesian analyses that include the official AMS data on this observable as well (see e.g. \cite{yuan2017ams}) confirm the presence of a mild discrepancy between the regions of the parameter space pointed by AMS antiproton and B/C data. 

However, spatial-dependent diffusion setups (e.g. the phenomenological two-zone models as those considered in \cite{feng2016prd} and \cite{Guo:2018wyf}, designed to capture both CR transport in pre-existing SNR-driven Galactic turbulence, and confinement by CR-driven turbulence via streaming instability) seem to solve the discrepancy, as well as the latest scenarios that include secondary production at the accelerator \cite{cholis2017}.

DM interpretations for the high-energy discrepancy are also still viable for quite large values of the DM particle mass (as shown, e.g., in \cite{lin2015dm,huang2017,feng2017}), in particular for light mediator scenarios. At the same time, the constraining power of AMS data in the energy range where no relevant feature or anomaly is present has been most recently explored in \cite{lin2017prd,cuoco2017prl}.

On the other hand, on the low-energy side, a possible indication of a DM signal for DM masses near 80 GeV has been found \cite{cuoco2017prl,cuoco2017jcap,cui2017prl}, with a hadronic annihilation cross section close to the thermal value: Interestingly, this tentative claim is compatible with the DM interpretation of the Galactic center gamma-ray excess (see section~7). Again, more investigation of the transport uncertainties (both in standard and beyond-standard scenarios), and more detailed combined studies of this signal together with constraints from other probes (e.g., the observation of dwarf spheroidals in the gamma-ray band) will be crucial in order to confirm the existence of this anomaly.


\section{The avenue for Antinuclei}\label{sec:antinuclei}

A milestone campaign for imprints of particle DM on the observable CR radiation may correspond to the discovery of Galactic light antinuclei such as antideuteron ($^{2}\overline{\textrm{H}}$) \cite{Donato:1999gy,Baer:2005tw,Duperray:2005si,Donato:2008yx} and antihelium-3 ($^{3}\overline{\textrm{He}}$) \cite{Carlson:2014ssa,Cirelli:2014qia}. At present, no compelling evidence for a detection of antimatter with mass number $A \geq 2$ has been experimentally corroborated in the measurement of Galactic CR fluxes. A notable upper limit on the flux of cosmic antideuteron has been set by the BESS facility -- reporting at the 95\% of confidence level $\Phi_{^{2}\overline{\textrm{H}}} \lesssim 2 \times 10^{-4}$ (m$^{2}$~s~sr~GeV/n)$^{-1}$ for kinetic energy per nucleon $0.17 \leq \textrm{T} \leq 1.15$ GeV/n \cite{Fuke:2005it} -- while the BESS-Polar collaboration currently constrains antihelium-to-helium flux to be smaller than 10$^{-7}$ in the interval of probed rigidities, 1.6~--~14~GV \cite{Abe:2012tz}.

Today, the AMS-02 mission is operating in the direction to perform the first historical observation of cosmic light antinuclei, with promising projected sensitivities \cite{Aramaki:2015pii}. A tentative claim of few antihelium events -- possibly measured by the AMS-02 collaboration -- might be already at hand, waiting for a firmer experimental response in the upcoming years; see, for instance, ref.~\cite{Coogan:2017pwt}.

The phenomenological relevance of the avenue for antinuclei in relation to DM indirect searches stems from the kinematics of spallation processes producing CR secondaries: the energy threshold associated to the production of one antideuteron is roughly 2.5 greater than the one required to produce a secondary antiproton. Moreover, the same energy threshold is a monotonic increasing function of the mass number $A$. Therefore, the low energy flux of cosmic antinuclei is expected to be small \cite{Chardonnet:1997dv}, opening a \textit{low-energy window}  -- related to kinetic energies per nucleon between $\sim$0.1 and few GeV/n -- for large signal-to-background ratios from exotica.

The physics of CR accelerators and of propagation impacting the prediction for antinuclei fluxes at Earth sits on the very similar grounds of the one discussed for the antiproton channel. On the one hand,
the unknowns stemming from Galactic CR propagation affecting the predicted Galactic fluxes for antideuteron and antihelium  should be correlated to the Galactic antiproton spectrum and boron-to-carbon ratio data. On the other one, the injection of antinuclei in the Galactic interstellar medium is intimately connected to the antiproton production mechanism. As discussed in the previous section, the latter may be sourced from in situ acceleration in the downstream region of supernova remnant shockwaves. Such mechanism generally yields a source term distinguished into a A-component, referred to CR secondaries accelerated by the shock, and a B-component, related to standard spallation processes sourced by CR primaries. Most recent analyses on the topic have included also these extra contributions in their predictions \cite{Herms:2016vop,Tomassetti:2017izg,Korsmeier:2017xzj}. 

\begin{figure}
\centering{
\includegraphics[scale = 0.36]{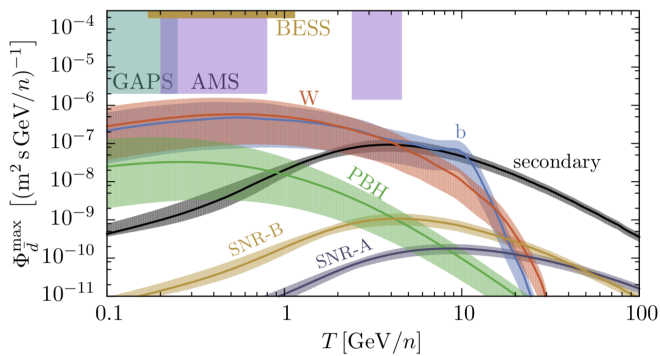}
}
\caption{Maximum $^{2}\overline{\textrm{H}}$ flux compatible with AMS-02 data from evaporation of primordial black holes (green), DM particle annihilation (blue and brown), compared with current and projected sensitivities from BESS, AMS and GAPS experiments. Maximum flux for the expected astrophysical background is also shown: in black color the contribution from CR spallation processes with the interstellar medium, and the subleading SNR-A and SNR-B ones.
Bands bracket uncertainties from force-field approximated solar modulation and an event-by-event coalescence model. Credit to~\cite{Herms:2016vop}, Figure~3.}
\label{fig:antideuteron}
\end{figure} 

Fig.~\ref{fig:antideuteron}, from \cite{Herms:2016vop}, reports the maximum contribution to the antideuteron flux with respect to the typical unknowns from CR physics, properly calibrated on antiproton AMS-02 data. The components from production in supernovae (SNR-A and SNR-B bands) are found to be subdominant, giving (at most) a 10\% effect with respect to the main standard component, obtained by considering the interactions of CR primaries with the interstellar medium (black line). Fig.~\ref{fig:antideuteron} also shows predictions for some benchmark exotica such as annihilating DM particles with masses of $\mathcal{O}(100)$~GeV and 100\% branching ratios into $b\bar{b}$ or $W^{+}W^{-}$ final states, together with the expectations from exotica that gained recent interest \cite{Abbott:2016blz,Bird:2016dcv} such as primordial black holes \cite{Kiraly:1981ci,Turner:1981ez}. The plot clearly highlights the importance in the aforementioned low-energy window. However, contrary to common wisdom \cite{Aramaki:2015pii}, Fig.~\ref{fig:antideuteron} also underlines the unlucky possibility that within the forthcoming years current and future experimental facilities may not be sensitive to antideuteron yields from DM/exotica production.

At this point, it is important to stress  that unknowns stemming from the physics of CR accelarators and Galactic transport should not be retained to be the major source of uncertainty in the prediction of cosmic antinuclei fluxes. The bands reported in Fig.~\ref{fig:antideuteron}, while including the effects of solar modulation -- typically studied within the force field approximation \cite{Gleeson:1968zza} and better investigated in \cite{Fornengo:2013osa} by means of numerical tools \cite{Maccione:2012cu,Vittino:2017fuh} -- are most importantly related to the coalescence model adopted to establish antideuteron formation. At present, the prediction of antinuclei fluxes seems to strongly depend on the assumptions made to describe antiproton-antineutron fusion \cite{Aramaki:2015pii}. Nowadays, state-of-the-art analyses can avoid to rely on simplistic analytical modeling \cite{Schwarzschild:1963zz}, making instead use of Monte Carlo event generators and available data from colliders. Large systematics on the formation of antideuteron and antihelium can be understood on the basis of the sensitivity of these studies on the hadronization model implemented and the experimental dataset considered, see e.g. \cite{Aramaki:2015pii,Lin:2018avl}. Interestingly, a recent work focused on the description of nuclear coalescence via a physical modeling for the fusion of nucleons into composite nuclei \cite{Blum:2017qnn}, and exploiting two-particle correlation measurements \cite{Lisa:2005dd}, has pointed out the possibility that the production cross section in pp collisions for antihelium-3 may have been underestimated by up to two orders of magnitude. Consequently, a putative detection of cosmic $^{3}\overline{\textrm{He}}$ events related to kinetic energies greater than 1 GeV/n may actually be within the reach of AMS-02 in few years \cite{Blum:2017iwq}.  
f
Therefore, we may conclude that searches for antinuclei, while being potentially exposed to CR propagation details, are mainly plagued by the assumptions and the systematics involved in the estimate of the poorly known production cross section. This can have a dramatic impact on our present ability to make projections for a concrete signal detection of antideuterons and antihelium-3. At the same time, these uncertainties leave us hope to foresee a spectacular discovery of exotica such as DM in the peculiar window of low-energy antinuclei events. 


\section{The positron channel}\label{sec:pos}

\begin{figure*}
\centering{
\includegraphics[scale = 0.37]{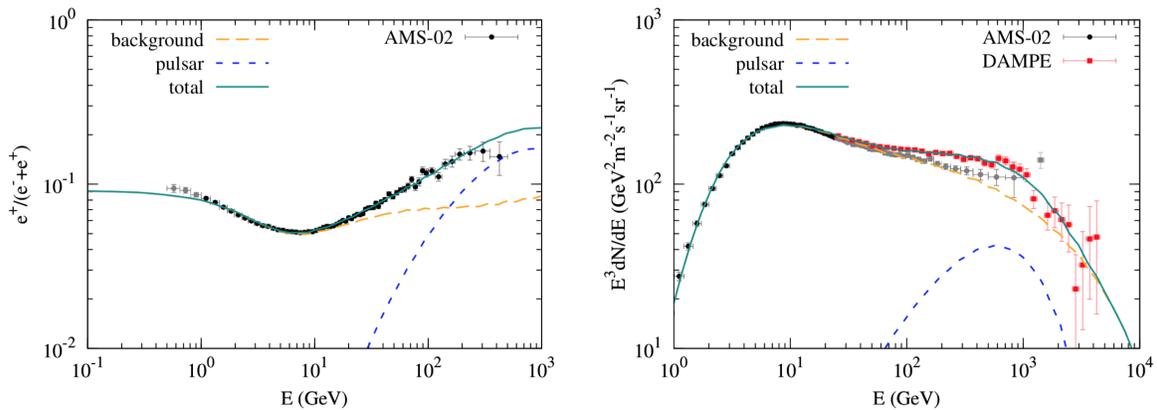}
}
\caption{The most recent leptonic data from AMS and DAMPE, interpreted within a pulsar scenario. Credit to \cite{yuan2017dampe}, Figure~2.}
\label{fig:positronDAMPE}
\end{figure*}

The positron channel has been under the spotlight for a long time in the DM indirect detection community. As mentioned in section~1, the paucity of those particles makes them an ideal target for DM searches, and the presence of a significant anomaly (with respect to the conventional expectations) has further increased the interest around this observable during the latest decade.

Let us start by clarifying, also in this case, what we mean by ``conventional'' predictions.
In the context of a simplified treatment of acceleration and transport -- as described in the mini-guide of CR physics -- a slightly different discussion is actually required for leptons in general.
In fact, high-energy leptons feature relevant energy losses especially at high energies, which implies a new timescale competing with the diffusion one. The simplified transport equation governing $e^{\pm}$ propagation is therefore of this kind:

\begin{equation}
\frac{\partial N}{\partial t} - D \, \Delta N - \frac{\partial }{\partial p} \left( \frac{d p}{d t} N \right) = Q
\label{eq:lepton_transport}
\end{equation}

where inverse Compton scattering on the photon background and synchrotron radiation define the typical timescale for energy losses, $\tau_{E_{loss}}(p)  \sim p/ |d p/dt| \propto p^{-1} $.
The solution of Eq.~(\ref{eq:lepton_transport}) is given by a Green function that boils down to:

\begin{equation}
N \simeq \frac{Q(p) \, \tau_{E_{loss}}}{\sqrt{D(p) \, \tau_{E_{loss}}}} \ .
\end{equation}
For primary electrons this result implies a scaling $N \propto$ p$^{\alpha_{e} -\frac{\delta}{2}-\frac{1}{2}}$, where $\alpha_{\textrm{CR}}$ is the CR injected spectral index.

In the standard scenarios, positrons in the Galaxy are believed to originate, like antiprotons, as an {\it entirely secondary} component arising from the collisions of relativistic protons with the ISM gas according to a chain like $p +  H \to  \dots \to \pi^{\pm} \to \mu^{\pm} + \dots \to e^{\pm} + \dots \ $. 
The source function of positrons is then expected to scale as:

\begin{equation}
q_{e^{+}}  \propto N_{p} \, n_{H} \, \sigma_{p \to e^{+}} \propto Q_{p} \, \tau_{diff} \propto p^{-\alpha_{p}-\delta} ,
\label{eq:pos_sources}
\end{equation}

where we have implemented the rather simplistic approximation of an energy independent cross section $\sigma_{p \to e^{+}}$, and used the fact that the relevant timescale for propagation of high-energy protons is the diffusion timescale, and defined $\alpha_{p}$ the spectral index for the proton injection source function. 
Plugging this result in Eq.~(\ref{eq:pos_sources}), we can predict the scaling of the propagated positron flux over the electron one:

\begin{equation}
 \frac{N_{e^{+}}}{N_{e^{-}}} \propto p^{- \alpha_{p} + \alpha_{e}  - \delta} \ , 
 \label{eq:pos_over_el}
\end{equation}

with $\alpha_{e}$ the spectral index at injection for the electron source distribution.

In the framework of diffusive shock acceleration,  the injected spectral index should not differ much among different species. Consequently, the ratio of secondary positrons over primary electrons is predicted to {\it decrease} with increasing energy, unless a (very unlikely) large difference between the source spectral indexes for protons and electrons is assumed {\it ad hoc}. 

The rise at high energy in the positron fraction originally discovered by PAMELA in 2009 \cite{PAMELApositron}, and subsequently confirmed by Fermi-LAT and AMS-02 \cite{AMSpositron} collaborations, constitutes then a substantial deviation from the standard prediction of Eq.~(\ref{eq:pos_over_el}) and appears robust with respect to uncertainties in CR transport models, implemented in a more realistic way (see, however, \cite{shaviv2009}). The release of the data on the absolute positron spectrum \cite{2014PhRvL.113l1102A} confirmed and strengthened this conclusion. 

The detection immediately triggered a debate in the community (see e.g. \cite{serpico2009,fermi2009} and references therein). 
A natural explanation in terms of nearby astrophysical accelerators of primary $e^{+} + e^{-}$ pairs, e.g. pulsar wind nebulae (already invoked in \cite{aharonian1995} as potential contributors to the leptonic flux), was soon considered as a very promising one; see Fig.~\ref{fig:positronDAMPE} for a recent realization of this scenario, compared to up-to-date experimental data. 
Other astrophysical interpretations were proposed (see, e.g., \cite{serpico2012}), including the already mentioned secondary production at accelerators \cite{blasi2009}. 

On the other hand, many DM scenarios were invoked as well: The tough challenges for model building are in this case the large annihilation cross section required to sustain the measured positron flux at high energy, and the strong constraints originating from other channels (including gamma rays, CMB, and antiprotons); we refer \cite{He2009} for an early review on the topic.

More interestingly, nowadays it is possible to challenge the widely debated pulsar hypothesis in several ways, and the uncertainties in CR transport play a major role. 

First of all, it is possible to look for an anisotropy in the arrival direction of high-energy leptons; moreover, the gamma-ray observatories may now allow to identify the emission from the leptons leaving nearby known pulsars. Along this track,  a detection of a TeV halo around Geminga has recently been reported in \cite{hawc2017}: In that paper a naive estimate of the diffusion coefficient in the vicinity of Geminga is presented, which turns out to be much smaller than the average Galactic one inferred by secondary-to-primary ratios, posing a challenge both to CR transport models and to the pulsar interpretation of the positron anomaly as well; see also the follow-up detailed discussion in \cite{giacinti2017}. 


Very recently, the antiproton and positron channels were critically re-examined in \cite{lipari2017}. In that paper it is noticed that the ratio between the positron (or antiproton) flux to the proton one is consistent with the secondary production rates in the conventional picture. Based on these considerations, the author suggests that Galactic positrons and antiprotons may have a common origin as secondaries in hadronic interactions, probably produced in the local interstellar environment, so that diffusion and energy loss do not act for enough time to leave an observable imprint on the spectrum. If confirmed, this would imply a completely different propagation scenario characterized by a much lower residence time ($\sim 1$ Myr) compared to current benchmark values: Such scenario would also accommodate the spectral break in the electron spectrum reported by H.E.S.S.\cite{Aharonian:2008aa}, but not the electron slope, which is actually steeper than the antiproton and positron one: In such alternative framework, the $e^-/p$ discrepancy could in principle be generated at source, by not-yet identified mechanisms. 

The take-home message of this discussion is that the positron channel is far from being understood, and the nature of the emission above $\simeq 30$ GeV remains mysterious. However, DM interpretations of this anomaly seems disfavored with respect to several alternative astrophysical scenarios, in particular the pulsar hypothesis. Possible avenues towards a clearer understanding of these issues are: {\bf 1)} More detailed studies of the leptonic CR anisotropy (that can in principle provide a smoking gun of the pulsar scenario, or in alternative strongly constrain the scenario itself); {\bf 2)} More data beyond TeV energies: experiments such as DAMPE\footnote{\url{http://dpnc.unige.ch/dampe/}} and CALET\footnote{\url{http://calet.phys.lsu.edu/}} are already operating, and the first results from DAMPE \cite{dampe2017nature} already showed some interesting features to be confirmed and further investigated \cite{yuan2017dampe} (as shown in Fig.~\ref{fig:positronDAMPE}); {\bf 3)} More investigations of the interplay with the high-energy gamma-ray observations, such as the TeV halo around Geminga, aimed at characterizing the diffusion properties of leptons in the vicinity of the sources; {\bf 4)} Also in this case, a better characterization of the diffusion halo size, for instance by means of analyses focused on the radio emission, following e.g. the approach of \cite{DiBernardo:2012zu,strong2011}.


\section{Gamma-ray opportunities}\label{sec:gamma}
The ubiquitous flux of high-energy CR nuclei and leptons may be able to transform the Galaxy into a huge pion factory and an efficient machine to up-scatter diffuse photons emitted by stars and reprocessed by dust grains. These processes -- namely $\pi^0$ decay and Inverse Compton scattering, with the addition of the (usually sub-dominant) bremsstrahlung emission -- yield a diffuse flux of high-energy photons from the MeV to the multi-TeV energy domain, reaching the current sensitivity of space missions such as Fermi-LAT and AGILE, and ground-based facilities such as H.E.S.S., MAGIC, VERITAS and HAWC. 

Therefore, the observable gamma-ray sky can give us today a quite unique diagnostics of CR transport far from the solar system environment. The \textit{Galactic diffuse gamma-ray emission} stemming from CR interactions with the ambient gas and radiation field constitute indeed the bulk of photons measured along the Galactic plane region \cite{Ackermann2012}. It depends on our observational knowledge of emitting targets \cite{Cholis:2011un}, namely the indirect tracing of gas column densities \cite{Pohl:2007dz,Tavakoli:2012jx} through, e.g., observed CO emissivities, or the characterization of the low-energy photon background, the so-called \textit{interstellar radiation field} \cite{Porter:2006tb}. Moreover, it crucially relies on the details about CR propagation across the Galaxy \cite{Tavakoli:2013zva}. 

In the last few years important progress has been pursued in the development of phenomenological viable models for Galactic CR propagation, able to match the observed GeV -- TeV photon data from the Galactic plane region, while reproducing local CR measurements \cite{Evoli2012,Gaggero:2014xla}. In the next future, a more systematic study of gamma-ray data in symbiosis with the analysis of local CR observables may offer us the most important chance to pin down the exact features underlying Eq.~(\ref{eq:general_CR_transport}) (see, for example, \cite{2013APh_Erl_Wolf,Cerri:2017joy}) in a data-driven fashion \cite{Gaggero:2015xza,Acero:2016qlg,Yang:2016jda,Guo:2018wyf}.

Looping over uncertainties both on the side of emitting targets and possibly also on the underlying CR transport properties is a challenging task \cite{Ackermann2012,Nezri:2012xu}. However, such an attempt is particularly welcome in order to constrain particle DM properties \cite{Tavakoli:2013zva,Charles:2016pgz,Conrad:2017pms}. 
As previously mentioned, anomalies in the gamma-ray sky may be extremely compelling for indirect DM searches. State-of-the-art N-body simulations (see, e.g., \cite{Springel:2008by, Schaller:2015mua}) predict an extended DM halo embedding and surrounding the Milky Way, with a central density peaking in correspondence to the Galactic center (GC). Then, within a scenario where DM particles pair-annihilate (or decay) eventually to gamma-ray photon yields, the GC region is likely the brightest possible target for DM indirect searches, being relatively close to us ($\sim 8$~kpc).

The kinematics of DM thermal relics annihilating today in the halo may actually give rise to peculiar photon energy spectra. In particular, gamma-ray photon lines would not have any well-known astrophysical counterpart. Therefore, mono-chromatic lines in the GeV~--~TeV range produced by DM particle pairs annihilating into two-body final state channels with one or two photons, give rise to a potential smoking-gun signature. However, in a standard scenario of electrically neutral particles \cite{Taoso:2007qk}, DM rate to mono-energetic photons will exhibit loop-suppression (see, e.g., \cite{Bergstrom:1997fh,Ullio:1997ke} for the benchmark of neutralino DM). Interestingly, a hint in favor of such a spectacular signature has been found in 2012 from dedicated analyses of Fermi-LAT data in an extended region of the Galactic Center \cite{Bringmann:2012vr,Weniger:2012tx}, showing a peak in the photon spectrum at an energy $\sim$130 GeV with significance of the excess at $\sim 3 \sigma$ level. After an optimal observational strategy has been carried out for the purpose \cite{Finkbeiner:2012ez,Weniger:2013tza}, the updated analysis from the Fermi-LAT collaboration does not support any longer the original evidence for such a spectral feature in the dataset \cite{Ackermann:2015lka}, suggesting previous claims to be related only to a statistical fluke. While current lack of detection of gamma-ray lines place important upper-limits on today's DM annihilation cross-section/decay rate into mono-chromatic photons in the GeV~--~TeV energy window \cite{Ackermann:2015lka,Abdalla:2016olq}, the search for pronounced spectral features in the gamma-ray sky remains one of the most tantalizing observational programs within the WIMP mass reach \cite{Bringmann:2012ez}.

\begin{figure*}
\centering{
\includegraphics[scale = 0.43]{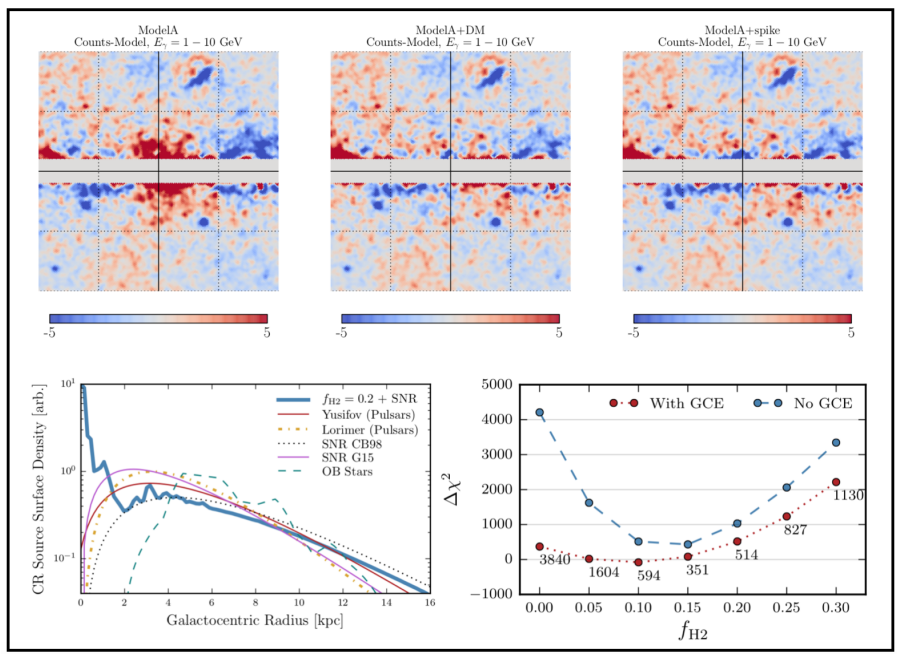}
}
\caption{Importance of the cosmic-ray contribution to the Galactic Center excess at few GeV from the analysis of Fermi-LAT data. The top three 
panels refer to the residual photon count maps shown in Fig.~3 of \cite{Gaggero:2015nsa}. In the left-bottom panel, different source distributions for primary CR injection (credit to \cite{Carlson:2015ona}, Fig.~1). The right-bottom panel shows how the quality of the fit to gamma-ray data can improve according to a better modeling of the CR primary source distribution (credit to \cite{Carlson:2016iis}, Fig.~4).}
\label{fig:GC_excess}
\end{figure*}

Interestingly enough, within almost an entire decade, an increasing number of studies focused on the GC region, repeatedly showing the existence of a statistically significant signal in Fermi-LAT data possibly correlated with spectral and morphological features of prompt gamma-ray emission from DM pair annihilation \cite{VitaleMorselli2009,Goodenough:2009gk,Hooper:2010mq, Boyarsky:2010dr,Hooper:2011ti,Abazajian:2012pn,Hooper:2013rwa,Gordon:2013vta,Huang:2013pda,Macias:2013vya,Abazajian:2014fta,Daylan:2014rsa,Zhou:2014lva,Calore:2014xka,Gaggero:2015nsa,Carlson:2015ona,Horiuchi:2016zwu,TheFermi-LAT:2017vmf,deBoer:2017sxb}. Originally, the analysis of Fermi-LAT data in 2009 concerning the innermost two degrees around the GC lead the authors of \cite{Goodenough:2009gk,Hooper:2010mq} to discover the existence of a bump-like feature in the photon spectrum exceeding the estimated astrophysical background with high statistical significance. Peaking around $\sim 3$ GeV and fitting an approximately spherical morphology, the signal found could be immediately associated to annihilating DM particles with mass range and cross-section remarkably within the WIMP ballpark. Strengthening such compelling interpretation, \cite{Hooper:2013rwa}, and successively \cite{Huang:2013pda}, focused on a region of interest (ROI) extended up to ten degrees in latitude, in correspondence to the low-latitude part of the Fermi bubbles \cite{2010ApJ...724.1044S}, finding new evidence for the gamma-ray excess even at few kpc of distance from the GC. Eventually, in \cite{Daylan:2014rsa} an optimized analysis implementing specific cuts to Fermi-LAT events -- improving the resolution of the gamma-ray maps -- could reach a statistical preference at the $\sim 30\sigma$ level for the inclusion of a WIMP-like template in the fit to the dataset. These claims triggered several phenomenological studies on the New Physics direction to undertake in order to explain this anomaly (see, e.g., \cite{Huang:2013apa,Alves:2014yha,Berlin:2014tja,Abdullah:2014lla,Agrawal:2014oha,Calore:2014nla,Kaplinghat:2015gha,Karwin:2016tsw}).

It is important to note that a close investigation about the impact of the Galactic diffuse emission -- related to CR physics -- on the robustness of the aforementioned evidence in favor of a DM indirect detection came only four years after \cite{Goodenough:2009gk}, with the studies in \cite{Zhou:2014lva,Calore:2014xka}. In particular, \cite{Calore:2014xka} analyzed the uncertainties related to the galactic diffuse modeling explicitly inspecting the systematics of 60 models with different characterization of CR transport physics and emitting target properties. Most importantly, the authors in \cite{Calore:2014xka} derived in a more data-driven way the overall systematics in the ROI of $20^{\circ} \times 20^{\circ}$ centered at the GC by looking at at the residuals obtained from a large number of control ROI along the whole Galactic plane, using their 60 galactic diffuse emission models for a principal component analysis. The systematics found in \cite{Calore:2014xka} associated to the GC excess signal enlarged the set of viable DM best-fit scenarios \cite{Agrawal:2014oha,Calore:2014nla}, and opened a new window for the interpretation of the anomaly in terms of a population of point sources such as  millisecond pulsars (MPs) \cite{Petrovic:2014xra}, a scenario originally proposed in \cite{Abazajian:2010zy} and successively supported in \cite{Abazajian:2012pn,Gordon:2013vta,Abazajian:2014fta,Yuan:2014rca}, while  criticized in \cite{Hooper:2013nhl,Cholis:2014lta}. Alternative astrophysical explanations in terms of outburst events of hadronic \cite{Carlson:2014cwa} or leptonic \cite{Petrovic:2014uda} origin turned out to be not favored by data \cite{Cholis:2015dea}.

Notably, the different CR injection terms ($Q_{0}$ appearing in Eq.~(\ref{eq:general_CR_transport}) adopted in all the aforementioned analyses, correspond to the radial distributions shown by the thin and dashed colored lines in the lower-left panel of Fig.~\ref{fig:GC_excess}. Their behavior in the GC proximity does not correlate well with the expected high-star formation rate present in the inmost few hundred parsecs around the GC \cite{Figer:2003tu}, the so-called Central Molecular Zone (CMZ). Multi-wavelength observations of the CMZ \cite{Crocker:2010qn} point indeed to an environment with large amount of molecular gas \cite{Ferriere:2007yq}, and hosting high-mass OB stars \cite{Figer:2002bi}, potential progenitors for standard CR acceleration sites such as supernova remnants \cite{1979ApJ...231...95M}. These pieces of informations motivated \cite{Gaggero:2015nsa} to re-analyze the GC excess implementing a novel steady-state CR source term capturing the CMZ star-forming activity. 
The three upper panels in Fig.~\ref{fig:GC_excess} from \cite{Gaggero:2015nsa} show the residual count maps of photons falling in the energy range of 1 -- 10 GeV, within a ROI of $10^{\circ} \times 10^{\circ}$ centered at the GC. From left to right, the result within a benchmark scenario originally identified in \cite{Calore:2014xka} to optimize the study of the GC anomaly at the GeV, and the remarkable improvement in the outcome of the fit to Fermi-LAT data when adding a DM component (central panel), or when re-visiting the whole Galactic diffuse emission background on the basis of the novel steady source term peaked at the GC (right panel). 

Notably, this result has been successively corroborated and refined in \cite{Carlson:2015ona,Carlson:2016iis}, with the implementation of the high-resolution Galactic gas distribution obtained in \cite{Pohl:2007dz}, able to resolve spiral arms and the Galactic bar, and most importantly providing kinematic resolution towards the GC, with the help of hydrodynamic simulations. Then, these HII density maps were correlated to the CR source injection term by means of a simple model of star formation \cite{1959ApJ...129..243S}. The thick light-blue line in the left-bottom panel of Fig.~\ref{fig:GC_excess}, taken from \cite{Carlson:2015ona}, shows a benchmark realization of the novel CR source radial distribution, that correctly does not fall off any longer at small Galactocentric radii. The red dashed (blue dot-dashed) line in the right-bottom panel, from Fig.~4 of \cite{Carlson:2016iis}, identifies the improvement, $\Delta \chi^{2} < 0$, of the description of gamma-ray data in the inner Galaxy region as a function of  the fraction $f_{H2}$ of CRs injected with a spatial distribution tracing the HII density maps of \cite{Pohl:2007dz}, in the presence (absence) of the DM component, against the best-fit scenario of a Galactic diffuse emission model built with a standard CR injection term plus the DM component. Inset numbers report the statistical preference of the red line over the blue one. 

The relevance of the Galactic diffuse emission modeling in assessing the significance of the GeV excess at the GC has been further acknowledged by the Fermi-LAT collaboration \cite{TheFermi-LAT:2015kwa}. In \cite{TheFermi-LAT:2015kwa}, injected CR electrons as in \cite{Carlson:2015ona,Carlson:2015ona}, together with the uncertainties on the interstellar radiation field, have been shown to play a major role in the characterization of the gamma-ray anomaly. As also marked by the right-bottom panel of Fig.~\ref{fig:GC_excess}, noisy residual photon counts around the GC in the analyses of \cite{Gaggero:2015nsa,Carlson:2015ona,TheFermi-LAT:2015kwa} still leave room for an extra-component in the description of gamma-ray data. Notably, two independent studies implementing different statistical techniques for clustering patterns in the observed photon count maps have followed, showing very high evidence in favor of a hitherto undetected population of point sources, able to fully account for the GeV gamma-ray excess \cite{Bartels:2015aea,Lee:2015fea}. The results of these two works reached remarkably similar conclusions, giving substantial credit to the MSP-like interpretation of the gamma-ray signal at the GC (see \cite{Hooper:2015jlu, Ploeg:2017vai} for discussions about the luminosity function of the putative MSPs at the GC) and triggering relevant dedicated searches \cite{Fermi-LAT:2017yoi}. Non-negligible mis-modeling of astrophysical backgrounds and foregrounds may affect the details on the prediction for such unresolved population of point sources within the Galactic bulge, see e.g. \cite{Horiuchi:2016zwu}. A recent novel tool, \texttt{SkyFACT},  developed in \cite{Storm:2017arh}, based on image reconstruction and adaptive spatio-spectral template regression, has allowed for dramatic improvements in the quality of the fits to gamma-ray data through fine-grained variations of galactic diffuse emission modeling. Exploiting this powerful package, a novel investigation on the morphology of the excess in connection to the stellar distribution of the bar/boxy bulge in the inner Galaxy provides strong support to the MSP hypothesis \cite{Bartels:2017vsx}. 

In conclusion, the GC GeV excess remains a widely studied signal in the astro-particle community. 
The updated comprehensive analysis carried out by the Fermi-LAT collaboration \cite{TheFermi-LAT:2017vmf} fairly summarizes the most relevant factors affecting the characterization of the signal: {\bf 1)} the details on CR production and propagation, especially in the GC proximity; {\bf 2)} the templates for the interstellar gas and radiation field in the inner Galaxy;
{\bf 3)} the emissivity and morphology of the Fermi bubbles towards low latitudes; {\bf 4)} the list of point sources near the GC identified within a given background model.

The take-home message of this long debate is likely twofold: The existence of an extended emission from the inner Galaxy peaked at few GeV is well established; however, the characterization of this emission, i.e. the morphology and its intensity, strongly depends on the assumptions of the CR source distribution. Interestingly, the interpretation in terms of unresolved point sources, possibly associated to a population of millisecond pulsars -- currently supported by wavelet and photon-count statistics analyses -- is testable in the future with more sensitive radio facilities \cite{Calore:2015bsx}.
As a final outlook, let us mention that a recent spatially extended gamma-ray signal from the center of M31 galaxy \cite{Ackermann:2017nya} has further renewed the interest on the GC anomaly and, for instance, may possibly shed new light on the existence of galactic bulge MSP populations \cite{Eckner:2017oul}. 

Let us now move away from the GC region, and consider another very relevant potential discovery window. Contrary to the complex astrophysical environment characterizing the CMZ and the GC, dwarf spheroidal galaxies of the Milky Way (dSphs) stand out as very promising targets in the gamma-ray band due to the corresponding low astrophysical background and foreground \cite{Lake:1990du,Evans:2003sc}. Being relatively close to us and associated to fairly large DM densities, the gamma-ray campaign on dSphs has been soon realized to be one of the most potentially sensitive probes to particle DM properties \cite{Strigari:2006rd,Strigari:2007at,Ackermann:2011wa}. At present, dSph gamma-ray upper-bounds are remarkably probing the benchmark thermal relic scenario within the WIMP mass window \cite{Ackermann:2015zua,Ahnen:2016qkx,Fermi-LAT:2016uux,Archambault:2017wyh}. Such upper-limits may be at odd with naive DM interpretations of the GC excess \cite{Abazajian:2015raa,Keeley:2017fbz}, while depend crucially on the estimated DM content in these galaxies,  potentially affected by several systematics \cite{Bonnivard:2014kza,Ullio:2016kvy,Hayashi:2016kcy,Ichikawa:2016nbi}. 

Interestingly, hints for a gamma-ray signal possibly compatible with the one observed at the GC have been found in the analysis of some of the most recently discovered Milky Way ultra-faint satellites, see, e.g., the case of Reticulum II \cite{Geringer-Sameth:2015lua,Hooper:2015ula,Bonnivard:2015tta}. According to the latest joint analysis of Fermi-LAT and Dark Energy Survey collaborations, the significance of these excesses remain at present well below the 3$\sigma$ level \cite{Fermi-LAT:2016uux}. Moreover, dedicated searches in the radio-band have not found any counterpart of the putative gamma-ray DM signal \cite{Regis:2017oet}. Note that -- from the perspective of a signal detection -- a broad multi-wavelength program for indirect DM searches in dSphs would be indeed promising \cite{Colafrancesco:2006he,Profumo:2010ya}. However, in opposition to the case of gamma-rays, constraints on DM annihilation/decay derived from the observation of dSphs in the radio and/or X band turn out to be subject to a larger set of astrophysical uncertainties, including CR transport physics \cite{Regis:2014tga}.


\section{Future prospects: From MeV to multi-TeV}

The future of indirect searches is particularly bright. 

In the gamma-ray band, two new regions of the spectrum will be explored. On the low-energy side, the MeV-GeV domain can be probed by planned experiments such as e-ASTROGAM \cite{astrogam2017}\footnote{e-ASTROGAM is proposed as ESA M5 mission.}, and AMEGO\footnote{See
\url{pcos.gsfc.nasa.gov/physpag/probe/AMEGO_probe.pdf}} (other
previously proposed missions include, \textit{e.g.}, COMPAIR \cite{Moiseev:2015lva} and ADEPT \cite{Hunter:2013wla}), which could be
realized in the mid- and long-term future in the late 2020s).  
All those experiments will feature a 2-3 order-of-magnitude increase in sensitivity~\cite{Essig:2013goa, Boddy:2015efa} with respect to previous instruments operating in this window like COMPTEL and EGRET
\cite{Strong:1998ck, Hunter:1997we}, and a remarkable energy resolution especially below $\simeq 10$ GeV, where the detection principle is based on Compton scattering instead of pair production. This will guarantee a high constraining power as far as sharp spectral features (e.g. lines originating from DM annihilation) are concerned \cite{bartels2017}.
On theoretical grounds,  vanilla WIMP scenario in this domain of energy scales may not be adequate due to the Lee-Weinberg bound \cite{Vysotsky:1977pe,Lee:1977ua}. Interesting models designed to offer an explanation for a $511$ keV emission line detected in the inner Galaxy may be instead well probed \cite{boehm2004PRD,boehm2004PRL}.

On the other end, in the TeV domain, while many Imaging Atmospheric Cherenkov Telescopes  (MAGIC \cite{Lorenz2004}, H.E.S.S.\footnote{\url{www.mpi-hd.mpg.de/hfm/HESS}} and VERITAS\footnote{\url{veritas.sao.arizona.edu/}}) and air-shower arrays (such as Milagro\footnote{\url{http://umdgrb.umd.edu/cosmic/milagro.html}} and HAWC\footnote{\url{umdgrb.umd.edu/hawc/index.php}} \cite{hawc2014}) have already been providing relevant  results, and have provided stringent constraints on DM annihilation \cite{hess2011DM}, CTA will provide a further, unprecedented increase in sensitivity above the TeV.
We want to emphasize that, as mentioned in Section~2, the most naive version of the {\it WIMP miracle} naturally yields a predicted mass for the DM candidate in that ballpark. 
CTA can probe the DM sector in different ways \cite{doro2013,silverwood2015,conrad2017,morselli2017}. 

In connection with the main topic of this review, we stress once again that the GC region appears promising also in this context, which implies that a more careful modeling of the $\gamma$-ray diffuse emission from the inner Galaxy will be required. The diffusion properties, again, play then a central role: In particular, the CR spectrum in the inner Galaxy is currently a matter or debate, and Fermi-LAT data seem to hint towards a hardening in the inner Galaxy \cite{Gaggero:2014xla,Acero:2016qlg,Yang:2016jda}. {\tt DRAGON}-based models \cite{Gaggero:2014xla,Cerri:2017joy} featuring inhomogeneous or anisotropic diffusion, which are key features of this specific numerical package, allow to reproduce Fermi-LAT data and provide TeV predictions \cite{Gaggero:2015xza,gaggero2017prl} which seem in accord with a collection of data from different experiments, including the bright, diffuse Galactic Ridge emission measured by H.E.S.S. over the latest decade \cite{hess2006,hess2016,hess2017}. These kind of studies will be even more important in the forthcoming years, due to the necessity, that we have stressed several times along this paper, of a good characterization of the astrophysical backgrounds. The study of dwarf spheroidal galaxies will also benefit from the expected increase in sensitivity in the TeV domain, and the expected performance of CTA in this channel is remarkable \cite{Hutten:2016jko,Lefranc:2016dgx,Lefranc:2016fgn}.

As far as charged particle channels are concerned, the antiparticle/antinuclear avenue is particularly promising despite the relevant uncertainties. 
In fact, new high-quality data are expected by many experiments: besides AMS-02, the already operating DAMPE \cite{dampe2017} space experiment and CALET \cite{calet2015}, on board the ISS,  have the opportunity to probe in particular the leptonic channel  all the way up to the multi-TeV range with unprecedented energy resolution and sensitivity; planned experiments such as HERD\footnote{\url{http://herd.ihep.ac.cn/}} will provide a further, very relevant extension in the covered energy range and sensitivity, and -- on the antinuclei side -- we are looking forward to experiments such as GAPS \cite{gaps2017}, a balloon experiment expected to operate in 2020 and optimized specifically for low-energy antinuclei signatures, thanks to a  novel detection technique based on exotic atom capture and decay.

To conclude this {\it grand tour}, let us emphasize once again the opportunity offered by the tremendous increase in experimental accuracy we are already witnessing in these days, and we will be further exploiting with the upcoming years. In our opinion, in order to take full advantage of these developments, interdisciplinarity is the main avenue. The guaranteed outcome of the research program aimed at indirectly detecting particle dark matter by means of astronomical data certainly involves a more profound understanding of the CR physics outlined across this review.  It is therefore crucial to deepen and broaden the connections between the research fields we have described above: Both the high-energy astrophysics and the particle physics community would unquestionably benefit from a cutting-edge increasing crossover. 

~

\textbf{Acknowledgements:} We are in debt with Dario Grasso, Gianfranco Bertone, Piero Ullio and Stefano Gabici for all their invaluable teachings about dark matter and cosmic-ray physics. We thank Alfredo Urbano and Carmelo Evoli for batting around a few interesting ideas on how to improve the manuscript. We wish to acknowledge all the authors of \cite{giesen2015,Gaggero:2015xza,Herms:2016vop,yuan2017dampe,Carlson:2015ona,Carlson:2016iis} for the figures reported in our review. We are also  grateful to Farinaldo Queiroz and Giorgio Arcadi for the opportunity of realizing this work.    

\bibliographystyle{apsrev}
\bibliography{refs}

\begin{thebibliography}{339}
\expandafter\ifx\csname natexlab\endcsname\relax\def\natexlab#1{#1}\fi
\expandafter\ifx\csname bibnamefont\endcsname\relax
  \def\bibnamefont#1{#1}\fi
\expandafter\ifx\csname bibfnamefont\endcsname\relax
  \def\bibfnamefont#1{#1}\fi
\expandafter\ifx\csname citenamefont\endcsname\relax
  \def\citenamefont#1{#1}\fi
\expandafter\ifx\csname url\endcsname\relax
  \def\url#1{\texttt{#1}}\fi
\expandafter\ifx\csname urlprefix\endcsname\relax\def\urlprefix{URL }\fi
\providecommand{\bibinfo}[2]{#2}
\providecommand{\eprint}[2][]{\url{#2}}

\bibitem[{\citenamefont{Silk et~al.}(2010)}]{Bertone:2010zza}
\bibinfo{author}{\bibfnamefont{J.}~\bibnamefont{Silk}} \bibnamefont{et~al.},
  \emph{\bibinfo{title}{{Particle Dark Matter: Observations, Models and
  Searches}}} (\bibinfo{publisher}{Cambridge Univ. Press},
  \bibinfo{address}{Cambridge}, \bibinfo{year}{2010}), ISBN
  \bibinfo{isbn}{9781107653924},
  \urlprefix\url{http://www.cambridge.org/uk/catalogue/catalogue.asp?isbn=9780521763684}.

\bibitem[{\citenamefont{Abdallah et~al.}(2015)}]{Abdallah:2015ter}
\bibinfo{author}{\bibfnamefont{J.}~\bibnamefont{Abdallah}}
  \bibnamefont{et~al.}, \bibinfo{journal}{Phys. Dark Univ.}
  \textbf{\bibinfo{volume}{9-10}}, \bibinfo{pages}{8} (\bibinfo{year}{2015}),
  \eprint{1506.03116}.

\bibitem[{\citenamefont{Marrodán~Undagoitia and
  Rauch}(2016)}]{Undagoitia:2015gya}
\bibinfo{author}{\bibfnamefont{T.}~\bibnamefont{Marrodán~Undagoitia}}
  \bibnamefont{and} \bibinfo{author}{\bibfnamefont{L.}~\bibnamefont{Rauch}},
  \bibinfo{journal}{J. Phys.} \textbf{\bibinfo{volume}{G43}},
  \bibinfo{pages}{013001} (\bibinfo{year}{2016}), \eprint{1509.08767}.

\bibitem[{\citenamefont{{Ellis} et~al.}(1984)\citenamefont{{Ellis}, {Hagelin},
  {Nanopoulos}, {Olive}, and {Srednicki}}}]{ellis1984}
\bibinfo{author}{\bibfnamefont{J.}~\bibnamefont{{Ellis}}},
  \bibinfo{author}{\bibfnamefont{J.~S.} \bibnamefont{{Hagelin}}},
  \bibinfo{author}{\bibfnamefont{D.~V.} \bibnamefont{{Nanopoulos}}},
  \bibinfo{author}{\bibfnamefont{K.}~\bibnamefont{{Olive}}}, \bibnamefont{and}
  \bibinfo{author}{\bibfnamefont{M.}~\bibnamefont{{Srednicki}}},
  \bibinfo{journal}{Nuclear Physics B} \textbf{\bibinfo{volume}{238}},
  \bibinfo{pages}{453} (\bibinfo{year}{1984}).

\bibitem[{\citenamefont{{Jungman} et~al.}(1996)\citenamefont{{Jungman},
  {Kamionkowski}, and {Griest}}}]{jungman1996}
\bibinfo{author}{\bibfnamefont{G.}~\bibnamefont{{Jungman}}},
  \bibinfo{author}{\bibfnamefont{M.}~\bibnamefont{{Kamionkowski}}},
  \bibnamefont{and} \bibinfo{author}{\bibfnamefont{K.}~\bibnamefont{{Griest}}},
  \bibinfo{journal}{Physics reports} \textbf{\bibinfo{volume}{267}},
  \bibinfo{pages}{195} (\bibinfo{year}{1996}), \eprint{hep-ph/9506380}.

\bibitem[{\citenamefont{Olive}(2003)}]{Olive:2003iq}
\bibinfo{author}{\bibfnamefont{K.~A.} \bibnamefont{Olive}}, in
  \emph{\bibinfo{booktitle}{{Particle physics and cosmology: The quest for
  physics beyond the standard model(s). Proceedings, Theoretical Advanced Study
  Institute, TASI 2002, Boulder, USA, June 3-28, 2002}}}
  (\bibinfo{year}{2003}), pp. \bibinfo{pages}{797--851},
  \eprint{astro-ph/0301505}.

\bibitem[{\citenamefont{Maurin et~al.}(2002)\citenamefont{Maurin, Taillet,
  Donato, Salati, Barrau, and Boudoul}}]{Maurin:2002ua}
\bibinfo{author}{\bibfnamefont{D.}~\bibnamefont{Maurin}},
  \bibinfo{author}{\bibfnamefont{R.}~\bibnamefont{Taillet}},
  \bibinfo{author}{\bibfnamefont{F.}~\bibnamefont{Donato}},
  \bibinfo{author}{\bibfnamefont{P.}~\bibnamefont{Salati}},
  \bibinfo{author}{\bibfnamefont{A.}~\bibnamefont{Barrau}}, \bibnamefont{and}
  \bibinfo{author}{\bibfnamefont{G.}~\bibnamefont{Boudoul}}
  (\bibinfo{year}{2002}), \eprint{astro-ph/0212111}.

\bibitem[{\citenamefont{Strigari}(2013)}]{Strigari:2013iaa}
\bibinfo{author}{\bibfnamefont{L.~E.} \bibnamefont{Strigari}},
  \bibinfo{journal}{Phys. Rept.} \textbf{\bibinfo{volume}{531}},
  \bibinfo{pages}{1} (\bibinfo{year}{2013}), \eprint{1211.7090}.

\bibitem[{\citenamefont{Bergström}(2000)}]{Bergstrom:2000pn}
\bibinfo{author}{\bibfnamefont{L.}~\bibnamefont{Bergström}},
  \bibinfo{journal}{Rept. Prog. Phys.} \textbf{\bibinfo{volume}{63}},
  \bibinfo{pages}{793} (\bibinfo{year}{2000}), \eprint{hep-ph/0002126}.

\bibitem[{\citenamefont{{Bertone} et~al.}(2005)\citenamefont{{Bertone},
  {Hooper}, and {Silk}}}]{bertone2005}
\bibinfo{author}{\bibfnamefont{G.}~\bibnamefont{{Bertone}}},
  \bibinfo{author}{\bibfnamefont{D.}~\bibnamefont{{Hooper}}}, \bibnamefont{and}
  \bibinfo{author}{\bibfnamefont{J.}~\bibnamefont{{Silk}}},
  \bibinfo{journal}{Physics reports} \textbf{\bibinfo{volume}{405}},
  \bibinfo{pages}{279} (\bibinfo{year}{2005}), \eprint{hep-ph/0404175}.

\bibitem[{\citenamefont{{Silk} and {Srednicki}}(1984)}]{silk1984}
\bibinfo{author}{\bibfnamefont{J.}~\bibnamefont{{Silk}}} \bibnamefont{and}
  \bibinfo{author}{\bibfnamefont{M.}~\bibnamefont{{Srednicki}}},
  \bibinfo{journal}{Physical Review Letters} \textbf{\bibinfo{volume}{53}},
  \bibinfo{pages}{624} (\bibinfo{year}{1984}).

\bibitem[{\citenamefont{{Stecker} et~al.}(1985)\citenamefont{{Stecker},
  {Rudaz}, and {Walsh}}}]{stecker1985}
\bibinfo{author}{\bibfnamefont{F.~W.} \bibnamefont{{Stecker}}},
  \bibinfo{author}{\bibfnamefont{S.}~\bibnamefont{{Rudaz}}}, \bibnamefont{and}
  \bibinfo{author}{\bibfnamefont{T.~F.} \bibnamefont{{Walsh}}},
  \bibinfo{journal}{Physical Review Letters} \textbf{\bibinfo{volume}{55}},
  \bibinfo{pages}{2622} (\bibinfo{year}{1985}).

\bibitem[{\citenamefont{Picozza et~al.}(2007)}]{pamela2006}
\bibinfo{author}{\bibfnamefont{P.}~\bibnamefont{Picozza}} \bibnamefont{et~al.},
  \bibinfo{journal}{Astropart. Phys.} \textbf{\bibinfo{volume}{27}},
  \bibinfo{pages}{296} (\bibinfo{year}{2007}), \eprint{astro-ph/0608697}.

\bibitem[{\citenamefont{{Kounine}}(2012)}]{ams2012}
\bibinfo{author}{\bibfnamefont{A.}~\bibnamefont{{Kounine}}},
  \bibinfo{journal}{International Journal of Modern Physics E}
  \textbf{\bibinfo{volume}{21}}, \bibinfo{eid}{1230005} (\bibinfo{year}{2012}).

\bibitem[{\citenamefont{Reinert and Winkler}(2017)}]{Reinert:2017aga}
\bibinfo{author}{\bibfnamefont{A.}~\bibnamefont{Reinert}} \bibnamefont{and}
  \bibinfo{author}{\bibfnamefont{M.~W.} \bibnamefont{Winkler}}
  (\bibinfo{year}{2017}), \eprint{1712.00002}.

\bibitem[{\citenamefont{Blum et~al.}(2017{\natexlab{a}})\citenamefont{Blum,
  Sato, and Waxman}}]{Blum:2017iwq}
\bibinfo{author}{\bibfnamefont{K.}~\bibnamefont{Blum}},
  \bibinfo{author}{\bibfnamefont{R.}~\bibnamefont{Sato}}, \bibnamefont{and}
  \bibinfo{author}{\bibfnamefont{E.}~\bibnamefont{Waxman}}
  (\bibinfo{year}{2017}{\natexlab{a}}), \eprint{1709.06507}.

\bibitem[{\citenamefont{{Adriani}
  et~al.}(2009{\natexlab{a}})\citenamefont{{Adriani}, {Barbarino},
  {Bazilevskaya}, {Bellotti}, {Boezio}, {Bogomolov}, {Bonechi}, {Bongi},
  {Bonvicini}, {Bottai} et~al.}}]{PAMELApositron}
\bibinfo{author}{\bibfnamefont{O.}~\bibnamefont{{Adriani}}},
  \bibinfo{author}{\bibfnamefont{G.~C.} \bibnamefont{{Barbarino}}},
  \bibinfo{author}{\bibfnamefont{G.~A.} \bibnamefont{{Bazilevskaya}}},
  \bibinfo{author}{\bibfnamefont{R.}~\bibnamefont{{Bellotti}}},
  \bibinfo{author}{\bibfnamefont{M.}~\bibnamefont{{Boezio}}},
  \bibinfo{author}{\bibfnamefont{E.~A.} \bibnamefont{{Bogomolov}}},
  \bibinfo{author}{\bibfnamefont{L.}~\bibnamefont{{Bonechi}}},
  \bibinfo{author}{\bibfnamefont{M.}~\bibnamefont{{Bongi}}},
  \bibinfo{author}{\bibfnamefont{V.}~\bibnamefont{{Bonvicini}}},
  \bibinfo{author}{\bibfnamefont{S.}~\bibnamefont{{Bottai}}},
  \bibnamefont{et~al.}, \bibinfo{journal}{Nature}
  \textbf{\bibinfo{volume}{458}}, \bibinfo{pages}{607}
  (\bibinfo{year}{2009}{\natexlab{a}}), \eprint{0810.4995}.

\bibitem[{\citenamefont{{Aguilar} et~al.}(2013)\citenamefont{{Aguilar},
  {Alberti}, {Alpat}, {Alvino}, {Ambrosi}, {Andeen}, {Anderhub}, {Arruda},
  {Azzarello}, {Bachlechner} et~al.}}]{AMSpositron}
\bibinfo{author}{\bibfnamefont{M.}~\bibnamefont{{Aguilar}}},
  \bibinfo{author}{\bibfnamefont{G.}~\bibnamefont{{Alberti}}},
  \bibinfo{author}{\bibfnamefont{B.}~\bibnamefont{{Alpat}}},
  \bibinfo{author}{\bibfnamefont{A.}~\bibnamefont{{Alvino}}},
  \bibinfo{author}{\bibfnamefont{G.}~\bibnamefont{{Ambrosi}}},
  \bibinfo{author}{\bibfnamefont{K.}~\bibnamefont{{Andeen}}},
  \bibinfo{author}{\bibfnamefont{H.}~\bibnamefont{{Anderhub}}},
  \bibinfo{author}{\bibfnamefont{L.}~\bibnamefont{{Arruda}}},
  \bibinfo{author}{\bibfnamefont{P.}~\bibnamefont{{Azzarello}}},
  \bibinfo{author}{\bibfnamefont{A.}~\bibnamefont{{Bachlechner}}},
  \bibnamefont{et~al.}, \bibinfo{journal}{Physical Review Letters}
  \textbf{\bibinfo{volume}{110}}, \bibinfo{eid}{141102} (\bibinfo{year}{2013}).

\bibitem[{\citenamefont{Aguilar
  et~al.}(2016{\natexlab{a}})\citenamefont{Aguilar, Ali~Cavasonza, Alpat,
  Ambrosi, Arruda, Attig, Aupetit, Azzarello, Bachlechner, Barao
  et~al.}}]{AMSantiproton}
\bibinfo{author}{\bibfnamefont{M.}~\bibnamefont{Aguilar}},
  \bibinfo{author}{\bibfnamefont{L.}~\bibnamefont{Ali~Cavasonza}},
  \bibinfo{author}{\bibfnamefont{B.}~\bibnamefont{Alpat}},
  \bibinfo{author}{\bibfnamefont{G.}~\bibnamefont{Ambrosi}},
  \bibinfo{author}{\bibfnamefont{L.}~\bibnamefont{Arruda}},
  \bibinfo{author}{\bibfnamefont{N.}~\bibnamefont{Attig}},
  \bibinfo{author}{\bibfnamefont{S.}~\bibnamefont{Aupetit}},
  \bibinfo{author}{\bibfnamefont{P.}~\bibnamefont{Azzarello}},
  \bibinfo{author}{\bibfnamefont{A.}~\bibnamefont{Bachlechner}},
  \bibinfo{author}{\bibfnamefont{F.}~\bibnamefont{Barao}}, \bibnamefont{et~al.}
  (\bibinfo{collaboration}{AMS Collaboration}), \bibinfo{journal}{Phys. Rev.
  Lett.} \textbf{\bibinfo{volume}{117}}, \bibinfo{pages}{091103}
  (\bibinfo{year}{2016}{\natexlab{a}}),
  \urlprefix\url{https://link.aps.org/doi/10.1103/PhysRevLett.117.091103}.

\bibitem[{\citenamefont{Cuoco et~al.}(2017{\natexlab{a}})\citenamefont{Cuoco,
  Krämer, and Korsmeier}}]{Cuoco:2016eej}
\bibinfo{author}{\bibfnamefont{A.}~\bibnamefont{Cuoco}},
  \bibinfo{author}{\bibfnamefont{M.}~\bibnamefont{Krämer}}, \bibnamefont{and}
  \bibinfo{author}{\bibfnamefont{M.}~\bibnamefont{Korsmeier}},
  \bibinfo{journal}{Phys. Rev. Lett.} \textbf{\bibinfo{volume}{118}},
  \bibinfo{pages}{191102} (\bibinfo{year}{2017}{\natexlab{a}}),
  \eprint{1610.03071}.

\bibitem[{\citenamefont{Cuoco et~al.}(2017{\natexlab{b}})\citenamefont{Cuoco,
  Heisig, Korsmeier, and Krämer}}]{Cuoco:2017okh}
\bibinfo{author}{\bibfnamefont{A.}~\bibnamefont{Cuoco}},
  \bibinfo{author}{\bibfnamefont{J.}~\bibnamefont{Heisig}},
  \bibinfo{author}{\bibfnamefont{M.}~\bibnamefont{Korsmeier}},
  \bibnamefont{and} \bibinfo{author}{\bibfnamefont{M.}~\bibnamefont{Krämer}},
  in \emph{\bibinfo{booktitle}{{2017 European Physical Society Conference on
  High Energy Physics (EPS-HEP 2017) Venice, Italy, July 5-12, 2017}}}
  (\bibinfo{year}{2017}{\natexlab{b}}), \eprint{1711.06460},
  \urlprefix\url{http://inspirehep.net/record/1636948/files/arXiv:1711.06460.pdf}.

\bibitem[{\citenamefont{{Bergstr{\"o}m} and {Snellman}}(1988)}]{bergstrom1988}
\bibinfo{author}{\bibfnamefont{L.}~\bibnamefont{{Bergstr{\"o}m}}}
  \bibnamefont{and}
  \bibinfo{author}{\bibfnamefont{H.}~\bibnamefont{{Snellman}}},
  \bibinfo{journal}{\prd} \textbf{\bibinfo{volume}{37}}, \bibinfo{pages}{3737}
  (\bibinfo{year}{1988}).

\bibitem[{\citenamefont{{Rudaz}}(1989)}]{rudaz1989}
\bibinfo{author}{\bibfnamefont{S.}~\bibnamefont{{Rudaz}}},
  \bibinfo{journal}{\prd} \textbf{\bibinfo{volume}{39}}, \bibinfo{pages}{3549}
  (\bibinfo{year}{1989}).

\bibitem[{\citenamefont{{Giudice} and {Griest}}(1989)}]{giudice1989}
\bibinfo{author}{\bibfnamefont{G.~F.} \bibnamefont{{Giudice}}}
  \bibnamefont{and} \bibinfo{author}{\bibfnamefont{K.}~\bibnamefont{{Griest}}},
  \bibinfo{journal}{\prd} \textbf{\bibinfo{volume}{40}}, \bibinfo{pages}{2549}
  (\bibinfo{year}{1989}).

\bibitem[{\citenamefont{{Bergstr{\"o}m}
  et~al.}(1998)\citenamefont{{Bergstr{\"o}m}, {Ullio}, and
  {Buckley}}}]{bergstrom1998}
\bibinfo{author}{\bibfnamefont{L.}~\bibnamefont{{Bergstr{\"o}m}}},
  \bibinfo{author}{\bibfnamefont{P.}~\bibnamefont{{Ullio}}}, \bibnamefont{and}
  \bibinfo{author}{\bibfnamefont{J.~H.} \bibnamefont{{Buckley}}},
  \bibinfo{journal}{Astroparticle Physics} \textbf{\bibinfo{volume}{9}},
  \bibinfo{pages}{137} (\bibinfo{year}{1998}), \eprint{astro-ph/9712318}.

\bibitem[{\citenamefont{Bringmann and Weniger}(2012)}]{Bringmann:2012ez}
\bibinfo{author}{\bibfnamefont{T.}~\bibnamefont{Bringmann}} \bibnamefont{and}
  \bibinfo{author}{\bibfnamefont{C.}~\bibnamefont{Weniger}},
  \bibinfo{journal}{Phys. Dark Univ.} \textbf{\bibinfo{volume}{1}},
  \bibinfo{pages}{194} (\bibinfo{year}{2012}), \eprint{1208.5481}.

\bibitem[{\citenamefont{Bertone and Hooper}(2016)}]{Bertone:2016nfn}
\bibinfo{author}{\bibfnamefont{G.}~\bibnamefont{Bertone}} \bibnamefont{and}
  \bibinfo{author}{\bibfnamefont{D.}~\bibnamefont{Hooper}},
  \bibinfo{journal}{Submitted to: Rev. Mod. Phys.}  (\bibinfo{year}{2016}),
  \eprint{1605.04909}.

\bibitem[{\citenamefont{de~Swart et~al.}(2017)\citenamefont{de~Swart, Bertone,
  and van Dongen}}]{deSwart:2017heh}
\bibinfo{author}{\bibfnamefont{J.}~\bibnamefont{de~Swart}},
  \bibinfo{author}{\bibfnamefont{G.}~\bibnamefont{Bertone}}, \bibnamefont{and}
  \bibinfo{author}{\bibfnamefont{J.}~\bibnamefont{van Dongen}}
  (\bibinfo{year}{2017}), \bibinfo{note}{[Nature Astron.1,0059(2017)]},
  \eprint{1703.00013}.

\bibitem[{\citenamefont{Hu et~al.}(2000)\citenamefont{Hu, Barkana, and
  Gruzinov}}]{Hu:2000ke}
\bibinfo{author}{\bibfnamefont{W.}~\bibnamefont{Hu}},
  \bibinfo{author}{\bibfnamefont{R.}~\bibnamefont{Barkana}}, \bibnamefont{and}
  \bibinfo{author}{\bibfnamefont{A.}~\bibnamefont{Gruzinov}},
  \bibinfo{journal}{Phys. Rev. Lett.} \textbf{\bibinfo{volume}{85}},
  \bibinfo{pages}{1158} (\bibinfo{year}{2000}), \eprint{astro-ph/0003365}.

\bibitem[{\citenamefont{Marsh}(2016)}]{Marsh:2015xka}
\bibinfo{author}{\bibfnamefont{D.~J.~E.} \bibnamefont{Marsh}},
  \bibinfo{journal}{Phys. Rept.} \textbf{\bibinfo{volume}{643}},
  \bibinfo{pages}{1} (\bibinfo{year}{2016}), \eprint{1510.07633}.

\bibitem[{\citenamefont{Hui et~al.}(2017)\citenamefont{Hui, Ostriker, Tremaine,
  and Witten}}]{Hui:2016ltb}
\bibinfo{author}{\bibfnamefont{L.}~\bibnamefont{Hui}},
  \bibinfo{author}{\bibfnamefont{J.~P.} \bibnamefont{Ostriker}},
  \bibinfo{author}{\bibfnamefont{S.}~\bibnamefont{Tremaine}}, \bibnamefont{and}
  \bibinfo{author}{\bibfnamefont{E.}~\bibnamefont{Witten}},
  \bibinfo{journal}{Phys. Rev.} \textbf{\bibinfo{volume}{D95}},
  \bibinfo{pages}{043541} (\bibinfo{year}{2017}), \eprint{1610.08297}.

\bibitem[{\citenamefont{Svrcek and Witten}(2006)}]{Svrcek:2006yi}
\bibinfo{author}{\bibfnamefont{P.}~\bibnamefont{Svrcek}} \bibnamefont{and}
  \bibinfo{author}{\bibfnamefont{E.}~\bibnamefont{Witten}},
  \bibinfo{journal}{JHEP} \textbf{\bibinfo{volume}{06}}, \bibinfo{pages}{051}
  (\bibinfo{year}{2006}), \eprint{hep-th/0605206}.

\bibitem[{\citenamefont{Arvanitaki et~al.}(2010)\citenamefont{Arvanitaki,
  Dimopoulos, Dubovsky, Kaloper, and March-Russell}}]{Arvanitaki:2009fg}
\bibinfo{author}{\bibfnamefont{A.}~\bibnamefont{Arvanitaki}},
  \bibinfo{author}{\bibfnamefont{S.}~\bibnamefont{Dimopoulos}},
  \bibinfo{author}{\bibfnamefont{S.}~\bibnamefont{Dubovsky}},
  \bibinfo{author}{\bibfnamefont{N.}~\bibnamefont{Kaloper}}, \bibnamefont{and}
  \bibinfo{author}{\bibfnamefont{J.}~\bibnamefont{March-Russell}},
  \bibinfo{journal}{Phys. Rev.} \textbf{\bibinfo{volume}{D81}},
  \bibinfo{pages}{123530} (\bibinfo{year}{2010}), \eprint{0905.4720}.

\bibitem[{\citenamefont{Abbott et~al.}(2016)}]{Abbott:2016blz}
\bibinfo{author}{\bibfnamefont{B.~P.} \bibnamefont{Abbott}}
  \bibnamefont{et~al.} (\bibinfo{collaboration}{Virgo, LIGO Scientific}),
  \bibinfo{journal}{Phys. Rev. Lett.} \textbf{\bibinfo{volume}{116}},
  \bibinfo{pages}{061102} (\bibinfo{year}{2016}), \eprint{1602.03837}.

\bibitem[{\citenamefont{Bird et~al.}(2016)\citenamefont{Bird, Cholis, Muñoz,
  Ali-Haïmoud, Kamionkowski, Kovetz, Raccanelli, and Riess}}]{Bird:2016dcv}
\bibinfo{author}{\bibfnamefont{S.}~\bibnamefont{Bird}},
  \bibinfo{author}{\bibfnamefont{I.}~\bibnamefont{Cholis}},
  \bibinfo{author}{\bibfnamefont{J.~B.} \bibnamefont{Muñoz}},
  \bibinfo{author}{\bibfnamefont{Y.}~\bibnamefont{Ali-Haïmoud}},
  \bibinfo{author}{\bibfnamefont{M.}~\bibnamefont{Kamionkowski}},
  \bibinfo{author}{\bibfnamefont{E.~D.} \bibnamefont{Kovetz}},
  \bibinfo{author}{\bibfnamefont{A.}~\bibnamefont{Raccanelli}},
  \bibnamefont{and} \bibinfo{author}{\bibfnamefont{A.~G.} \bibnamefont{Riess}},
  \bibinfo{journal}{Phys. Rev. Lett.} \textbf{\bibinfo{volume}{116}},
  \bibinfo{pages}{201301} (\bibinfo{year}{2016}), \eprint{1603.00464}.

\bibitem[{\citenamefont{{Chapline}}(1975)}]{1975Natur.253..251C}
\bibinfo{author}{\bibfnamefont{G.~F.} \bibnamefont{{Chapline}}},
  \bibinfo{journal}{\nat} \textbf{\bibinfo{volume}{253}}, \bibinfo{pages}{251}
  (\bibinfo{year}{1975}).

\bibitem[{\citenamefont{Carr et~al.}(2016)\citenamefont{Carr, Kuhnel, and
  Sandstad}}]{Carr:2016drx}
\bibinfo{author}{\bibfnamefont{B.}~\bibnamefont{Carr}},
  \bibinfo{author}{\bibfnamefont{F.}~\bibnamefont{Kuhnel}}, \bibnamefont{and}
  \bibinfo{author}{\bibfnamefont{M.}~\bibnamefont{Sandstad}},
  \bibinfo{journal}{Phys. Rev.} \textbf{\bibinfo{volume}{D94}},
  \bibinfo{pages}{083504} (\bibinfo{year}{2016}), \eprint{1607.06077}.

\bibitem[{\citenamefont{Feng}(2010)}]{Feng:2010gw}
\bibinfo{author}{\bibfnamefont{J.~L.} \bibnamefont{Feng}},
  \bibinfo{journal}{Ann. Rev. Astron. Astrophys.}
  \textbf{\bibinfo{volume}{48}}, \bibinfo{pages}{495} (\bibinfo{year}{2010}),
  \eprint{1003.0904}.

\bibitem[{\citenamefont{Giudice}(2008)}]{Giudice:2008bi}
\bibinfo{author}{\bibfnamefont{G.~F.} \bibnamefont{Giudice}}
  (\bibinfo{year}{2008}), \eprint{0801.2562}.

\bibitem[{\citenamefont{Regis et~al.}(2007)\citenamefont{Regis, Serone, and
  Ullio}}]{Regis:2006hc}
\bibinfo{author}{\bibfnamefont{M.}~\bibnamefont{Regis}},
  \bibinfo{author}{\bibfnamefont{M.}~\bibnamefont{Serone}}, \bibnamefont{and}
  \bibinfo{author}{\bibfnamefont{P.}~\bibnamefont{Ullio}},
  \bibinfo{journal}{JHEP} \textbf{\bibinfo{volume}{03}}, \bibinfo{pages}{084}
  (\bibinfo{year}{2007}), \eprint{hep-ph/0612286}.

\bibitem[{\citenamefont{Panico et~al.}(2008)\citenamefont{Panico, Ponton,
  Santiago, and Serone}}]{Panico:2008bx}
\bibinfo{author}{\bibfnamefont{G.}~\bibnamefont{Panico}},
  \bibinfo{author}{\bibfnamefont{E.}~\bibnamefont{Ponton}},
  \bibinfo{author}{\bibfnamefont{J.}~\bibnamefont{Santiago}}, \bibnamefont{and}
  \bibinfo{author}{\bibfnamefont{M.}~\bibnamefont{Serone}},
  \bibinfo{journal}{Phys. Rev.} \textbf{\bibinfo{volume}{D77}},
  \bibinfo{pages}{115012} (\bibinfo{year}{2008}), \eprint{0801.1645}.

\bibitem[{\citenamefont{Ryttov and Sannino}(2008)}]{Ryttov:2008xe}
\bibinfo{author}{\bibfnamefont{T.~A.} \bibnamefont{Ryttov}} \bibnamefont{and}
  \bibinfo{author}{\bibfnamefont{F.}~\bibnamefont{Sannino}},
  \bibinfo{journal}{Phys. Rev.} \textbf{\bibinfo{volume}{D78}},
  \bibinfo{pages}{115010} (\bibinfo{year}{2008}), \eprint{0809.0713}.

\bibitem[{\citenamefont{Belyaev et~al.}(2011)\citenamefont{Belyaev, Frandsen,
  Sarkar, and Sannino}}]{Belyaev:2010kp}
\bibinfo{author}{\bibfnamefont{A.}~\bibnamefont{Belyaev}},
  \bibinfo{author}{\bibfnamefont{M.~T.} \bibnamefont{Frandsen}},
  \bibinfo{author}{\bibfnamefont{S.}~\bibnamefont{Sarkar}}, \bibnamefont{and}
  \bibinfo{author}{\bibfnamefont{F.}~\bibnamefont{Sannino}},
  \bibinfo{journal}{Phys. Rev.} \textbf{\bibinfo{volume}{D83}},
  \bibinfo{pages}{015007} (\bibinfo{year}{2011}), \eprint{1007.4839}.

\bibitem[{\citenamefont{Frigerio et~al.}(2012)\citenamefont{Frigerio, Pomarol,
  Riva, and Urbano}}]{Frigerio:2012uc}
\bibinfo{author}{\bibfnamefont{M.}~\bibnamefont{Frigerio}},
  \bibinfo{author}{\bibfnamefont{A.}~\bibnamefont{Pomarol}},
  \bibinfo{author}{\bibfnamefont{F.}~\bibnamefont{Riva}}, \bibnamefont{and}
  \bibinfo{author}{\bibfnamefont{A.}~\bibnamefont{Urbano}},
  \bibinfo{journal}{JHEP} \textbf{\bibinfo{volume}{07}}, \bibinfo{pages}{015}
  (\bibinfo{year}{2012}), \eprint{1204.2808}.

\bibitem[{\citenamefont{Bruggisser et~al.}(2017)\citenamefont{Bruggisser, Riva,
  and Urbano}}]{Bruggisser:2016ixa}
\bibinfo{author}{\bibfnamefont{S.}~\bibnamefont{Bruggisser}},
  \bibinfo{author}{\bibfnamefont{F.}~\bibnamefont{Riva}}, \bibnamefont{and}
  \bibinfo{author}{\bibfnamefont{A.}~\bibnamefont{Urbano}},
  \bibinfo{journal}{SciPost Phys.} \textbf{\bibinfo{volume}{3}},
  \bibinfo{pages}{017} (\bibinfo{year}{2017}), \eprint{1607.02474}.

\bibitem[{\citenamefont{Giudice and Romanino}(2004)}]{Giudice:2004tc}
\bibinfo{author}{\bibfnamefont{G.~F.} \bibnamefont{Giudice}} \bibnamefont{and}
  \bibinfo{author}{\bibfnamefont{A.}~\bibnamefont{Romanino}},
  \bibinfo{journal}{Nucl. Phys.} \textbf{\bibinfo{volume}{B699}},
  \bibinfo{pages}{65} (\bibinfo{year}{2004}), \bibinfo{note}{[Erratum: Nucl.
  Phys.B706,487(2005)]}, \eprint{hep-ph/0406088}.

\bibitem[{\citenamefont{Arkani-Hamed et~al.}(2005)\citenamefont{Arkani-Hamed,
  Dimopoulos, Giudice, and Romanino}}]{ArkaniHamed:2004yi}
\bibinfo{author}{\bibfnamefont{N.}~\bibnamefont{Arkani-Hamed}},
  \bibinfo{author}{\bibfnamefont{S.}~\bibnamefont{Dimopoulos}},
  \bibinfo{author}{\bibfnamefont{G.~F.} \bibnamefont{Giudice}},
  \bibnamefont{and} \bibinfo{author}{\bibfnamefont{A.}~\bibnamefont{Romanino}},
  \bibinfo{journal}{Nucl. Phys.} \textbf{\bibinfo{volume}{B709}},
  \bibinfo{pages}{3} (\bibinfo{year}{2005}), \eprint{hep-ph/0409232}.

\bibitem[{\citenamefont{Masiero et~al.}(2005)\citenamefont{Masiero, Profumo,
  and Ullio}}]{Masiero:2004ft}
\bibinfo{author}{\bibfnamefont{A.}~\bibnamefont{Masiero}},
  \bibinfo{author}{\bibfnamefont{S.}~\bibnamefont{Profumo}}, \bibnamefont{and}
  \bibinfo{author}{\bibfnamefont{P.}~\bibnamefont{Ullio}},
  \bibinfo{journal}{Nucl. Phys.} \textbf{\bibinfo{volume}{B712}},
  \bibinfo{pages}{86} (\bibinfo{year}{2005}), \eprint{hep-ph/0412058}.

\bibitem[{\citenamefont{Servant and Tait}(2003)}]{Servant:2002aq}
\bibinfo{author}{\bibfnamefont{G.}~\bibnamefont{Servant}} \bibnamefont{and}
  \bibinfo{author}{\bibfnamefont{T.~M.~P.} \bibnamefont{Tait}},
  \bibinfo{journal}{Nucl. Phys.} \textbf{\bibinfo{volume}{B650}},
  \bibinfo{pages}{391} (\bibinfo{year}{2003}), \eprint{hep-ph/0206071}.

\bibitem[{\citenamefont{Agashe and Servant}(2004)}]{Agashe:2004ci}
\bibinfo{author}{\bibfnamefont{K.}~\bibnamefont{Agashe}} \bibnamefont{and}
  \bibinfo{author}{\bibfnamefont{G.}~\bibnamefont{Servant}},
  \bibinfo{journal}{Phys. Rev. Lett.} \textbf{\bibinfo{volume}{93}},
  \bibinfo{pages}{231805} (\bibinfo{year}{2004}), \eprint{hep-ph/0403143}.

\bibitem[{\citenamefont{Agashe and Servant}(2005)}]{Agashe:2004bm}
\bibinfo{author}{\bibfnamefont{K.}~\bibnamefont{Agashe}} \bibnamefont{and}
  \bibinfo{author}{\bibfnamefont{G.}~\bibnamefont{Servant}},
  \bibinfo{journal}{JCAP} \textbf{\bibinfo{volume}{0502}}, \bibinfo{pages}{002}
  (\bibinfo{year}{2005}), \eprint{hep-ph/0411254}.

\bibitem[{\citenamefont{Hooper and Profumo}(2007)}]{Hooper:2007qk}
\bibinfo{author}{\bibfnamefont{D.}~\bibnamefont{Hooper}} \bibnamefont{and}
  \bibinfo{author}{\bibfnamefont{S.}~\bibnamefont{Profumo}},
  \bibinfo{journal}{Phys. Rept.} \textbf{\bibinfo{volume}{453}},
  \bibinfo{pages}{29} (\bibinfo{year}{2007}), \eprint{hep-ph/0701197}.

\bibitem[{\citenamefont{Gondolo and Gelmini}(1991)}]{Gondolo:1990dk}
\bibinfo{author}{\bibfnamefont{P.}~\bibnamefont{Gondolo}} \bibnamefont{and}
  \bibinfo{author}{\bibfnamefont{G.}~\bibnamefont{Gelmini}},
  \bibinfo{journal}{Nucl. Phys.} \textbf{\bibinfo{volume}{B360}},
  \bibinfo{pages}{145} (\bibinfo{year}{1991}).

\bibitem[{\citenamefont{Profumo}(2017)}]{Profumo:2017hqp}
\bibinfo{author}{\bibfnamefont{S.}~\bibnamefont{Profumo}},
  \emph{\bibinfo{title}{{An Introduction to Particle Dark Matter}}}
  (\bibinfo{publisher}{World Scientific}, \bibinfo{year}{2017}), ISBN
  \bibinfo{isbn}{9781786340009, 9781786340016, 9781786340009, 9781786340016}.

\bibitem[{\citenamefont{Ade et~al.}(2016)}]{Ade:2015xua}
\bibinfo{author}{\bibfnamefont{P.~A.~R.} \bibnamefont{Ade}}
  \bibnamefont{et~al.} (\bibinfo{collaboration}{Planck}),
  \bibinfo{journal}{Astron. Astrophys.} \textbf{\bibinfo{volume}{594}},
  \bibinfo{pages}{A13} (\bibinfo{year}{2016}), \eprint{1502.01589}.

\bibitem[{\citenamefont{Vysotsky et~al.}(1977)\citenamefont{Vysotsky, Dolgov,
  and Zeldovich}}]{Vysotsky:1977pe}
\bibinfo{author}{\bibfnamefont{M.~I.} \bibnamefont{Vysotsky}},
  \bibinfo{author}{\bibfnamefont{A.~D.} \bibnamefont{Dolgov}},
  \bibnamefont{and} \bibinfo{author}{\bibfnamefont{{\relax Ya}.~B.}
  \bibnamefont{Zeldovich}}, \bibinfo{journal}{JETP Lett.}
  \textbf{\bibinfo{volume}{26}}, \bibinfo{pages}{188} (\bibinfo{year}{1977}),
  \bibinfo{note}{[Pisma Zh. Eksp. Teor. Fiz.26,200(1977)]}.

\bibitem[{\citenamefont{Lee and Weinberg}(1977)}]{Lee:1977ua}
\bibinfo{author}{\bibfnamefont{B.~W.} \bibnamefont{Lee}} \bibnamefont{and}
  \bibinfo{author}{\bibfnamefont{S.}~\bibnamefont{Weinberg}},
  \bibinfo{journal}{Phys. Rev. Lett.} \textbf{\bibinfo{volume}{39}},
  \bibinfo{pages}{165} (\bibinfo{year}{1977}).

\bibitem[{\citenamefont{Griest and Kamionkowski}(1990)}]{Griest:1989wd}
\bibinfo{author}{\bibfnamefont{K.}~\bibnamefont{Griest}} \bibnamefont{and}
  \bibinfo{author}{\bibfnamefont{M.}~\bibnamefont{Kamionkowski}},
  \bibinfo{journal}{Phys. Rev. Lett.} \textbf{\bibinfo{volume}{64}},
  \bibinfo{pages}{615} (\bibinfo{year}{1990}).

\bibitem[{\citenamefont{Harigaya et~al.}(2016)\citenamefont{Harigaya, Ibe,
  Kaneta, Nakano, and Suzuki}}]{Harigaya:2016nlg}
\bibinfo{author}{\bibfnamefont{K.}~\bibnamefont{Harigaya}},
  \bibinfo{author}{\bibfnamefont{M.}~\bibnamefont{Ibe}},
  \bibinfo{author}{\bibfnamefont{K.}~\bibnamefont{Kaneta}},
  \bibinfo{author}{\bibfnamefont{W.}~\bibnamefont{Nakano}}, \bibnamefont{and}
  \bibinfo{author}{\bibfnamefont{M.}~\bibnamefont{Suzuki}},
  \bibinfo{journal}{JHEP} \textbf{\bibinfo{volume}{08}}, \bibinfo{pages}{151}
  (\bibinfo{year}{2016}), \eprint{1606.00159}.

\bibitem[{\citenamefont{Griest and Seckel}(1991)}]{Griest:1990kh}
\bibinfo{author}{\bibfnamefont{K.}~\bibnamefont{Griest}} \bibnamefont{and}
  \bibinfo{author}{\bibfnamefont{D.}~\bibnamefont{Seckel}},
  \bibinfo{journal}{Phys. Rev.} \textbf{\bibinfo{volume}{D43}},
  \bibinfo{pages}{3191} (\bibinfo{year}{1991}).

\bibitem[{\citenamefont{Carlson et~al.}(1992)\citenamefont{Carlson, Machacek,
  and Hall}}]{Carlson:1992fn}
\bibinfo{author}{\bibfnamefont{E.~D.} \bibnamefont{Carlson}},
  \bibinfo{author}{\bibfnamefont{M.~E.} \bibnamefont{Machacek}},
  \bibnamefont{and} \bibinfo{author}{\bibfnamefont{L.~J.} \bibnamefont{Hall}},
  \bibinfo{journal}{Astrophys. J.} \textbf{\bibinfo{volume}{398}},
  \bibinfo{pages}{43} (\bibinfo{year}{1992}).

\bibitem[{\citenamefont{Pospelov et~al.}(2008)\citenamefont{Pospelov, Ritz, and
  Voloshin}}]{Pospelov:2007mp}
\bibinfo{author}{\bibfnamefont{M.}~\bibnamefont{Pospelov}},
  \bibinfo{author}{\bibfnamefont{A.}~\bibnamefont{Ritz}}, \bibnamefont{and}
  \bibinfo{author}{\bibfnamefont{M.~B.} \bibnamefont{Voloshin}},
  \bibinfo{journal}{Phys. Lett.} \textbf{\bibinfo{volume}{B662}},
  \bibinfo{pages}{53} (\bibinfo{year}{2008}), \eprint{0711.4866}.

\bibitem[{\citenamefont{Hochberg et~al.}(2014)\citenamefont{Hochberg, Kuflik,
  Volansky, and Wacker}}]{Hochberg:2014dra}
\bibinfo{author}{\bibfnamefont{Y.}~\bibnamefont{Hochberg}},
  \bibinfo{author}{\bibfnamefont{E.}~\bibnamefont{Kuflik}},
  \bibinfo{author}{\bibfnamefont{T.}~\bibnamefont{Volansky}}, \bibnamefont{and}
  \bibinfo{author}{\bibfnamefont{J.~G.} \bibnamefont{Wacker}},
  \bibinfo{journal}{Phys. Rev. Lett.} \textbf{\bibinfo{volume}{113}},
  \bibinfo{pages}{171301} (\bibinfo{year}{2014}), \eprint{1402.5143}.

\bibitem[{\citenamefont{D'Eramo and Thaler}(2010)}]{DEramo:2010keq}
\bibinfo{author}{\bibfnamefont{F.}~\bibnamefont{D'Eramo}} \bibnamefont{and}
  \bibinfo{author}{\bibfnamefont{J.}~\bibnamefont{Thaler}},
  \bibinfo{journal}{JHEP} \textbf{\bibinfo{volume}{06}}, \bibinfo{pages}{109}
  (\bibinfo{year}{2010}), \eprint{1003.5912}.

\bibitem[{\citenamefont{D'Agnolo and Ruderman}(2015)}]{DAgnolo:2015ujb}
\bibinfo{author}{\bibfnamefont{R.~T.} \bibnamefont{D'Agnolo}} \bibnamefont{and}
  \bibinfo{author}{\bibfnamefont{J.~T.} \bibnamefont{Ruderman}},
  \bibinfo{journal}{Phys. Rev. Lett.} \textbf{\bibinfo{volume}{115}},
  \bibinfo{pages}{061301} (\bibinfo{year}{2015}), \eprint{1505.07107}.

\bibitem[{\citenamefont{Pappadopulo et~al.}(2016)\citenamefont{Pappadopulo,
  Ruderman, and Trevisan}}]{Pappadopulo:2016pkp}
\bibinfo{author}{\bibfnamefont{D.}~\bibnamefont{Pappadopulo}},
  \bibinfo{author}{\bibfnamefont{J.~T.} \bibnamefont{Ruderman}},
  \bibnamefont{and} \bibinfo{author}{\bibfnamefont{G.}~\bibnamefont{Trevisan}},
  \bibinfo{journal}{Phys. Rev.} \textbf{\bibinfo{volume}{D94}},
  \bibinfo{pages}{035005} (\bibinfo{year}{2016}), \eprint{1602.04219}.

\bibitem[{\citenamefont{Dror et~al.}(2016)\citenamefont{Dror, Kuflik, and
  Ng}}]{Dror:2016rxc}
\bibinfo{author}{\bibfnamefont{J.~A.} \bibnamefont{Dror}},
  \bibinfo{author}{\bibfnamefont{E.}~\bibnamefont{Kuflik}}, \bibnamefont{and}
  \bibinfo{author}{\bibfnamefont{W.~H.} \bibnamefont{Ng}},
  \bibinfo{journal}{Phys. Rev. Lett.} \textbf{\bibinfo{volume}{117}},
  \bibinfo{pages}{211801} (\bibinfo{year}{2016}), \eprint{1607.03110}.

\bibitem[{\citenamefont{D'Agnolo et~al.}(2017)\citenamefont{D'Agnolo,
  Pappadopulo, and Ruderman}}]{DAgnolo:2017dbv}
\bibinfo{author}{\bibfnamefont{R.~T.} \bibnamefont{D'Agnolo}},
  \bibinfo{author}{\bibfnamefont{D.}~\bibnamefont{Pappadopulo}},
  \bibnamefont{and} \bibinfo{author}{\bibfnamefont{J.~T.}
  \bibnamefont{Ruderman}}, \bibinfo{journal}{Phys. Rev. Lett.}
  \textbf{\bibinfo{volume}{119}}, \bibinfo{pages}{061102}
  (\bibinfo{year}{2017}), \eprint{1705.08450}.

\bibitem[{\citenamefont{D'Eramo et~al.}(2017)\citenamefont{D'Eramo, Fernandez,
  and Profumo}}]{DEramo:2017gpl}
\bibinfo{author}{\bibfnamefont{F.}~\bibnamefont{D'Eramo}},
  \bibinfo{author}{\bibfnamefont{N.}~\bibnamefont{Fernandez}},
  \bibnamefont{and} \bibinfo{author}{\bibfnamefont{S.}~\bibnamefont{Profumo}},
  \bibinfo{journal}{JCAP} \textbf{\bibinfo{volume}{1705}}, \bibinfo{pages}{012}
  (\bibinfo{year}{2017}), \eprint{1703.04793}.

\bibitem[{\citenamefont{Arkani-Hamed et~al.}(2009)\citenamefont{Arkani-Hamed,
  Finkbeiner, Slatyer, and Weiner}}]{ArkaniHamed:2008qn}
\bibinfo{author}{\bibfnamefont{N.}~\bibnamefont{Arkani-Hamed}},
  \bibinfo{author}{\bibfnamefont{D.~P.} \bibnamefont{Finkbeiner}},
  \bibinfo{author}{\bibfnamefont{T.~R.} \bibnamefont{Slatyer}},
  \bibnamefont{and} \bibinfo{author}{\bibfnamefont{N.}~\bibnamefont{Weiner}},
  \bibinfo{journal}{Phys. Rev.} \textbf{\bibinfo{volume}{D79}},
  \bibinfo{pages}{015014} (\bibinfo{year}{2009}), \eprint{0810.0713}.

\bibitem[{\citenamefont{Feng et~al.}(2010{\natexlab{a}})\citenamefont{Feng,
  Kaplinghat, and Yu}}]{Feng:2009hw}
\bibinfo{author}{\bibfnamefont{J.~L.} \bibnamefont{Feng}},
  \bibinfo{author}{\bibfnamefont{M.}~\bibnamefont{Kaplinghat}},
  \bibnamefont{and} \bibinfo{author}{\bibfnamefont{H.-B.} \bibnamefont{Yu}},
  \bibinfo{journal}{Phys. Rev. Lett.} \textbf{\bibinfo{volume}{104}},
  \bibinfo{pages}{151301} (\bibinfo{year}{2010}{\natexlab{a}}),
  \eprint{0911.0422}.

\bibitem[{\citenamefont{Feng et~al.}(2010{\natexlab{b}})\citenamefont{Feng,
  Kaplinghat, and Yu}}]{Feng:2010zp}
\bibinfo{author}{\bibfnamefont{J.~L.} \bibnamefont{Feng}},
  \bibinfo{author}{\bibfnamefont{M.}~\bibnamefont{Kaplinghat}},
  \bibnamefont{and} \bibinfo{author}{\bibfnamefont{H.-B.} \bibnamefont{Yu}},
  \bibinfo{journal}{Phys. Rev.} \textbf{\bibinfo{volume}{D82}},
  \bibinfo{pages}{083525} (\bibinfo{year}{2010}{\natexlab{b}}),
  \eprint{1005.4678}.

\bibitem[{\citenamefont{Hryczuk et~al.}(2011)\citenamefont{Hryczuk, Iengo, and
  Ullio}}]{Hryczuk:2010zi}
\bibinfo{author}{\bibfnamefont{A.}~\bibnamefont{Hryczuk}},
  \bibinfo{author}{\bibfnamefont{R.}~\bibnamefont{Iengo}}, \bibnamefont{and}
  \bibinfo{author}{\bibfnamefont{P.}~\bibnamefont{Ullio}},
  \bibinfo{journal}{JHEP} \textbf{\bibinfo{volume}{03}}, \bibinfo{pages}{069}
  (\bibinfo{year}{2011}), \eprint{1010.2172}.

\bibitem[{\citenamefont{Shepherd et~al.}(2009)\citenamefont{Shepherd, Tait, and
  Zaharijas}}]{Shepherd:2009sa}
\bibinfo{author}{\bibfnamefont{W.}~\bibnamefont{Shepherd}},
  \bibinfo{author}{\bibfnamefont{T.~M.~P.} \bibnamefont{Tait}},
  \bibnamefont{and}
  \bibinfo{author}{\bibfnamefont{G.}~\bibnamefont{Zaharijas}},
  \bibinfo{journal}{Phys. Rev.} \textbf{\bibinfo{volume}{D79}},
  \bibinfo{pages}{055022} (\bibinfo{year}{2009}), \eprint{0901.2125}.

\bibitem[{\citenamefont{von Harling and Petraki}(2014)}]{vonHarling:2014kha}
\bibinfo{author}{\bibfnamefont{B.}~\bibnamefont{von Harling}} \bibnamefont{and}
  \bibinfo{author}{\bibfnamefont{K.}~\bibnamefont{Petraki}},
  \bibinfo{journal}{JCAP} \textbf{\bibinfo{volume}{1412}}, \bibinfo{pages}{033}
  (\bibinfo{year}{2014}), \eprint{1407.7874}.

\bibitem[{\citenamefont{An et~al.}(2016)\citenamefont{An, Wise, and
  Zhang}}]{An:2016gad}
\bibinfo{author}{\bibfnamefont{H.}~\bibnamefont{An}},
  \bibinfo{author}{\bibfnamefont{M.~B.} \bibnamefont{Wise}}, \bibnamefont{and}
  \bibinfo{author}{\bibfnamefont{Y.}~\bibnamefont{Zhang}},
  \bibinfo{journal}{Phys. Rev.} \textbf{\bibinfo{volume}{D93}},
  \bibinfo{pages}{115020} (\bibinfo{year}{2016}), \eprint{1604.01776}.

\bibitem[{\citenamefont{Cirelli et~al.}(2017)\citenamefont{Cirelli, Panci,
  Petraki, Sala, and Taoso}}]{Cirelli:2016rnw}
\bibinfo{author}{\bibfnamefont{M.}~\bibnamefont{Cirelli}},
  \bibinfo{author}{\bibfnamefont{P.}~\bibnamefont{Panci}},
  \bibinfo{author}{\bibfnamefont{K.}~\bibnamefont{Petraki}},
  \bibinfo{author}{\bibfnamefont{F.}~\bibnamefont{Sala}}, \bibnamefont{and}
  \bibinfo{author}{\bibfnamefont{M.}~\bibnamefont{Taoso}},
  \bibinfo{journal}{JCAP} \textbf{\bibinfo{volume}{1705}}, \bibinfo{pages}{036}
  (\bibinfo{year}{2017}), \eprint{1612.07295}.

\bibitem[{\citenamefont{Mitridate et~al.}(2017)\citenamefont{Mitridate, Redi,
  Smirnov, and Strumia}}]{Mitridate:2017izz}
\bibinfo{author}{\bibfnamefont{A.}~\bibnamefont{Mitridate}},
  \bibinfo{author}{\bibfnamefont{M.}~\bibnamefont{Redi}},
  \bibinfo{author}{\bibfnamefont{J.}~\bibnamefont{Smirnov}}, \bibnamefont{and}
  \bibinfo{author}{\bibfnamefont{A.}~\bibnamefont{Strumia}},
  \bibinfo{journal}{JCAP} \textbf{\bibinfo{volume}{1705}}, \bibinfo{pages}{006}
  (\bibinfo{year}{2017}), \eprint{1702.01141}.

\bibitem[{\citenamefont{Feng and Kumar}(2008)}]{Feng:2008ya}
\bibinfo{author}{\bibfnamefont{J.~L.} \bibnamefont{Feng}} \bibnamefont{and}
  \bibinfo{author}{\bibfnamefont{J.}~\bibnamefont{Kumar}},
  \bibinfo{journal}{Phys. Rev. Lett.} \textbf{\bibinfo{volume}{101}},
  \bibinfo{pages}{231301} (\bibinfo{year}{2008}), \eprint{0803.4196}.

\bibitem[{\citenamefont{Cirelli et~al.}(2011)\citenamefont{Cirelli, Corcella,
  Hektor, Hutsi, Kadastik, Panci, Raidal, Sala, and Strumia}}]{Cirelli:2010xx}
\bibinfo{author}{\bibfnamefont{M.}~\bibnamefont{Cirelli}},
  \bibinfo{author}{\bibfnamefont{G.}~\bibnamefont{Corcella}},
  \bibinfo{author}{\bibfnamefont{A.}~\bibnamefont{Hektor}},
  \bibinfo{author}{\bibfnamefont{G.}~\bibnamefont{Hutsi}},
  \bibinfo{author}{\bibfnamefont{M.}~\bibnamefont{Kadastik}},
  \bibinfo{author}{\bibfnamefont{P.}~\bibnamefont{Panci}},
  \bibinfo{author}{\bibfnamefont{M.}~\bibnamefont{Raidal}},
  \bibinfo{author}{\bibfnamefont{F.}~\bibnamefont{Sala}}, \bibnamefont{and}
  \bibinfo{author}{\bibfnamefont{A.}~\bibnamefont{Strumia}},
  \bibinfo{journal}{JCAP} \textbf{\bibinfo{volume}{1103}}, \bibinfo{pages}{051}
  (\bibinfo{year}{2011}), \bibinfo{note}{[Erratum: JCAP1210,E01(2012)]},
  \eprint{1012.4515}.

\bibitem[{\citenamefont{Buch et~al.}(2015)\citenamefont{Buch, Cirelli, Giesen,
  and Taoso}}]{Buch:2015iya}
\bibinfo{author}{\bibfnamefont{J.}~\bibnamefont{Buch}},
  \bibinfo{author}{\bibfnamefont{M.}~\bibnamefont{Cirelli}},
  \bibinfo{author}{\bibfnamefont{G.}~\bibnamefont{Giesen}}, \bibnamefont{and}
  \bibinfo{author}{\bibfnamefont{M.}~\bibnamefont{Taoso}},
  \bibinfo{journal}{JCAP} \textbf{\bibinfo{volume}{1509}}, \bibinfo{pages}{037}
  (\bibinfo{year}{2015}), \eprint{1505.01049}.

\bibitem[{\citenamefont{Arcadi et~al.}(2017)\citenamefont{Arcadi, Dutra, Ghosh,
  Lindner, Mambrini, Pierre, Profumo, and Queiroz}}]{Arcadi:2017kky}
\bibinfo{author}{\bibfnamefont{G.}~\bibnamefont{Arcadi}},
  \bibinfo{author}{\bibfnamefont{M.}~\bibnamefont{Dutra}},
  \bibinfo{author}{\bibfnamefont{P.}~\bibnamefont{Ghosh}},
  \bibinfo{author}{\bibfnamefont{M.}~\bibnamefont{Lindner}},
  \bibinfo{author}{\bibfnamefont{Y.}~\bibnamefont{Mambrini}},
  \bibinfo{author}{\bibfnamefont{M.}~\bibnamefont{Pierre}},
  \bibinfo{author}{\bibfnamefont{S.}~\bibnamefont{Profumo}}, \bibnamefont{and}
  \bibinfo{author}{\bibfnamefont{F.~S.} \bibnamefont{Queiroz}}
  (\bibinfo{year}{2017}), \eprint{1703.07364}.

\bibitem[{\citenamefont{{Bell}}(2013)}]{bell2013}
\bibinfo{author}{\bibfnamefont{A.~R.} \bibnamefont{{Bell}}},
  \bibinfo{journal}{Astroparticle Physics} \textbf{\bibinfo{volume}{43}},
  \bibinfo{pages}{56} (\bibinfo{year}{2013}).

\bibitem[{\citenamefont{{Baade} and {Zwicky}}(1934{\natexlab{a}})}]{baade1934}
\bibinfo{author}{\bibfnamefont{W.}~\bibnamefont{{Baade}}} \bibnamefont{and}
  \bibinfo{author}{\bibfnamefont{F.}~\bibnamefont{{Zwicky}}},
  \bibinfo{journal}{Contributions from the Mount Wilson Observatory, vol.~3,
  pp.79-83} \textbf{\bibinfo{volume}{3}}, \bibinfo{pages}{79}
  (\bibinfo{year}{1934}{\natexlab{a}}).

\bibitem[{\citenamefont{{Baade} and
  {Zwicky}}(1934{\natexlab{b}})}]{baade1934PR}
\bibinfo{author}{\bibfnamefont{W.}~\bibnamefont{{Baade}}} \bibnamefont{and}
  \bibinfo{author}{\bibfnamefont{F.}~\bibnamefont{{Zwicky}}},
  \bibinfo{journal}{Physical Review} \textbf{\bibinfo{volume}{46}},
  \bibinfo{pages}{76} (\bibinfo{year}{1934}{\natexlab{b}}).

\bibitem[{\citenamefont{{Morrison}}(1957)}]{morrison1957}
\bibinfo{author}{\bibfnamefont{P.}~\bibnamefont{{Morrison}}},
  \bibinfo{journal}{Reviews of Modern Physics} \textbf{\bibinfo{volume}{29}},
  \bibinfo{pages}{235} (\bibinfo{year}{1957}).

\bibitem[{\citenamefont{{Ginzburg}}(1956)}]{ginzburg1956}
\bibinfo{author}{\bibfnamefont{V.~L.} \bibnamefont{{Ginzburg}}},
  \bibinfo{journal}{Il Nuovo Cimento} \textbf{\bibinfo{volume}{3}},
  \bibinfo{pages}{38} (\bibinfo{year}{1956}).

\bibitem[{\citenamefont{{Blandford} and {Ostriker}}(1978)}]{blandford1978}
\bibinfo{author}{\bibfnamefont{R.~D.} \bibnamefont{{Blandford}}}
  \bibnamefont{and} \bibinfo{author}{\bibfnamefont{J.~P.}
  \bibnamefont{{Ostriker}}}, \bibinfo{journal}{The Astrophysical Journal
  Letters} \textbf{\bibinfo{volume}{221}}, \bibinfo{pages}{L29}
  (\bibinfo{year}{1978}).

\bibitem[{\citenamefont{{Bell}}(1978)}]{bell1978}
\bibinfo{author}{\bibfnamefont{A.~R.} \bibnamefont{{Bell}}},
  \bibinfo{journal}{MNRAS} \textbf{\bibinfo{volume}{182}}, \bibinfo{pages}{147}
  (\bibinfo{year}{1978}).

\bibitem[{\citenamefont{{Axford} et~al.}(1977)\citenamefont{{Axford}, {Leer},
  and {Skadron}}}]{axford1977}
\bibinfo{author}{\bibfnamefont{W.~I.} \bibnamefont{{Axford}}},
  \bibinfo{author}{\bibfnamefont{E.}~\bibnamefont{{Leer}}}, \bibnamefont{and}
  \bibinfo{author}{\bibfnamefont{G.}~\bibnamefont{{Skadron}}},
  \bibinfo{journal}{International Cosmic Ray Conference}
  \textbf{\bibinfo{volume}{11}}, \bibinfo{pages}{132} (\bibinfo{year}{1977}).

\bibitem[{\citenamefont{{Krymskii}}(1977)}]{krymskii1977}
\bibinfo{author}{\bibfnamefont{G.~F.} \bibnamefont{{Krymskii}}},
  \bibinfo{journal}{Akademiia Nauk SSSR Doklady}
  \textbf{\bibinfo{volume}{234}}, \bibinfo{pages}{1306} (\bibinfo{year}{1977}).

\bibitem[{\citenamefont{{Aharonian} et~al.}(1995)\citenamefont{{Aharonian},
  {Atoyan}, and {Voelk}}}]{aharonian1995}
\bibinfo{author}{\bibfnamefont{F.~A.} \bibnamefont{{Aharonian}}},
  \bibinfo{author}{\bibfnamefont{A.~M.} \bibnamefont{{Atoyan}}},
  \bibnamefont{and} \bibinfo{author}{\bibfnamefont{H.~J.}
  \bibnamefont{{Voelk}}}, \bibinfo{journal}{Astronomy and astrophysics}
  \textbf{\bibinfo{volume}{294}}, \bibinfo{pages}{L41} (\bibinfo{year}{1995}).

\bibitem[{\citenamefont{{Murphy} et~al.}(2016)\citenamefont{{Murphy}, {Sasaki},
  {Binns}, {Brandt}, {Hams}, {Israel}, {Labrador}, {Link}, {Mewaldt},
  {Mitchell} et~al.}}]{murphy2016}
\bibinfo{author}{\bibfnamefont{R.~P.} \bibnamefont{{Murphy}}},
  \bibinfo{author}{\bibfnamefont{M.}~\bibnamefont{{Sasaki}}},
  \bibinfo{author}{\bibfnamefont{W.~R.} \bibnamefont{{Binns}}},
  \bibinfo{author}{\bibfnamefont{T.~J.} \bibnamefont{{Brandt}}},
  \bibinfo{author}{\bibfnamefont{T.}~\bibnamefont{{Hams}}},
  \bibinfo{author}{\bibfnamefont{M.~H.} \bibnamefont{{Israel}}},
  \bibinfo{author}{\bibfnamefont{A.~W.} \bibnamefont{{Labrador}}},
  \bibinfo{author}{\bibfnamefont{J.~T.} \bibnamefont{{Link}}},
  \bibinfo{author}{\bibfnamefont{R.~A.} \bibnamefont{{Mewaldt}}},
  \bibinfo{author}{\bibfnamefont{J.~W.} \bibnamefont{{Mitchell}}},
  \bibnamefont{et~al.}, \bibinfo{journal}{The Astrophysical Journal}
  \textbf{\bibinfo{volume}{831}}, \bibinfo{eid}{148} (\bibinfo{year}{2016}),
  \eprint{1608.08183}.

\bibitem[{\citenamefont{{Heinz} and {Sunyaev}}(2002)}]{heinz2002}
\bibinfo{author}{\bibfnamefont{S.}~\bibnamefont{{Heinz}}} \bibnamefont{and}
  \bibinfo{author}{\bibfnamefont{R.}~\bibnamefont{{Sunyaev}}},
  \bibinfo{journal}{Astronomy and Astrophysics} \textbf{\bibinfo{volume}{390}},
  \bibinfo{pages}{751} (\bibinfo{year}{2002}), \eprint{astro-ph/0204183}.

\bibitem[{\citenamefont{Antoni et~al.}(2004)\citenamefont{Antoni, Apel, Badea,
  Bekk, Bercuci, Blümer, Bozdog, Brancus, Büttner, Daumiller
  et~al.}}]{anisoKASCADE}
\bibinfo{author}{\bibfnamefont{T.}~\bibnamefont{Antoni}},
  \bibinfo{author}{\bibfnamefont{W.~D.} \bibnamefont{Apel}},
  \bibinfo{author}{\bibfnamefont{A.~F.} \bibnamefont{Badea}},
  \bibinfo{author}{\bibfnamefont{K.}~\bibnamefont{Bekk}},
  \bibinfo{author}{\bibfnamefont{A.}~\bibnamefont{Bercuci}},
  \bibinfo{author}{\bibfnamefont{H.}~\bibnamefont{Blümer}},
  \bibinfo{author}{\bibfnamefont{H.}~\bibnamefont{Bozdog}},
  \bibinfo{author}{\bibfnamefont{I.~M.} \bibnamefont{Brancus}},
  \bibinfo{author}{\bibfnamefont{C.}~\bibnamefont{Büttner}},
  \bibinfo{author}{\bibfnamefont{K.}~\bibnamefont{Daumiller}},
  \bibnamefont{et~al.}, \bibinfo{journal}{The Astrophysical Journal}
  \textbf{\bibinfo{volume}{604}}, \bibinfo{pages}{687} (\bibinfo{year}{2004}),
  \urlprefix\url{http://stacks.iop.org/0004-637X/604/i=2/a=687}.

\bibitem[{\citenamefont{Abdo et~al.}(2009)\citenamefont{Abdo, Allen, Aune,
  Berley, Casanova, Chen, Dingus, Ellsworth, Fleysher, Fleysher
  et~al.}}]{anisoMILAGRO}
\bibinfo{author}{\bibfnamefont{A.~A.} \bibnamefont{Abdo}},
  \bibinfo{author}{\bibfnamefont{B.~T.} \bibnamefont{Allen}},
  \bibinfo{author}{\bibfnamefont{T.}~\bibnamefont{Aune}},
  \bibinfo{author}{\bibfnamefont{D.}~\bibnamefont{Berley}},
  \bibinfo{author}{\bibfnamefont{S.}~\bibnamefont{Casanova}},
  \bibinfo{author}{\bibfnamefont{C.}~\bibnamefont{Chen}},
  \bibinfo{author}{\bibfnamefont{B.~L.} \bibnamefont{Dingus}},
  \bibinfo{author}{\bibfnamefont{R.~W.} \bibnamefont{Ellsworth}},
  \bibinfo{author}{\bibfnamefont{L.}~\bibnamefont{Fleysher}},
  \bibinfo{author}{\bibfnamefont{R.}~\bibnamefont{Fleysher}},
  \bibnamefont{et~al.}, \bibinfo{journal}{The Astrophysical Journal}
  \textbf{\bibinfo{volume}{698}}, \bibinfo{pages}{2121} (\bibinfo{year}{2009}),
  \urlprefix\url{http://stacks.iop.org/0004-637X/698/i=2/a=2121}.

\bibitem[{\citenamefont{Aglietta et~al.}(2009)\citenamefont{Aglietta,
  Alekseenko, Alessandro, Antonioli, Arneodo, Bergamasco, Bertaina, Bonino,
  Castellina, Chiavassa et~al.}}]{anisoEASTOP}
\bibinfo{author}{\bibfnamefont{M.}~\bibnamefont{Aglietta}},
  \bibinfo{author}{\bibfnamefont{V.~V.} \bibnamefont{Alekseenko}},
  \bibinfo{author}{\bibfnamefont{B.}~\bibnamefont{Alessandro}},
  \bibinfo{author}{\bibfnamefont{P.}~\bibnamefont{Antonioli}},
  \bibinfo{author}{\bibfnamefont{F.}~\bibnamefont{Arneodo}},
  \bibinfo{author}{\bibfnamefont{L.}~\bibnamefont{Bergamasco}},
  \bibinfo{author}{\bibfnamefont{M.}~\bibnamefont{Bertaina}},
  \bibinfo{author}{\bibfnamefont{R.}~\bibnamefont{Bonino}},
  \bibinfo{author}{\bibfnamefont{A.}~\bibnamefont{Castellina}},
  \bibinfo{author}{\bibfnamefont{A.}~\bibnamefont{Chiavassa}},
  \bibnamefont{et~al.}, \bibinfo{journal}{The Astrophysical Journal Letters}
  \textbf{\bibinfo{volume}{692}}, \bibinfo{pages}{L130} (\bibinfo{year}{2009}),
  \urlprefix\url{http://stacks.iop.org/1538-4357/692/i=2/a=L130}.

\bibitem[{\citenamefont{Amenomori et~al.}(2017)\citenamefont{Amenomori, Bi,
  Chen, Chen, Chen, Cui, Danzengluobu, Ding, Feng, Feng et~al.}}]{anisoTIBET}
\bibinfo{author}{\bibfnamefont{M.}~\bibnamefont{Amenomori}},
  \bibinfo{author}{\bibfnamefont{X.~J.} \bibnamefont{Bi}},
  \bibinfo{author}{\bibfnamefont{D.}~\bibnamefont{Chen}},
  \bibinfo{author}{\bibfnamefont{T.~L.} \bibnamefont{Chen}},
  \bibinfo{author}{\bibfnamefont{W.~Y.} \bibnamefont{Chen}},
  \bibinfo{author}{\bibfnamefont{S.~W.} \bibnamefont{Cui}},
  \bibinfo{author}{\bibnamefont{Danzengluobu}},
  \bibinfo{author}{\bibfnamefont{L.~K.} \bibnamefont{Ding}},
  \bibinfo{author}{\bibfnamefont{C.~F.} \bibnamefont{Feng}},
  \bibinfo{author}{\bibfnamefont{Z.}~\bibnamefont{Feng}}, \bibnamefont{et~al.},
  \bibinfo{journal}{The Astrophysical Journal} \textbf{\bibinfo{volume}{836}},
  \bibinfo{pages}{153} (\bibinfo{year}{2017}),
  \urlprefix\url{http://stacks.iop.org/0004-637X/836/i=2/a=153}.

\bibitem[{\citenamefont{Aartsen et~al.}(2013)\citenamefont{Aartsen, Abbasi,
  Abdou, Ackermann, Adams, Aguilar, Ahlers, Altmann, Andeen, Auffenberg
  et~al.}}]{anisoIceTop}
\bibinfo{author}{\bibfnamefont{M.~G.} \bibnamefont{Aartsen}},
  \bibinfo{author}{\bibfnamefont{R.}~\bibnamefont{Abbasi}},
  \bibinfo{author}{\bibfnamefont{Y.}~\bibnamefont{Abdou}},
  \bibinfo{author}{\bibfnamefont{M.}~\bibnamefont{Ackermann}},
  \bibinfo{author}{\bibfnamefont{J.}~\bibnamefont{Adams}},
  \bibinfo{author}{\bibfnamefont{J.~A.} \bibnamefont{Aguilar}},
  \bibinfo{author}{\bibfnamefont{M.}~\bibnamefont{Ahlers}},
  \bibinfo{author}{\bibfnamefont{D.}~\bibnamefont{Altmann}},
  \bibinfo{author}{\bibfnamefont{K.}~\bibnamefont{Andeen}},
  \bibinfo{author}{\bibfnamefont{J.}~\bibnamefont{Auffenberg}},
  \bibnamefont{et~al.}, \bibinfo{journal}{The Astrophysical Journal}
  \textbf{\bibinfo{volume}{765}}, \bibinfo{pages}{55} (\bibinfo{year}{2013}),
  \urlprefix\url{http://stacks.iop.org/0004-637X/765/i=1/a=55}.

\bibitem[{\citenamefont{Abbasi et~al.}(2012)\citenamefont{Abbasi, Abdou,
  Abu-Zayyad, Ackermann, Adams, Aguilar, Ahlers, Allen, Altmann, Andeen
  et~al.}}]{anisoIceCube}
\bibinfo{author}{\bibfnamefont{R.}~\bibnamefont{Abbasi}},
  \bibinfo{author}{\bibfnamefont{Y.}~\bibnamefont{Abdou}},
  \bibinfo{author}{\bibfnamefont{T.}~\bibnamefont{Abu-Zayyad}},
  \bibinfo{author}{\bibfnamefont{M.}~\bibnamefont{Ackermann}},
  \bibinfo{author}{\bibfnamefont{J.}~\bibnamefont{Adams}},
  \bibinfo{author}{\bibfnamefont{J.~A.} \bibnamefont{Aguilar}},
  \bibinfo{author}{\bibfnamefont{M.}~\bibnamefont{Ahlers}},
  \bibinfo{author}{\bibfnamefont{M.~M.} \bibnamefont{Allen}},
  \bibinfo{author}{\bibfnamefont{D.}~\bibnamefont{Altmann}},
  \bibinfo{author}{\bibfnamefont{K.}~\bibnamefont{Andeen}},
  \bibnamefont{et~al.}, \bibinfo{journal}{The Astrophysical Journal}
  \textbf{\bibinfo{volume}{746}}, \bibinfo{pages}{33} (\bibinfo{year}{2012}),
  \urlprefix\url{http://stacks.iop.org/0004-637X/746/i=1/a=33}.

\bibitem[{\citenamefont{Pollack and Fazio}(1963)}]{pollack1963}
\bibinfo{author}{\bibfnamefont{J.~B.} \bibnamefont{Pollack}} \bibnamefont{and}
  \bibinfo{author}{\bibfnamefont{G.~G.} \bibnamefont{Fazio}},
  \bibinfo{journal}{Phys. Rev.} \textbf{\bibinfo{volume}{131}},
  \bibinfo{pages}{2684} (\bibinfo{year}{1963}),
  \urlprefix\url{https://link.aps.org/doi/10.1103/PhysRev.131.2684}.

\bibitem[{\citenamefont{{Ginzburg} and {Syrovatskii}}(1964)}]{ginzburg1964}
\bibinfo{author}{\bibfnamefont{V.~L.} \bibnamefont{{Ginzburg}}}
  \bibnamefont{and} \bibinfo{author}{\bibfnamefont{S.~I.}
  \bibnamefont{{Syrovatskii}}}, \emph{\bibinfo{title}{{The Origin of Cosmic
  Rays}}} (\bibinfo{year}{1964}).

\bibitem[{\citenamefont{{Berezinskii} et~al.}(1990)\citenamefont{{Berezinskii},
  {Bulanov}, {Dogiel}, and {Ptuskin}}}]{Berezinskii1990}
\bibinfo{author}{\bibfnamefont{V.~S.} \bibnamefont{{Berezinskii}}},
  \bibinfo{author}{\bibfnamefont{S.~V.} \bibnamefont{{Bulanov}}},
  \bibinfo{author}{\bibfnamefont{V.~A.} \bibnamefont{{Dogiel}}},
  \bibnamefont{and} \bibinfo{author}{\bibfnamefont{V.~S.}
  \bibnamefont{{Ptuskin}}}, \emph{\bibinfo{title}{{Astrophysics of cosmic
  rays}}} (\bibinfo{publisher}{North-Holland}, \bibinfo{year}{1990}).

\bibitem[{\citenamefont{{Jokipii}}(1966)}]{Jokipii1966}
\bibinfo{author}{\bibfnamefont{J.~R.} \bibnamefont{{Jokipii}}},
  \bibinfo{journal}{\apj} \textbf{\bibinfo{volume}{146}}, \bibinfo{pages}{480}
  (\bibinfo{year}{1966}).

\bibitem[{\citenamefont{{Jokipii} and {Parker}}(1968)}]{Jokipii1968}
\bibinfo{author}{\bibfnamefont{J.~R.} \bibnamefont{{Jokipii}}}
  \bibnamefont{and} \bibinfo{author}{\bibfnamefont{E.~N.}
  \bibnamefont{{Parker}}}, \bibinfo{journal}{Physical Review Letters}
  \textbf{\bibinfo{volume}{21}}, \bibinfo{pages}{44} (\bibinfo{year}{1968}).

\bibitem[{\citenamefont{{Armstrong} et~al.}(1995)\citenamefont{{Armstrong},
  {Rickett}, and {Spangler}}}]{armstrong1995}
\bibinfo{author}{\bibfnamefont{J.~W.} \bibnamefont{{Armstrong}}},
  \bibinfo{author}{\bibfnamefont{B.~J.} \bibnamefont{{Rickett}}},
  \bibnamefont{and} \bibinfo{author}{\bibfnamefont{S.~R.}
  \bibnamefont{{Spangler}}}, \bibinfo{journal}{\apj}
  \textbf{\bibinfo{volume}{443}}, \bibinfo{pages}{209} (\bibinfo{year}{1995}).

\bibitem[{\citenamefont{{Jansson} and {Farrar}}(2012)}]{farrar2012}
\bibinfo{author}{\bibfnamefont{R.}~\bibnamefont{{Jansson}}} \bibnamefont{and}
  \bibinfo{author}{\bibfnamefont{G.~R.} \bibnamefont{{Farrar}}},
  \bibinfo{journal}{The Astrophysical Journal Letters}
  \textbf{\bibinfo{volume}{761}}, \bibinfo{eid}{L11} (\bibinfo{year}{2012}),
  \eprint{1210.7820}.

\bibitem[{\citenamefont{{Terral} and {Ferri{\`e}re}}(2017)}]{ferriere2017}
\bibinfo{author}{\bibfnamefont{P.}~\bibnamefont{{Terral}}} \bibnamefont{and}
  \bibinfo{author}{\bibfnamefont{K.}~\bibnamefont{{Ferri{\`e}re}}},
  \bibinfo{journal}{Astronomy and Astrophysics} \textbf{\bibinfo{volume}{600}},
  \bibinfo{eid}{A29} (\bibinfo{year}{2017}), \eprint{1611.10222}.

\bibitem[{\citenamefont{{Elmegreen} and {Scalo}}(2004)}]{scalo2004}
\bibinfo{author}{\bibfnamefont{B.~G.} \bibnamefont{{Elmegreen}}}
  \bibnamefont{and} \bibinfo{author}{\bibfnamefont{J.}~\bibnamefont{{Scalo}}},
  \bibinfo{journal}{Annual Review of Astronomy and Astrophysics}
  \textbf{\bibinfo{volume}{42}}, \bibinfo{pages}{211} (\bibinfo{year}{2004}),
  \eprint{astro-ph/0404451}.

\bibitem[{\citenamefont{{Cesarsky}}(1980)}]{cesarsky1980}
\bibinfo{author}{\bibfnamefont{C.~J.} \bibnamefont{{Cesarsky}}},
  \bibinfo{journal}{Annual review of Astronomy and Astrophysics}
  \textbf{\bibinfo{volume}{18}}, \bibinfo{pages}{289} (\bibinfo{year}{1980}).

\bibitem[{\citenamefont{{Blasi} et~al.}(2012)\citenamefont{{Blasi}, {Amato},
  and {Serpico}}}]{blasi2012}
\bibinfo{author}{\bibfnamefont{P.}~\bibnamefont{{Blasi}}},
  \bibinfo{author}{\bibfnamefont{E.}~\bibnamefont{{Amato}}}, \bibnamefont{and}
  \bibinfo{author}{\bibfnamefont{P.~D.} \bibnamefont{{Serpico}}},
  \bibinfo{journal}{Physical Review Letters} \textbf{\bibinfo{volume}{109}},
  \bibinfo{eid}{061101} (\bibinfo{year}{2012}), \eprint{1207.3706}.

\bibitem[{\citenamefont{{Ptuskin} et~al.}(2006)\citenamefont{{Ptuskin},
  {Moskalenko}, {Jones}, {Strong}, and {Zirakashvili}}}]{Ptuskin2006}
\bibinfo{author}{\bibfnamefont{V.~S.} \bibnamefont{{Ptuskin}}},
  \bibinfo{author}{\bibfnamefont{I.~V.} \bibnamefont{{Moskalenko}}},
  \bibinfo{author}{\bibfnamefont{F.~C.} \bibnamefont{{Jones}}},
  \bibinfo{author}{\bibfnamefont{A.~W.} \bibnamefont{{Strong}}},
  \bibnamefont{and} \bibinfo{author}{\bibfnamefont{V.~N.}
  \bibnamefont{{Zirakashvili}}}, \bibinfo{journal}{\apj}
  \textbf{\bibinfo{volume}{642}}, \bibinfo{pages}{902} (\bibinfo{year}{2006}),
  \eprint{astro-ph/0510335}.

\bibitem[{\citenamefont{{Blasi}}(2013)}]{Blasi2013Rev}
\bibinfo{author}{\bibfnamefont{P.}~\bibnamefont{{Blasi}}},
  \bibinfo{journal}{Astronomy and Astrophysics Reviews}
  \textbf{\bibinfo{volume}{21}}, \bibinfo{eid}{70} (\bibinfo{year}{2013}),
  \eprint{1311.7346}.

\bibitem[{\citenamefont{Skilling}(1975)}]{Skilling1975}
\bibinfo{author}{\bibfnamefont{J.}~\bibnamefont{Skilling}},
  \textbf{\bibinfo{volume}{172}}, \bibinfo{pages}{557} (\bibinfo{year}{1975}).

\bibitem[{\citenamefont{Cerri et~al.}(2017)\citenamefont{Cerri, Gaggero,
  Vittino, Evoli, and Grasso}}]{Cerri:2017joy}
\bibinfo{author}{\bibfnamefont{S.~S.} \bibnamefont{Cerri}},
  \bibinfo{author}{\bibfnamefont{D.}~\bibnamefont{Gaggero}},
  \bibinfo{author}{\bibfnamefont{A.}~\bibnamefont{Vittino}},
  \bibinfo{author}{\bibfnamefont{C.}~\bibnamefont{Evoli}}, \bibnamefont{and}
  \bibinfo{author}{\bibfnamefont{D.}~\bibnamefont{Grasso}},
  \bibinfo{journal}{JCAP} \textbf{\bibinfo{volume}{1710}}, \bibinfo{pages}{019}
  (\bibinfo{year}{2017}), \eprint{1707.07694}.

\bibitem[{\citenamefont{{Evoli} et~al.}(2012)\citenamefont{{Evoli}, {Gaggero},
  {Grasso}, and {Maccione}}}]{Evoli2012}
\bibinfo{author}{\bibfnamefont{C.}~\bibnamefont{{Evoli}}},
  \bibinfo{author}{\bibfnamefont{D.}~\bibnamefont{{Gaggero}}},
  \bibinfo{author}{\bibfnamefont{D.}~\bibnamefont{{Grasso}}}, \bibnamefont{and}
  \bibinfo{author}{\bibfnamefont{L.}~\bibnamefont{{Maccione}}},
  \bibinfo{journal}{Physical Review Letters} \textbf{\bibinfo{volume}{108}},
  \bibinfo{eid}{211102} (\bibinfo{year}{2012}), \eprint{1203.0570}.

\bibitem[{\citenamefont{Gaggero
  et~al.}(2015{\natexlab{a}})\citenamefont{Gaggero, Urbano, Valli, and
  Ullio}}]{Gaggero:2014xla}
\bibinfo{author}{\bibfnamefont{D.}~\bibnamefont{Gaggero}},
  \bibinfo{author}{\bibfnamefont{A.}~\bibnamefont{Urbano}},
  \bibinfo{author}{\bibfnamefont{M.}~\bibnamefont{Valli}}, \bibnamefont{and}
  \bibinfo{author}{\bibfnamefont{P.}~\bibnamefont{Ullio}},
  \bibinfo{journal}{Phys. Rev.} \textbf{\bibinfo{volume}{D91}},
  \bibinfo{pages}{083012} (\bibinfo{year}{2015}{\natexlab{a}}),
  \eprint{1411.7623}.

\bibitem[{\citenamefont{{Gaggero} et~al.}(2017)\citenamefont{{Gaggero},
  {Grasso}, {Marinelli}, {Taoso}, and {Urbano}}}]{gaggero2017prl}
\bibinfo{author}{\bibfnamefont{D.}~\bibnamefont{{Gaggero}}},
  \bibinfo{author}{\bibfnamefont{D.}~\bibnamefont{{Grasso}}},
  \bibinfo{author}{\bibfnamefont{A.}~\bibnamefont{{Marinelli}}},
  \bibinfo{author}{\bibfnamefont{M.}~\bibnamefont{{Taoso}}}, \bibnamefont{and}
  \bibinfo{author}{\bibfnamefont{A.}~\bibnamefont{{Urbano}}},
  \bibinfo{journal}{Physical Review Letters} \textbf{\bibinfo{volume}{119}},
  \bibinfo{eid}{031101} (\bibinfo{year}{2017}), \eprint{1702.01124}.

\bibitem[{\citenamefont{{Tomassetti}}(2015)}]{Tomassetti2015twoHalo}
\bibinfo{author}{\bibfnamefont{N.}~\bibnamefont{{Tomassetti}}},
  \bibinfo{journal}{Physical Review D} \textbf{\bibinfo{volume}{92}},
  \bibinfo{eid}{081301} (\bibinfo{year}{2015}), \eprint{1509.05775}.

\bibitem[{\citenamefont{{Feng} et~al.}(2016{\natexlab{a}})\citenamefont{{Feng},
  {Tomassetti}, and {Oliva}}}]{feng2016}
\bibinfo{author}{\bibfnamefont{J.}~\bibnamefont{{Feng}}},
  \bibinfo{author}{\bibfnamefont{N.}~\bibnamefont{{Tomassetti}}},
  \bibnamefont{and} \bibinfo{author}{\bibfnamefont{A.}~\bibnamefont{{Oliva}}},
  \bibinfo{journal}{Physical Review D} \textbf{\bibinfo{volume}{94}},
  \bibinfo{eid}{123007} (\bibinfo{year}{2016}{\natexlab{a}}),
  \eprint{1610.06182}.

\bibitem[{\citenamefont{{Strong} et~al.}(2007)\citenamefont{{Strong},
  {Moskalenko}, and {Ptuskin}}}]{StrongMoskalenko2007}
\bibinfo{author}{\bibfnamefont{A.~W.} \bibnamefont{{Strong}}},
  \bibinfo{author}{\bibfnamefont{I.~V.} \bibnamefont{{Moskalenko}}},
  \bibnamefont{and} \bibinfo{author}{\bibfnamefont{V.~S.}
  \bibnamefont{{Ptuskin}}}, \bibinfo{journal}{Annual Review of Nuclear and
  Particle Science} \textbf{\bibinfo{volume}{57}}, \bibinfo{pages}{285}
  (\bibinfo{year}{2007}), \eprint{astro-ph/0701517}.

\bibitem[{\citenamefont{{Evoli}
  et~al.}(2017{\natexlab{a}})\citenamefont{{Evoli}, {Gaggero}, {Vittino}, {Di
  Bernardo}, {Di Mauro}, {Ligorini}, {Ullio}, and {Grasso}}}]{Evoli2017I}
\bibinfo{author}{\bibfnamefont{C.}~\bibnamefont{{Evoli}}},
  \bibinfo{author}{\bibfnamefont{D.}~\bibnamefont{{Gaggero}}},
  \bibinfo{author}{\bibfnamefont{A.}~\bibnamefont{{Vittino}}},
  \bibinfo{author}{\bibfnamefont{G.}~\bibnamefont{{Di Bernardo}}},
  \bibinfo{author}{\bibfnamefont{M.}~\bibnamefont{{Di Mauro}}},
  \bibinfo{author}{\bibfnamefont{A.}~\bibnamefont{{Ligorini}}},
  \bibinfo{author}{\bibfnamefont{P.}~\bibnamefont{{Ullio}}}, \bibnamefont{and}
  \bibinfo{author}{\bibfnamefont{D.}~\bibnamefont{{Grasso}}},
  \bibinfo{journal}{JCAP} \textbf{\bibinfo{volume}{2}}, \bibinfo{eid}{015}
  (\bibinfo{year}{2017}{\natexlab{a}}), \eprint{1607.07886}.

\bibitem[{\citenamefont{{Sridhar} and {Goldreich}}(1994)}]{GS1994}
\bibinfo{author}{\bibfnamefont{S.}~\bibnamefont{{Sridhar}}} \bibnamefont{and}
  \bibinfo{author}{\bibfnamefont{P.}~\bibnamefont{{Goldreich}}},
  \bibinfo{journal}{The Astrophysical Journal} \textbf{\bibinfo{volume}{432}},
  \bibinfo{pages}{612} (\bibinfo{year}{1994}).

\bibitem[{\citenamefont{{Goldreich} and {Sridhar}}(1995)}]{GS1995}
\bibinfo{author}{\bibfnamefont{P.}~\bibnamefont{{Goldreich}}} \bibnamefont{and}
  \bibinfo{author}{\bibfnamefont{S.}~\bibnamefont{{Sridhar}}},
  \bibinfo{journal}{The Astrophysical Journal} \textbf{\bibinfo{volume}{438}},
  \bibinfo{pages}{763} (\bibinfo{year}{1995}).

\bibitem[{\citenamefont{{Chandran}}(2000)}]{Chandran2000}
\bibinfo{author}{\bibfnamefont{B.~D.~G.} \bibnamefont{{Chandran}}},
  \bibinfo{journal}{Physical Review Letters} \textbf{\bibinfo{volume}{85}},
  \bibinfo{pages}{4656} (\bibinfo{year}{2000}), \eprint{astro-ph/0008498}.

\bibitem[{\citenamefont{{Yan} and {Lazarian}}(2002)}]{Yan2002}
\bibinfo{author}{\bibfnamefont{H.}~\bibnamefont{{Yan}}} \bibnamefont{and}
  \bibinfo{author}{\bibfnamefont{A.}~\bibnamefont{{Lazarian}}},
  \bibinfo{journal}{Physical Review Letters} \textbf{\bibinfo{volume}{89}},
  \bibinfo{eid}{281102} (\bibinfo{year}{2002}), \eprint{astro-ph/0205285}.

\bibitem[{\citenamefont{{Yan} and {Lazarian}}(2008)}]{Yan2008}
\bibinfo{author}{\bibfnamefont{H.}~\bibnamefont{{Yan}}} \bibnamefont{and}
  \bibinfo{author}{\bibfnamefont{A.}~\bibnamefont{{Lazarian}}},
  \bibinfo{journal}{The Astrophysical Journal} \textbf{\bibinfo{volume}{673}},
  \bibinfo{eid}{942-953} (\bibinfo{year}{2008}), \eprint{0710.2617}.

\bibitem[{\citenamefont{{Strong} and {Youssefi}}(1995)}]{Galprop1}
\bibinfo{author}{\bibfnamefont{A.~W.} \bibnamefont{{Strong}}} \bibnamefont{and}
  \bibinfo{author}{\bibfnamefont{G.}~\bibnamefont{{Youssefi}}},
  \bibinfo{journal}{International Cosmic Ray Conference}
  \textbf{\bibinfo{volume}{3}}, \bibinfo{pages}{48} (\bibinfo{year}{1995}).

\bibitem[{\citenamefont{Strong and Moskalenko}(1998)}]{Galprop2}
\bibinfo{author}{\bibfnamefont{A.~W.} \bibnamefont{Strong}} \bibnamefont{and}
  \bibinfo{author}{\bibfnamefont{I.~V.} \bibnamefont{Moskalenko}},
  \bibinfo{journal}{Astrophys. J.} \textbf{\bibinfo{volume}{509}},
  \bibinfo{pages}{212} (\bibinfo{year}{1998}), \eprint{astro-ph/9807150}.

\bibitem[{\citenamefont{Moskalenko and Strong}(1998)}]{Galprop3}
\bibinfo{author}{\bibfnamefont{I.~V.} \bibnamefont{Moskalenko}}
  \bibnamefont{and} \bibinfo{author}{\bibfnamefont{A.~W.}
  \bibnamefont{Strong}}, \bibinfo{journal}{Astrophys. J.}
  \textbf{\bibinfo{volume}{493}}, \bibinfo{pages}{694} (\bibinfo{year}{1998}),
  \eprint{astro-ph/9710124}.

\bibitem[{\citenamefont{{Strong} and {Moskalenko}}(2001)}]{2001ICRC....5.1942S}
\bibinfo{author}{\bibfnamefont{A.~W.} \bibnamefont{{Strong}}} \bibnamefont{and}
  \bibinfo{author}{\bibfnamefont{I.~V.} \bibnamefont{{Moskalenko}}},
  \bibinfo{journal}{International Cosmic Ray Conference}
  \textbf{\bibinfo{volume}{5}}, \bibinfo{pages}{1942} (\bibinfo{year}{2001}),
  \eprint{astro-ph/0106504}.

\bibitem[{\citenamefont{Evoli et~al.}(2008)\citenamefont{Evoli, Gaggero,
  Grasso, and Maccione}}]{Evoli:2008dv}
\bibinfo{author}{\bibfnamefont{C.}~\bibnamefont{Evoli}},
  \bibinfo{author}{\bibfnamefont{D.}~\bibnamefont{Gaggero}},
  \bibinfo{author}{\bibfnamefont{D.}~\bibnamefont{Grasso}}, \bibnamefont{and}
  \bibinfo{author}{\bibfnamefont{L.}~\bibnamefont{Maccione}},
  \bibinfo{journal}{JCAP} \textbf{\bibinfo{volume}{0810}}, \bibinfo{pages}{018}
  (\bibinfo{year}{2008}), \eprint{0807.4730}.

\bibitem[{\citenamefont{Gaggero et~al.}(2013)\citenamefont{Gaggero, Maccione,
  Di~Bernardo, Evoli, and Grasso}}]{Gaggero:2013rya}
\bibinfo{author}{\bibfnamefont{D.}~\bibnamefont{Gaggero}},
  \bibinfo{author}{\bibfnamefont{L.}~\bibnamefont{Maccione}},
  \bibinfo{author}{\bibfnamefont{G.}~\bibnamefont{Di~Bernardo}},
  \bibinfo{author}{\bibfnamefont{C.}~\bibnamefont{Evoli}}, \bibnamefont{and}
  \bibinfo{author}{\bibfnamefont{D.}~\bibnamefont{Grasso}},
  \bibinfo{journal}{Phys. Rev. Lett.} \textbf{\bibinfo{volume}{111}},
  \bibinfo{pages}{021102} (\bibinfo{year}{2013}), \eprint{1304.6718}.

\bibitem[{\citenamefont{{Evoli}
  et~al.}(2017{\natexlab{b}})\citenamefont{{Evoli}, {Gaggero}, {Vittino}, {Di
  Mauro}, {Grasso}, and {Mazziotta}}}]{Evoli2017II}
\bibinfo{author}{\bibfnamefont{C.}~\bibnamefont{{Evoli}}},
  \bibinfo{author}{\bibfnamefont{D.}~\bibnamefont{{Gaggero}}},
  \bibinfo{author}{\bibfnamefont{A.}~\bibnamefont{{Vittino}}},
  \bibinfo{author}{\bibfnamefont{M.}~\bibnamefont{{Di Mauro}}},
  \bibinfo{author}{\bibfnamefont{D.}~\bibnamefont{{Grasso}}}, \bibnamefont{and}
  \bibinfo{author}{\bibfnamefont{M.~N.} \bibnamefont{{Mazziotta}}},
  \bibinfo{journal}{ArXiv e-prints}  (\bibinfo{year}{2017}{\natexlab{b}}),
  \eprint{1711.09616}.

\bibitem[{\citenamefont{Kissmann}(2014)}]{Picard1}
\bibinfo{author}{\bibfnamefont{R.}~\bibnamefont{Kissmann}},
  \bibinfo{journal}{Astropart. Phys.} \textbf{\bibinfo{volume}{55}},
  \bibinfo{pages}{37} (\bibinfo{year}{2014}), \eprint{1401.4035}.

\bibitem[{\citenamefont{Werner et~al.}(2015)\citenamefont{Werner, Kissmann,
  Strong, and Reimer}}]{Picard2}
\bibinfo{author}{\bibfnamefont{M.}~\bibnamefont{Werner}},
  \bibinfo{author}{\bibfnamefont{R.}~\bibnamefont{Kissmann}},
  \bibinfo{author}{\bibfnamefont{A.~W.} \bibnamefont{Strong}},
  \bibnamefont{and} \bibinfo{author}{\bibfnamefont{O.}~\bibnamefont{Reimer}},
  \bibinfo{journal}{Astropart. Phys.} \textbf{\bibinfo{volume}{64}},
  \bibinfo{pages}{18} (\bibinfo{year}{2015}), \eprint{1410.5266}.

\bibitem[{\citenamefont{Maurin et~al.}(2001)\citenamefont{Maurin, Donato,
  Taillet, and Salati}}]{Usine}
\bibinfo{author}{\bibfnamefont{D.}~\bibnamefont{Maurin}},
  \bibinfo{author}{\bibfnamefont{F.}~\bibnamefont{Donato}},
  \bibinfo{author}{\bibfnamefont{R.}~\bibnamefont{Taillet}}, \bibnamefont{and}
  \bibinfo{author}{\bibfnamefont{P.}~\bibnamefont{Salati}},
  \bibinfo{journal}{Astrophys. J.} \textbf{\bibinfo{volume}{555}},
  \bibinfo{pages}{585} (\bibinfo{year}{2001}), \eprint{astro-ph/0101231}.

\bibitem[{\citenamefont{Aguilar
  et~al.}(2016{\natexlab{b}})\citenamefont{Aguilar, Ali~Cavasonza, Ambrosi,
  Arruda, Attig, Aupetit, Azzarello, Bachlechner, Barao, Barrau
  et~al.}}]{PhysRevLett.117.231102}
\bibinfo{author}{\bibfnamefont{M.}~\bibnamefont{Aguilar}},
  \bibinfo{author}{\bibfnamefont{L.}~\bibnamefont{Ali~Cavasonza}},
  \bibinfo{author}{\bibfnamefont{G.}~\bibnamefont{Ambrosi}},
  \bibinfo{author}{\bibfnamefont{L.}~\bibnamefont{Arruda}},
  \bibinfo{author}{\bibfnamefont{N.}~\bibnamefont{Attig}},
  \bibinfo{author}{\bibfnamefont{S.}~\bibnamefont{Aupetit}},
  \bibinfo{author}{\bibfnamefont{P.}~\bibnamefont{Azzarello}},
  \bibinfo{author}{\bibfnamefont{A.}~\bibnamefont{Bachlechner}},
  \bibinfo{author}{\bibfnamefont{F.}~\bibnamefont{Barao}},
  \bibinfo{author}{\bibfnamefont{A.}~\bibnamefont{Barrau}},
  \bibnamefont{et~al.} (\bibinfo{collaboration}{AMS Collaboration}),
  \bibinfo{journal}{Phys. Rev. Lett.} \textbf{\bibinfo{volume}{117}},
  \bibinfo{pages}{231102} (\bibinfo{year}{2016}{\natexlab{b}}),
  \urlprefix\url{https://link.aps.org/doi/10.1103/PhysRevLett.117.231102}.

\bibitem[{\citenamefont{{Trotta}
  et~al.}(2011{\natexlab{a}})\citenamefont{{Trotta}, {J{\'o}hannesson},
  {Moskalenko}, {Porter}, {Ruiz de Austri}, and {Strong}}}]{trotta2011}
\bibinfo{author}{\bibfnamefont{R.}~\bibnamefont{{Trotta}}},
  \bibinfo{author}{\bibfnamefont{G.}~\bibnamefont{{J{\'o}hannesson}}},
  \bibinfo{author}{\bibfnamefont{I.~V.} \bibnamefont{{Moskalenko}}},
  \bibinfo{author}{\bibfnamefont{T.~A.} \bibnamefont{{Porter}}},
  \bibinfo{author}{\bibfnamefont{R.}~\bibnamefont{{Ruiz de Austri}}},
  \bibnamefont{and} \bibinfo{author}{\bibfnamefont{A.~W.}
  \bibnamefont{{Strong}}}, \bibinfo{journal}{The Astrophysical Journal}
  \textbf{\bibinfo{volume}{729}}, \bibinfo{eid}{106}
  (\bibinfo{year}{2011}{\natexlab{a}}), \eprint{1011.0037}.

\bibitem[{\citenamefont{{J{\'o}hannesson}
  et~al.}(2016)\citenamefont{{J{\'o}hannesson}, {Ruiz de Austri}, {Vincent},
  {Moskalenko}, {Orlando}, {Porter}, {Strong}, {Trotta}, {Feroz}, {Graff}
  et~al.}}]{trotta2016}
\bibinfo{author}{\bibfnamefont{G.}~\bibnamefont{{J{\'o}hannesson}}},
  \bibinfo{author}{\bibfnamefont{R.}~\bibnamefont{{Ruiz de Austri}}},
  \bibinfo{author}{\bibfnamefont{A.~C.} \bibnamefont{{Vincent}}},
  \bibinfo{author}{\bibfnamefont{I.~V.} \bibnamefont{{Moskalenko}}},
  \bibinfo{author}{\bibfnamefont{E.}~\bibnamefont{{Orlando}}},
  \bibinfo{author}{\bibfnamefont{T.~A.} \bibnamefont{{Porter}}},
  \bibinfo{author}{\bibfnamefont{A.~W.} \bibnamefont{{Strong}}},
  \bibinfo{author}{\bibfnamefont{R.}~\bibnamefont{{Trotta}}},
  \bibinfo{author}{\bibfnamefont{F.}~\bibnamefont{{Feroz}}},
  \bibinfo{author}{\bibfnamefont{P.}~\bibnamefont{{Graff}}},
  \bibnamefont{et~al.}, \bibinfo{journal}{The Astrophysical Journal}
  \textbf{\bibinfo{volume}{824}}, \bibinfo{eid}{16} (\bibinfo{year}{2016}),
  \eprint{1602.02243}.

\bibitem[{\citenamefont{Evoli et~al.}(2015)\citenamefont{Evoli, Gaggero, and
  Grasso}}]{Evoli:2015vaa}
\bibinfo{author}{\bibfnamefont{C.}~\bibnamefont{Evoli}},
  \bibinfo{author}{\bibfnamefont{D.}~\bibnamefont{Gaggero}}, \bibnamefont{and}
  \bibinfo{author}{\bibfnamefont{D.}~\bibnamefont{Grasso}},
  \bibinfo{journal}{JCAP} \textbf{\bibinfo{volume}{1512}}, \bibinfo{pages}{039}
  (\bibinfo{year}{2015}), \eprint{1504.05175}.

\bibitem[{\citenamefont{{Yuan} et~al.}(2017{\natexlab{a}})\citenamefont{{Yuan},
  {Lin}, {Fang}, and {Bi}}}]{yuan2017ams}
\bibinfo{author}{\bibfnamefont{Q.}~\bibnamefont{{Yuan}}},
  \bibinfo{author}{\bibfnamefont{S.-J.} \bibnamefont{{Lin}}},
  \bibinfo{author}{\bibfnamefont{K.}~\bibnamefont{{Fang}}}, \bibnamefont{and}
  \bibinfo{author}{\bibfnamefont{X.-J.} \bibnamefont{{Bi}}},
  \bibinfo{journal}{Physical Review D} \textbf{\bibinfo{volume}{95}},
  \bibinfo{eid}{083007} (\bibinfo{year}{2017}{\natexlab{a}}),
  \eprint{1701.06149}.

\bibitem[{\citenamefont{{Niu} and {Li}}(2017)}]{niu2017}
\bibinfo{author}{\bibfnamefont{J.-S.} \bibnamefont{{Niu}}} \bibnamefont{and}
  \bibinfo{author}{\bibfnamefont{T.}~\bibnamefont{{Li}}},
  \bibinfo{journal}{ArXiv e-prints}  (\bibinfo{year}{2017}),
  \eprint{1705.11089}.

\bibitem[{\citenamefont{{Giesen} et~al.}(2015)\citenamefont{{Giesen},
  {Boudaud}, {G{\'e}nolini}, {Poulin}, {Cirelli}, {Salati}, and
  {Serpico}}}]{giesen2015}
\bibinfo{author}{\bibfnamefont{G.}~\bibnamefont{{Giesen}}},
  \bibinfo{author}{\bibfnamefont{M.}~\bibnamefont{{Boudaud}}},
  \bibinfo{author}{\bibfnamefont{Y.}~\bibnamefont{{G{\'e}nolini}}},
  \bibinfo{author}{\bibfnamefont{V.}~\bibnamefont{{Poulin}}},
  \bibinfo{author}{\bibfnamefont{M.}~\bibnamefont{{Cirelli}}},
  \bibinfo{author}{\bibfnamefont{P.}~\bibnamefont{{Salati}}}, \bibnamefont{and}
  \bibinfo{author}{\bibfnamefont{P.~D.} \bibnamefont{{Serpico}}},
  \bibinfo{journal}{JCAP} \textbf{\bibinfo{volume}{9}}, \bibinfo{eid}{023}
  (\bibinfo{year}{2015}), \eprint{1504.04276}.

\bibitem[{\citenamefont{{Buffington} and {Schindler}}(1981)}]{buffington1981}
\bibinfo{author}{\bibfnamefont{A.}~\bibnamefont{{Buffington}}}
  \bibnamefont{and} \bibinfo{author}{\bibfnamefont{S.~M.}
  \bibnamefont{{Schindler}}}, \bibinfo{journal}{The Astrophysical journal
  letters} \textbf{\bibinfo{volume}{247}}, \bibinfo{pages}{L105}
  (\bibinfo{year}{1981}).

\bibitem[{\citenamefont{{Ellis} et~al.}(1988)\citenamefont{{Ellis}, {Flores},
  {Freese}, {Ritz}, {Seckel}, and {Silk}}}]{ellis1988}
\bibinfo{author}{\bibfnamefont{J.}~\bibnamefont{{Ellis}}},
  \bibinfo{author}{\bibfnamefont{R.~A.} \bibnamefont{{Flores}}},
  \bibinfo{author}{\bibfnamefont{K.}~\bibnamefont{{Freese}}},
  \bibinfo{author}{\bibfnamefont{S.}~\bibnamefont{{Ritz}}},
  \bibinfo{author}{\bibfnamefont{D.}~\bibnamefont{{Seckel}}}, \bibnamefont{and}
  \bibinfo{author}{\bibfnamefont{J.}~\bibnamefont{{Silk}}},
  \bibinfo{journal}{Physics Letters B} \textbf{\bibinfo{volume}{214}},
  \bibinfo{pages}{403} (\bibinfo{year}{1988}).

\bibitem[{\citenamefont{{Rudaz} and {Stecker}}(1988)}]{rudaz1988}
\bibinfo{author}{\bibfnamefont{S.}~\bibnamefont{{Rudaz}}} \bibnamefont{and}
  \bibinfo{author}{\bibfnamefont{F.~W.} \bibnamefont{{Stecker}}},
  \bibinfo{journal}{\apj} \textbf{\bibinfo{volume}{325}}, \bibinfo{pages}{16}
  (\bibinfo{year}{1988}).

\bibitem[{\citenamefont{{Bergstr{\"o}m}
  et~al.}(1999)\citenamefont{{Bergstr{\"o}m}, {Edsj{\"o}}, and
  {Ullio}}}]{bergstrom1999}
\bibinfo{author}{\bibfnamefont{L.}~\bibnamefont{{Bergstr{\"o}m}}},
  \bibinfo{author}{\bibfnamefont{J.}~\bibnamefont{{Edsj{\"o}}}},
  \bibnamefont{and} \bibinfo{author}{\bibfnamefont{P.}~\bibnamefont{{Ullio}}},
  \bibinfo{journal}{\apj} \textbf{\bibinfo{volume}{526}}, \bibinfo{pages}{215}
  (\bibinfo{year}{1999}), \eprint{astro-ph/9902012}.

\bibitem[{\citenamefont{{Adriani}
  et~al.}(2009{\natexlab{b}})\citenamefont{{Adriani}, {Barbarino},
  {Bazilevskaya}, {Bellotti}, {Boezio}, {Bogomolov}, {Bonechi}, {Bongi},
  {Bonvicini}, {Bottai} et~al.}}]{pamela2009pbar}
\bibinfo{author}{\bibfnamefont{O.}~\bibnamefont{{Adriani}}},
  \bibinfo{author}{\bibfnamefont{G.~C.} \bibnamefont{{Barbarino}}},
  \bibinfo{author}{\bibfnamefont{G.~A.} \bibnamefont{{Bazilevskaya}}},
  \bibinfo{author}{\bibfnamefont{R.}~\bibnamefont{{Bellotti}}},
  \bibinfo{author}{\bibfnamefont{M.}~\bibnamefont{{Boezio}}},
  \bibinfo{author}{\bibfnamefont{E.~A.} \bibnamefont{{Bogomolov}}},
  \bibinfo{author}{\bibfnamefont{L.}~\bibnamefont{{Bonechi}}},
  \bibinfo{author}{\bibfnamefont{M.}~\bibnamefont{{Bongi}}},
  \bibinfo{author}{\bibfnamefont{V.}~\bibnamefont{{Bonvicini}}},
  \bibinfo{author}{\bibfnamefont{S.}~\bibnamefont{{Bottai}}},
  \bibnamefont{et~al.}, \bibinfo{journal}{Physical Review Letters}
  \textbf{\bibinfo{volume}{102}}, \bibinfo{eid}{051101}
  (\bibinfo{year}{2009}{\natexlab{b}}), \eprint{0810.4994}.

\bibitem[{\citenamefont{{Adriani} et~al.}(2010)\citenamefont{{Adriani},
  {Barbarino}, {Bazilevskaya}, {Bellotti}, {Boezio}, {Bogomolov}, {Bonechi},
  {Bongi}, {Bonvicini}, {Borisov} et~al.}}]{pamela2010pbar}
\bibinfo{author}{\bibfnamefont{O.}~\bibnamefont{{Adriani}}},
  \bibinfo{author}{\bibfnamefont{G.~C.} \bibnamefont{{Barbarino}}},
  \bibinfo{author}{\bibfnamefont{G.~A.} \bibnamefont{{Bazilevskaya}}},
  \bibinfo{author}{\bibfnamefont{R.}~\bibnamefont{{Bellotti}}},
  \bibinfo{author}{\bibfnamefont{M.}~\bibnamefont{{Boezio}}},
  \bibinfo{author}{\bibfnamefont{E.~A.} \bibnamefont{{Bogomolov}}},
  \bibinfo{author}{\bibfnamefont{L.}~\bibnamefont{{Bonechi}}},
  \bibinfo{author}{\bibfnamefont{M.}~\bibnamefont{{Bongi}}},
  \bibinfo{author}{\bibfnamefont{V.}~\bibnamefont{{Bonvicini}}},
  \bibinfo{author}{\bibfnamefont{S.}~\bibnamefont{{Borisov}}},
  \bibnamefont{et~al.}, \bibinfo{journal}{Physical Review Letters}
  \textbf{\bibinfo{volume}{105}}, \bibinfo{eid}{121101} (\bibinfo{year}{2010}),
  \eprint{1007.0821}.

\bibitem[{\citenamefont{{Di Bernardo} et~al.}(2010)\citenamefont{{Di Bernardo},
  {Evoli}, {Gaggero}, {Grasso}, and {Maccione}}}]{dibernardo2010pbar}
\bibinfo{author}{\bibfnamefont{G.}~\bibnamefont{{Di Bernardo}}},
  \bibinfo{author}{\bibfnamefont{C.}~\bibnamefont{{Evoli}}},
  \bibinfo{author}{\bibfnamefont{D.}~\bibnamefont{{Gaggero}}},
  \bibinfo{author}{\bibfnamefont{D.}~\bibnamefont{{Grasso}}}, \bibnamefont{and}
  \bibinfo{author}{\bibfnamefont{L.}~\bibnamefont{{Maccione}}},
  \bibinfo{journal}{Astroparticle Physics} \textbf{\bibinfo{volume}{34}},
  \bibinfo{pages}{274} (\bibinfo{year}{2010}), \eprint{0909.4548}.

\bibitem[{\citenamefont{{Trotta}
  et~al.}(2011{\natexlab{b}})\citenamefont{{Trotta}, {J{\'o}hannesson},
  {Moskalenko}, {Porter}, {Ruiz de Austri}, and {Strong}}}]{trotta2011pbar}
\bibinfo{author}{\bibfnamefont{R.}~\bibnamefont{{Trotta}}},
  \bibinfo{author}{\bibfnamefont{G.}~\bibnamefont{{J{\'o}hannesson}}},
  \bibinfo{author}{\bibfnamefont{I.~V.} \bibnamefont{{Moskalenko}}},
  \bibinfo{author}{\bibfnamefont{T.~A.} \bibnamefont{{Porter}}},
  \bibinfo{author}{\bibfnamefont{R.}~\bibnamefont{{Ruiz de Austri}}},
  \bibnamefont{and} \bibinfo{author}{\bibfnamefont{A.~W.}
  \bibnamefont{{Strong}}}, \bibinfo{journal}{\apj}
  \textbf{\bibinfo{volume}{729}}, \bibinfo{eid}{106}
  (\bibinfo{year}{2011}{\natexlab{b}}), \eprint{1011.0037}.

\bibitem[{\citenamefont{{Bringmann} and {Salati}}(2007)}]{bringmann2007}
\bibinfo{author}{\bibfnamefont{T.}~\bibnamefont{{Bringmann}}} \bibnamefont{and}
  \bibinfo{author}{\bibfnamefont{P.}~\bibnamefont{{Salati}}},
  \bibinfo{journal}{Physical Review D} \textbf{\bibinfo{volume}{75}},
  \bibinfo{eid}{083006} (\bibinfo{year}{2007}), \eprint{astro-ph/0612514}.

\bibitem[{\citenamefont{{Donato} et~al.}(2009)\citenamefont{{Donato}, {Maurin},
  {Brun}, {Delahaye}, and {Salati}}}]{donato2009}
\bibinfo{author}{\bibfnamefont{F.}~\bibnamefont{{Donato}}},
  \bibinfo{author}{\bibfnamefont{D.}~\bibnamefont{{Maurin}}},
  \bibinfo{author}{\bibfnamefont{P.}~\bibnamefont{{Brun}}},
  \bibinfo{author}{\bibfnamefont{T.}~\bibnamefont{{Delahaye}}},
  \bibnamefont{and} \bibinfo{author}{\bibfnamefont{P.}~\bibnamefont{{Salati}}},
  \bibinfo{journal}{Physical Review Letters} \textbf{\bibinfo{volume}{102}},
  \bibinfo{eid}{071301} (\bibinfo{year}{2009}), \eprint{0810.5292}.

\bibitem[{\citenamefont{{Cirelli} et~al.}(2014)\citenamefont{{Cirelli},
  {Gaggero}, {Giesen}, {Taoso}, and {Urbano}}}]{cirelli2014}
\bibinfo{author}{\bibfnamefont{M.}~\bibnamefont{{Cirelli}}},
  \bibinfo{author}{\bibfnamefont{D.}~\bibnamefont{{Gaggero}}},
  \bibinfo{author}{\bibfnamefont{G.}~\bibnamefont{{Giesen}}},
  \bibinfo{author}{\bibfnamefont{M.}~\bibnamefont{{Taoso}}}, \bibnamefont{and}
  \bibinfo{author}{\bibfnamefont{A.}~\bibnamefont{{Urbano}}},
  \bibinfo{journal}{JCAP} \textbf{\bibinfo{volume}{12}}, \bibinfo{eid}{045}
  (\bibinfo{year}{2014}), \eprint{1407.2173}.

\bibitem[{\citenamefont{{Strong} et~al.}(2011)\citenamefont{{Strong},
  {Orlando}, and {Jaffe}}}]{strong2011}
\bibinfo{author}{\bibfnamefont{A.~W.} \bibnamefont{{Strong}}},
  \bibinfo{author}{\bibfnamefont{E.}~\bibnamefont{{Orlando}}},
  \bibnamefont{and} \bibinfo{author}{\bibfnamefont{T.~R.}
  \bibnamefont{{Jaffe}}}, \bibinfo{journal}{Astronomy and astrophysics}
  \textbf{\bibinfo{volume}{534}}, \bibinfo{eid}{A54} (\bibinfo{year}{2011}),
  \eprint{1108.4822}.

\bibitem[{\citenamefont{Di~Bernardo et~al.}(2013)\citenamefont{Di~Bernardo,
  Evoli, Gaggero, Grasso, and Maccione}}]{DiBernardo:2012zu}
\bibinfo{author}{\bibfnamefont{G.}~\bibnamefont{Di~Bernardo}},
  \bibinfo{author}{\bibfnamefont{C.}~\bibnamefont{Evoli}},
  \bibinfo{author}{\bibfnamefont{D.}~\bibnamefont{Gaggero}},
  \bibinfo{author}{\bibfnamefont{D.}~\bibnamefont{Grasso}}, \bibnamefont{and}
  \bibinfo{author}{\bibfnamefont{L.}~\bibnamefont{Maccione}},
  \bibinfo{journal}{JCAP} \textbf{\bibinfo{volume}{1303}}, \bibinfo{pages}{036}
  (\bibinfo{year}{2013}), \eprint{1210.4546}.

\bibitem[{\citenamefont{{Evoli} et~al.}(2015)\citenamefont{{Evoli}, {Gaggero},
  and {Grasso}}}]{evoli2015pbar}
\bibinfo{author}{\bibfnamefont{C.}~\bibnamefont{{Evoli}}},
  \bibinfo{author}{\bibfnamefont{D.}~\bibnamefont{{Gaggero}}},
  \bibnamefont{and} \bibinfo{author}{\bibfnamefont{D.}~\bibnamefont{{Grasso}}},
  \bibinfo{journal}{JCAP} \textbf{\bibinfo{volume}{12}}, \bibinfo{eid}{039}
  (\bibinfo{year}{2015}), \eprint{1504.05175}.

\bibitem[{\citenamefont{{Duperray} et~al.}(2003)\citenamefont{{Duperray},
  {Huang}, {Protasov}, and {Bu{\'e}nerd}}}]{Duperray:2003bd}
\bibinfo{author}{\bibfnamefont{R.~P.} \bibnamefont{{Duperray}}},
  \bibinfo{author}{\bibfnamefont{C.-Y.} \bibnamefont{{Huang}}},
  \bibinfo{author}{\bibfnamefont{K.~V.} \bibnamefont{{Protasov}}},
  \bibnamefont{and}
  \bibinfo{author}{\bibfnamefont{M.}~\bibnamefont{{Bu{\'e}nerd}}},
  \bibinfo{journal}{\prd} \textbf{\bibinfo{volume}{68}}, \bibinfo{eid}{094017}
  (\bibinfo{year}{2003}), \eprint{astro-ph/0305274}.

\bibitem[{\citenamefont{di~Mauro et~al.}(2014)\citenamefont{di~Mauro, Donato,
  Goudelis, and Serpico}}]{diMauro:2014zea}
\bibinfo{author}{\bibfnamefont{M.}~\bibnamefont{di~Mauro}},
  \bibinfo{author}{\bibfnamefont{F.}~\bibnamefont{Donato}},
  \bibinfo{author}{\bibfnamefont{A.}~\bibnamefont{Goudelis}}, \bibnamefont{and}
  \bibinfo{author}{\bibfnamefont{P.~D.} \bibnamefont{Serpico}},
  \bibinfo{journal}{Phys. Rev.} \textbf{\bibinfo{volume}{D90}},
  \bibinfo{pages}{085017} (\bibinfo{year}{2014}), \eprint{1408.0288}.

\bibitem[{\citenamefont{Kappl and Winkler}(2014)}]{Kappl:2014hha}
\bibinfo{author}{\bibfnamefont{R.}~\bibnamefont{Kappl}} \bibnamefont{and}
  \bibinfo{author}{\bibfnamefont{M.~W.} \bibnamefont{Winkler}},
  \bibinfo{journal}{JCAP} \textbf{\bibinfo{volume}{1409}}, \bibinfo{pages}{051}
  (\bibinfo{year}{2014}), \eprint{1408.0299}.

\bibitem[{\citenamefont{Winkler}(2017)}]{Winkler:2017xor}
\bibinfo{author}{\bibfnamefont{M.~W.} \bibnamefont{Winkler}},
  \bibinfo{journal}{JCAP} \textbf{\bibinfo{volume}{1702}}, \bibinfo{pages}{048}
  (\bibinfo{year}{2017}), \eprint{1701.04866}.

\bibitem[{\citenamefont{Pierog et~al.}(2015)\citenamefont{Pierog, Karpenko,
  Katzy, Yatsenko, and Werner}}]{PhysRevC.92.034906}
\bibinfo{author}{\bibfnamefont{T.}~\bibnamefont{Pierog}},
  \bibinfo{author}{\bibfnamefont{I.}~\bibnamefont{Karpenko}},
  \bibinfo{author}{\bibfnamefont{J.~M.} \bibnamefont{Katzy}},
  \bibinfo{author}{\bibfnamefont{E.}~\bibnamefont{Yatsenko}}, \bibnamefont{and}
  \bibinfo{author}{\bibfnamefont{K.}~\bibnamefont{Werner}},
  \bibinfo{journal}{Phys. Rev. C} \textbf{\bibinfo{volume}{92}},
  \bibinfo{pages}{034906} (\bibinfo{year}{2015}),
  \urlprefix\url{http://link.aps.org/doi/10.1103/PhysRevC.92.034906}.

\bibitem[{\citenamefont{Engel et~al.}(1999)\citenamefont{Engel, Gaisser,
  Stanev, and Lipari}}]{Engel:1999db}
\bibinfo{author}{\bibfnamefont{R.}~\bibnamefont{Engel}},
  \bibinfo{author}{\bibfnamefont{T.~K.} \bibnamefont{Gaisser}},
  \bibinfo{author}{\bibfnamefont{T.}~\bibnamefont{Stanev}}, \bibnamefont{and}
  \bibinfo{author}{\bibfnamefont{P.}~\bibnamefont{Lipari}}, in
  \emph{\bibinfo{booktitle}{{Proceedings, 26th International Cosmic Ray
  Conference, August 17-25, 1999, Salt Lake City: Invited, Rapporteur, and
  Highlight Papers}}} (\bibinfo{year}{1999}), vol.~\bibinfo{volume}{1}, pp.
  \bibinfo{pages}{415--418},
  \urlprefix\url{http://krusty.physics.utah.edu/~icrc1999/root/vol1/h2_5_03.pdf}.

\bibitem[{\citenamefont{Ostapchenko}(2011)}]{PhysRevD.83.014018}
\bibinfo{author}{\bibfnamefont{S.}~\bibnamefont{Ostapchenko}},
  \bibinfo{journal}{Phys. Rev. D} \textbf{\bibinfo{volume}{83}},
  \bibinfo{pages}{014018} (\bibinfo{year}{2011}),
  \urlprefix\url{http://link.aps.org/doi/10.1103/PhysRevD.83.014018}.

\bibitem[{\citenamefont{{Blasi}}(2009)}]{blasi2009}
\bibinfo{author}{\bibfnamefont{P.}~\bibnamefont{{Blasi}}},
  \bibinfo{journal}{Physical Review Letters} \textbf{\bibinfo{volume}{103}},
  \bibinfo{eid}{051104} (\bibinfo{year}{2009}), \eprint{0903.2794}.

\bibitem[{\citenamefont{{Mertsch} and {Sarkar}}(2014)}]{mertsch2014}
\bibinfo{author}{\bibfnamefont{P.}~\bibnamefont{{Mertsch}}} \bibnamefont{and}
  \bibinfo{author}{\bibfnamefont{S.}~\bibnamefont{{Sarkar}}},
  \bibinfo{journal}{Physical Review D} \textbf{\bibinfo{volume}{90}},
  \bibinfo{eid}{061301} (\bibinfo{year}{2014}), \eprint{1402.0855}.

\bibitem[{\citenamefont{{Feng} et~al.}(2016{\natexlab{b}})\citenamefont{{Feng},
  {Tomassetti}, and {Oliva}}}]{feng2016prd}
\bibinfo{author}{\bibfnamefont{J.}~\bibnamefont{{Feng}}},
  \bibinfo{author}{\bibfnamefont{N.}~\bibnamefont{{Tomassetti}}},
  \bibnamefont{and} \bibinfo{author}{\bibfnamefont{A.}~\bibnamefont{{Oliva}}},
  \bibinfo{journal}{Physical Review D} \textbf{\bibinfo{volume}{94}},
  \bibinfo{eid}{123007} (\bibinfo{year}{2016}{\natexlab{b}}),
  \eprint{1610.06182}.

\bibitem[{\citenamefont{Guo and Yuan}(2018)}]{Guo:2018wyf}
\bibinfo{author}{\bibfnamefont{Y.-Q.} \bibnamefont{Guo}} \bibnamefont{and}
  \bibinfo{author}{\bibfnamefont{Q.}~\bibnamefont{Yuan}}
  (\bibinfo{year}{2018}), \eprint{1801.05904}.

\bibitem[{\citenamefont{{Cholis} et~al.}(2017)\citenamefont{{Cholis}, {Hooper},
  and {Linden}}}]{cholis2017}
\bibinfo{author}{\bibfnamefont{I.}~\bibnamefont{{Cholis}}},
  \bibinfo{author}{\bibfnamefont{D.}~\bibnamefont{{Hooper}}}, \bibnamefont{and}
  \bibinfo{author}{\bibfnamefont{T.}~\bibnamefont{{Linden}}},
  \bibinfo{journal}{Physical Review D} \textbf{\bibinfo{volume}{95}},
  \bibinfo{eid}{123007} (\bibinfo{year}{2017}), \eprint{1701.04406}.

\bibitem[{\citenamefont{{Lin} et~al.}(2015)\citenamefont{{Lin}, {Bi}, {Yin},
  and {Yu}}}]{lin2015dm}
\bibinfo{author}{\bibfnamefont{S.-J.} \bibnamefont{{Lin}}},
  \bibinfo{author}{\bibfnamefont{X.-J.} \bibnamefont{{Bi}}},
  \bibinfo{author}{\bibfnamefont{P.-F.} \bibnamefont{{Yin}}}, \bibnamefont{and}
  \bibinfo{author}{\bibfnamefont{Z.-H.} \bibnamefont{{Yu}}},
  \bibinfo{journal}{ArXiv e-prints}  (\bibinfo{year}{2015}),
  \eprint{1504.07230}.

\bibitem[{\citenamefont{{Huang} et~al.}(2017)\citenamefont{{Huang}, {Wei},
  {Wu}, {Zhang}, and {Zhou}}}]{huang2017}
\bibinfo{author}{\bibfnamefont{X.-J.} \bibnamefont{{Huang}}},
  \bibinfo{author}{\bibfnamefont{C.-C.} \bibnamefont{{Wei}}},
  \bibinfo{author}{\bibfnamefont{Y.-L.} \bibnamefont{{Wu}}},
  \bibinfo{author}{\bibfnamefont{W.-H.} \bibnamefont{{Zhang}}},
  \bibnamefont{and} \bibinfo{author}{\bibfnamefont{Y.-F.}
  \bibnamefont{{Zhou}}}, \bibinfo{journal}{Physical Review D}
  \textbf{\bibinfo{volume}{95}}, \bibinfo{eid}{063021} (\bibinfo{year}{2017}),
  \eprint{1611.01983}.

\bibitem[{\citenamefont{{Feng} and {Zhang}}(2017)}]{feng2017}
\bibinfo{author}{\bibfnamefont{J.}~\bibnamefont{{Feng}}} \bibnamefont{and}
  \bibinfo{author}{\bibfnamefont{H.-H.} \bibnamefont{{Zhang}}},
  \bibinfo{journal}{ArXiv e-prints}  (\bibinfo{year}{2017}),
  \eprint{1701.02263}.

\bibitem[{\citenamefont{{Lin} et~al.}(2017)\citenamefont{{Lin}, {Bi}, {Feng},
  {Yin}, and {Yu}}}]{lin2017prd}
\bibinfo{author}{\bibfnamefont{S.-J.} \bibnamefont{{Lin}}},
  \bibinfo{author}{\bibfnamefont{X.-J.} \bibnamefont{{Bi}}},
  \bibinfo{author}{\bibfnamefont{J.}~\bibnamefont{{Feng}}},
  \bibinfo{author}{\bibfnamefont{P.-F.} \bibnamefont{{Yin}}}, \bibnamefont{and}
  \bibinfo{author}{\bibfnamefont{Z.-H.} \bibnamefont{{Yu}}},
  \bibinfo{journal}{Physical review D} \textbf{\bibinfo{volume}{96}},
  \bibinfo{eid}{123010} (\bibinfo{year}{2017}), \eprint{1612.04001}.

\bibitem[{\citenamefont{Cuoco et~al.}(2017{\natexlab{c}})\citenamefont{Cuoco,
  Kramer, and Korsmeier}}]{cuoco2017prl}
\bibinfo{author}{\bibfnamefont{A.}~\bibnamefont{Cuoco}},
  \bibinfo{author}{\bibfnamefont{M.}~\bibnamefont{Kramer}}, \bibnamefont{and}
  \bibinfo{author}{\bibfnamefont{M.}~\bibnamefont{Korsmeier}},
  \bibinfo{journal}{Phys. Rev. Lett.} \textbf{\bibinfo{volume}{118}},
  \bibinfo{pages}{191102} (\bibinfo{year}{2017}{\natexlab{c}}),
  \eprint{1610.03071}.

\bibitem[{\citenamefont{{Cuoco} et~al.}(2017)\citenamefont{{Cuoco}, {Heisig},
  {Korsmeier}, and {Kr{\"a}mer}}}]{cuoco2017jcap}
\bibinfo{author}{\bibfnamefont{A.}~\bibnamefont{{Cuoco}}},
  \bibinfo{author}{\bibfnamefont{J.}~\bibnamefont{{Heisig}}},
  \bibinfo{author}{\bibfnamefont{M.}~\bibnamefont{{Korsmeier}}},
  \bibnamefont{and}
  \bibinfo{author}{\bibfnamefont{M.}~\bibnamefont{{Kr{\"a}mer}}},
  \bibinfo{journal}{JCAP} \textbf{\bibinfo{volume}{10}}, \bibinfo{eid}{053}
  (\bibinfo{year}{2017}), \eprint{1704.08258}.

\bibitem[{\citenamefont{{Cui} et~al.}(2017)\citenamefont{{Cui}, {Yuan}, {Tsai},
  and {Fan}}}]{cui2017prl}
\bibinfo{author}{\bibfnamefont{M.-Y.} \bibnamefont{{Cui}}},
  \bibinfo{author}{\bibfnamefont{Q.}~\bibnamefont{{Yuan}}},
  \bibinfo{author}{\bibfnamefont{Y.-L.~S.} \bibnamefont{{Tsai}}},
  \bibnamefont{and} \bibinfo{author}{\bibfnamefont{Y.-Z.} \bibnamefont{{Fan}}},
  \bibinfo{journal}{Physical Review Letters} \textbf{\bibinfo{volume}{118}},
  \bibinfo{eid}{191101} (\bibinfo{year}{2017}), \eprint{1610.03840}.

\bibitem[{\citenamefont{Donato et~al.}(2000)\citenamefont{Donato, Fornengo, and
  Salati}}]{Donato:1999gy}
\bibinfo{author}{\bibfnamefont{F.}~\bibnamefont{Donato}},
  \bibinfo{author}{\bibfnamefont{N.}~\bibnamefont{Fornengo}}, \bibnamefont{and}
  \bibinfo{author}{\bibfnamefont{P.}~\bibnamefont{Salati}},
  \bibinfo{journal}{Phys. Rev.} \textbf{\bibinfo{volume}{D62}},
  \bibinfo{pages}{043003} (\bibinfo{year}{2000}), \eprint{hep-ph/9904481}.

\bibitem[{\citenamefont{Baer and Profumo}(2005)}]{Baer:2005tw}
\bibinfo{author}{\bibfnamefont{H.}~\bibnamefont{Baer}} \bibnamefont{and}
  \bibinfo{author}{\bibfnamefont{S.}~\bibnamefont{Profumo}},
  \bibinfo{journal}{JCAP} \textbf{\bibinfo{volume}{0512}}, \bibinfo{pages}{008}
  (\bibinfo{year}{2005}), \eprint{astro-ph/0510722}.

\bibitem[{\citenamefont{Duperray et~al.}(2005)\citenamefont{Duperray, Baret,
  Maurin, Boudoul, Barrau, Derome, Protasov, and Buenerd}}]{Duperray:2005si}
\bibinfo{author}{\bibfnamefont{R.}~\bibnamefont{Duperray}},
  \bibinfo{author}{\bibfnamefont{B.}~\bibnamefont{Baret}},
  \bibinfo{author}{\bibfnamefont{D.}~\bibnamefont{Maurin}},
  \bibinfo{author}{\bibfnamefont{G.}~\bibnamefont{Boudoul}},
  \bibinfo{author}{\bibfnamefont{A.}~\bibnamefont{Barrau}},
  \bibinfo{author}{\bibfnamefont{L.}~\bibnamefont{Derome}},
  \bibinfo{author}{\bibfnamefont{K.}~\bibnamefont{Protasov}}, \bibnamefont{and}
  \bibinfo{author}{\bibfnamefont{M.}~\bibnamefont{Buenerd}},
  \bibinfo{journal}{Phys. Rev.} \textbf{\bibinfo{volume}{D71}},
  \bibinfo{pages}{083013} (\bibinfo{year}{2005}), \eprint{astro-ph/0503544}.

\bibitem[{\citenamefont{Donato et~al.}(2008)\citenamefont{Donato, Fornengo, and
  Maurin}}]{Donato:2008yx}
\bibinfo{author}{\bibfnamefont{F.}~\bibnamefont{Donato}},
  \bibinfo{author}{\bibfnamefont{N.}~\bibnamefont{Fornengo}}, \bibnamefont{and}
  \bibinfo{author}{\bibfnamefont{D.}~\bibnamefont{Maurin}},
  \bibinfo{journal}{Phys. Rev.} \textbf{\bibinfo{volume}{D78}},
  \bibinfo{pages}{043506} (\bibinfo{year}{2008}), \eprint{0803.2640}.

\bibitem[{\citenamefont{Carlson et~al.}(2014)\citenamefont{Carlson, Coogan,
  Linden, Profumo, Ibarra, and Wild}}]{Carlson:2014ssa}
\bibinfo{author}{\bibfnamefont{E.}~\bibnamefont{Carlson}},
  \bibinfo{author}{\bibfnamefont{A.}~\bibnamefont{Coogan}},
  \bibinfo{author}{\bibfnamefont{T.}~\bibnamefont{Linden}},
  \bibinfo{author}{\bibfnamefont{S.}~\bibnamefont{Profumo}},
  \bibinfo{author}{\bibfnamefont{A.}~\bibnamefont{Ibarra}}, \bibnamefont{and}
  \bibinfo{author}{\bibfnamefont{S.}~\bibnamefont{Wild}},
  \bibinfo{journal}{Phys. Rev.} \textbf{\bibinfo{volume}{D89}},
  \bibinfo{pages}{076005} (\bibinfo{year}{2014}), \eprint{1401.2461}.

\bibitem[{\citenamefont{Cirelli et~al.}(2014)\citenamefont{Cirelli, Fornengo,
  Taoso, and Vittino}}]{Cirelli:2014qia}
\bibinfo{author}{\bibfnamefont{M.}~\bibnamefont{Cirelli}},
  \bibinfo{author}{\bibfnamefont{N.}~\bibnamefont{Fornengo}},
  \bibinfo{author}{\bibfnamefont{M.}~\bibnamefont{Taoso}}, \bibnamefont{and}
  \bibinfo{author}{\bibfnamefont{A.}~\bibnamefont{Vittino}},
  \bibinfo{journal}{JHEP} \textbf{\bibinfo{volume}{08}}, \bibinfo{pages}{009}
  (\bibinfo{year}{2014}), \eprint{1401.4017}.

\bibitem[{\citenamefont{Fuke et~al.}(2005)}]{Fuke:2005it}
\bibinfo{author}{\bibfnamefont{H.}~\bibnamefont{Fuke}} \bibnamefont{et~al.},
  \bibinfo{journal}{Phys. Rev. Lett.} \textbf{\bibinfo{volume}{95}},
  \bibinfo{pages}{081101} (\bibinfo{year}{2005}), \eprint{astro-ph/0504361}.

\bibitem[{\citenamefont{Abe et~al.}(2012)}]{Abe:2012tz}
\bibinfo{author}{\bibfnamefont{K.}~\bibnamefont{Abe}} \bibnamefont{et~al.},
  \bibinfo{journal}{Phys. Rev. Lett.} \textbf{\bibinfo{volume}{108}},
  \bibinfo{pages}{131301} (\bibinfo{year}{2012}), \eprint{1201.2967}.

\bibitem[{\citenamefont{Aramaki et~al.}(2016)}]{Aramaki:2015pii}
\bibinfo{author}{\bibfnamefont{T.}~\bibnamefont{Aramaki}} \bibnamefont{et~al.},
  \bibinfo{journal}{Phys. Rept.} \textbf{\bibinfo{volume}{618}},
  \bibinfo{pages}{1} (\bibinfo{year}{2016}), \eprint{1505.07785}.

\bibitem[{\citenamefont{Coogan and Profumo}(2017)}]{Coogan:2017pwt}
\bibinfo{author}{\bibfnamefont{A.}~\bibnamefont{Coogan}} \bibnamefont{and}
  \bibinfo{author}{\bibfnamefont{S.}~\bibnamefont{Profumo}},
  \bibinfo{journal}{Phys. Rev.} \textbf{\bibinfo{volume}{D96}},
  \bibinfo{pages}{083020} (\bibinfo{year}{2017}), \eprint{1705.09664}.

\bibitem[{\citenamefont{Chardonnet et~al.}(1997)\citenamefont{Chardonnet,
  Orloff, and Salati}}]{Chardonnet:1997dv}
\bibinfo{author}{\bibfnamefont{P.}~\bibnamefont{Chardonnet}},
  \bibinfo{author}{\bibfnamefont{J.}~\bibnamefont{Orloff}}, \bibnamefont{and}
  \bibinfo{author}{\bibfnamefont{P.}~\bibnamefont{Salati}},
  \bibinfo{journal}{Phys. Lett.} \textbf{\bibinfo{volume}{B409}},
  \bibinfo{pages}{313} (\bibinfo{year}{1997}), \eprint{astro-ph/9705110}.

\bibitem[{\citenamefont{Herms et~al.}(2017)\citenamefont{Herms, Ibarra,
  Vittino, and Wild}}]{Herms:2016vop}
\bibinfo{author}{\bibfnamefont{J.}~\bibnamefont{Herms}},
  \bibinfo{author}{\bibfnamefont{A.}~\bibnamefont{Ibarra}},
  \bibinfo{author}{\bibfnamefont{A.}~\bibnamefont{Vittino}}, \bibnamefont{and}
  \bibinfo{author}{\bibfnamefont{S.}~\bibnamefont{Wild}},
  \bibinfo{journal}{JCAP} \textbf{\bibinfo{volume}{1702}}, \bibinfo{pages}{018}
  (\bibinfo{year}{2017}), \eprint{1610.00699}.

\bibitem[{\citenamefont{Tomassetti and Oliva}(2017)}]{Tomassetti:2017izg}
\bibinfo{author}{\bibfnamefont{N.}~\bibnamefont{Tomassetti}} \bibnamefont{and}
  \bibinfo{author}{\bibfnamefont{A.}~\bibnamefont{Oliva}},
  \bibinfo{journal}{Astrophys. J.} \textbf{\bibinfo{volume}{844}},
  \bibinfo{pages}{L26} (\bibinfo{year}{2017}), \eprint{1707.06915}.

\bibitem[{\citenamefont{Korsmeier et~al.}(2017)\citenamefont{Korsmeier, Donato,
  and Fornengo}}]{Korsmeier:2017xzj}
\bibinfo{author}{\bibfnamefont{M.}~\bibnamefont{Korsmeier}},
  \bibinfo{author}{\bibfnamefont{F.}~\bibnamefont{Donato}}, \bibnamefont{and}
  \bibinfo{author}{\bibfnamefont{N.}~\bibnamefont{Fornengo}}
  (\bibinfo{year}{2017}), \eprint{1711.08465}.

\bibitem[{\citenamefont{Kiraly et~al.}(1981)\citenamefont{Kiraly, Szabelski,
  Wdowczyk, and Wolfendale}}]{Kiraly:1981ci}
\bibinfo{author}{\bibfnamefont{P.}~\bibnamefont{Kiraly}},
  \bibinfo{author}{\bibfnamefont{J.}~\bibnamefont{Szabelski}},
  \bibinfo{author}{\bibfnamefont{J.}~\bibnamefont{Wdowczyk}}, \bibnamefont{and}
  \bibinfo{author}{\bibfnamefont{A.~W.} \bibnamefont{Wolfendale}},
  \bibinfo{journal}{Nature} \textbf{\bibinfo{volume}{293}},
  \bibinfo{pages}{120} (\bibinfo{year}{1981}).

\bibitem[{\citenamefont{Turner}(1982)}]{Turner:1981ez}
\bibinfo{author}{\bibfnamefont{M.~S.} \bibnamefont{Turner}},
  \bibinfo{journal}{Nature} \textbf{\bibinfo{volume}{297}},
  \bibinfo{pages}{379} (\bibinfo{year}{1982}).

\bibitem[{\citenamefont{Gleeson and Axford}(1968)}]{Gleeson:1968zza}
\bibinfo{author}{\bibfnamefont{L.~J.} \bibnamefont{Gleeson}} \bibnamefont{and}
  \bibinfo{author}{\bibfnamefont{W.~I.} \bibnamefont{Axford}},
  \bibinfo{journal}{Astrophys. J.} \textbf{\bibinfo{volume}{154}},
  \bibinfo{pages}{1011} (\bibinfo{year}{1968}).

\bibitem[{\citenamefont{Fornengo et~al.}(2013)\citenamefont{Fornengo, Maccione,
  and Vittino}}]{Fornengo:2013osa}
\bibinfo{author}{\bibfnamefont{N.}~\bibnamefont{Fornengo}},
  \bibinfo{author}{\bibfnamefont{L.}~\bibnamefont{Maccione}}, \bibnamefont{and}
  \bibinfo{author}{\bibfnamefont{A.}~\bibnamefont{Vittino}},
  \bibinfo{journal}{JCAP} \textbf{\bibinfo{volume}{1309}}, \bibinfo{pages}{031}
  (\bibinfo{year}{2013}), \eprint{1306.4171}.

\bibitem[{\citenamefont{Maccione}(2013)}]{Maccione:2012cu}
\bibinfo{author}{\bibfnamefont{L.}~\bibnamefont{Maccione}},
  \bibinfo{journal}{Phys. Rev. Lett.} \textbf{\bibinfo{volume}{110}},
  \bibinfo{pages}{081101} (\bibinfo{year}{2013}), \eprint{1211.6905}.

\bibitem[{\citenamefont{Vittino et~al.}(2017)\citenamefont{Vittino, Evoli, and
  Gaggero}}]{Vittino:2017fuh}
\bibinfo{author}{\bibfnamefont{A.}~\bibnamefont{Vittino}},
  \bibinfo{author}{\bibfnamefont{C.}~\bibnamefont{Evoli}}, \bibnamefont{and}
  \bibinfo{author}{\bibfnamefont{D.}~\bibnamefont{Gaggero}}, in
  \emph{\bibinfo{booktitle}{{Proceedings, 35th International Cosmic Ray
  Conference (ICRC 2017): Bexco, Busan, Korea, July 12-20, 2017}}}
  (\bibinfo{year}{2017}), \eprint{1707.09003},
  \urlprefix\url{http://inspirehep.net/record/1613480/files/arXiv:1707.09003.pdf}.

\bibitem[{\citenamefont{Schwarzschild and
  Zupancic}(1963)}]{Schwarzschild:1963zz}
\bibinfo{author}{\bibfnamefont{A.}~\bibnamefont{Schwarzschild}}
  \bibnamefont{and} \bibinfo{author}{\bibfnamefont{C.}~\bibnamefont{Zupancic}},
  \bibinfo{journal}{Phys. Rev.} \textbf{\bibinfo{volume}{129}},
  \bibinfo{pages}{854} (\bibinfo{year}{1963}).

\bibitem[{\citenamefont{Lin et~al.}(2018)\citenamefont{Lin, Bi, and
  Yin}}]{Lin:2018avl}
\bibinfo{author}{\bibfnamefont{S.-J.} \bibnamefont{Lin}},
  \bibinfo{author}{\bibfnamefont{X.-J.} \bibnamefont{Bi}}, \bibnamefont{and}
  \bibinfo{author}{\bibfnamefont{P.-F.} \bibnamefont{Yin}}
  (\bibinfo{year}{2018}), \eprint{1801.00997}.

\bibitem[{\citenamefont{Blum et~al.}(2017{\natexlab{b}})\citenamefont{Blum, Ng,
  Sato, and Takimoto}}]{Blum:2017qnn}
\bibinfo{author}{\bibfnamefont{K.}~\bibnamefont{Blum}},
  \bibinfo{author}{\bibfnamefont{K.~C.~Y.} \bibnamefont{Ng}},
  \bibinfo{author}{\bibfnamefont{R.}~\bibnamefont{Sato}}, \bibnamefont{and}
  \bibinfo{author}{\bibfnamefont{M.}~\bibnamefont{Takimoto}},
  \bibinfo{journal}{Phys. Rev.} \textbf{\bibinfo{volume}{D96}},
  \bibinfo{pages}{103021} (\bibinfo{year}{2017}{\natexlab{b}}),
  \eprint{1704.05431}.

\bibitem[{\citenamefont{Lisa et~al.}(2005)\citenamefont{Lisa, Pratt, Soltz, and
  Wiedemann}}]{Lisa:2005dd}
\bibinfo{author}{\bibfnamefont{M.~A.} \bibnamefont{Lisa}},
  \bibinfo{author}{\bibfnamefont{S.}~\bibnamefont{Pratt}},
  \bibinfo{author}{\bibfnamefont{R.}~\bibnamefont{Soltz}}, \bibnamefont{and}
  \bibinfo{author}{\bibfnamefont{U.}~\bibnamefont{Wiedemann}},
  \bibinfo{journal}{Ann. Rev. Nucl. Part. Sci.} \textbf{\bibinfo{volume}{55}},
  \bibinfo{pages}{357} (\bibinfo{year}{2005}), \eprint{nucl-ex/0505014}.

\bibitem[{\citenamefont{{Yuan} et~al.}(2017{\natexlab{b}})\citenamefont{{Yuan},
  {Feng}, {Yin}, {Fan}, {Bi}, {Cui}, {Dong}, {Guo}, {Fang}, {Hu}
  et~al.}}]{yuan2017dampe}
\bibinfo{author}{\bibfnamefont{Q.}~\bibnamefont{{Yuan}}},
  \bibinfo{author}{\bibfnamefont{L.}~\bibnamefont{{Feng}}},
  \bibinfo{author}{\bibfnamefont{P.-F.} \bibnamefont{{Yin}}},
  \bibinfo{author}{\bibfnamefont{Y.-Z.} \bibnamefont{{Fan}}},
  \bibinfo{author}{\bibfnamefont{X.-J.} \bibnamefont{{Bi}}},
  \bibinfo{author}{\bibfnamefont{M.-Y.} \bibnamefont{{Cui}}},
  \bibinfo{author}{\bibfnamefont{T.-K.} \bibnamefont{{Dong}}},
  \bibinfo{author}{\bibfnamefont{Y.-Q.} \bibnamefont{{Guo}}},
  \bibinfo{author}{\bibfnamefont{K.}~\bibnamefont{{Fang}}},
  \bibinfo{author}{\bibfnamefont{H.-B.} \bibnamefont{{Hu}}},
  \bibnamefont{et~al.}, \bibinfo{journal}{ArXiv e-prints}
  (\bibinfo{year}{2017}{\natexlab{b}}), \eprint{1711.10989}.

\bibitem[{\citenamefont{{Shaviv} et~al.}(2009)\citenamefont{{Shaviv}, {Nakar},
  and {Piran}}}]{shaviv2009}
\bibinfo{author}{\bibfnamefont{N.~J.} \bibnamefont{{Shaviv}}},
  \bibinfo{author}{\bibfnamefont{E.}~\bibnamefont{{Nakar}}}, \bibnamefont{and}
  \bibinfo{author}{\bibfnamefont{T.}~\bibnamefont{{Piran}}},
  \bibinfo{journal}{Physical Review Letters} \textbf{\bibinfo{volume}{103}},
  \bibinfo{eid}{111302} (\bibinfo{year}{2009}), \eprint{0902.0376}.

\bibitem[{\citenamefont{{Aguilar} et~al.}(2014)}]{2014PhRvL.113l1102A}
\bibinfo{author}{\bibfnamefont{M.}~\bibnamefont{{Aguilar}}}
  \bibnamefont{et~al.}, \bibinfo{journal}{Physical Review Letters}
  \textbf{\bibinfo{volume}{113}}, \bibinfo{eid}{121102} (\bibinfo{year}{2014}).

\bibitem[{\citenamefont{{Serpico}}(2009)}]{serpico2009}
\bibinfo{author}{\bibfnamefont{P.~D.} \bibnamefont{{Serpico}}},
  \bibinfo{journal}{Physical Review D} \textbf{\bibinfo{volume}{79}},
  \bibinfo{eid}{021302} (\bibinfo{year}{2009}), \eprint{0810.4846}.

\bibitem[{\citenamefont{{Grasso} et~al.}(2009)\citenamefont{{Grasso},
  {Profumo}, {Strong}, {Baldini}, {Bellazzini}, {Bloom}, {Bregeon}, {Di
  Bernardo}, {Gaggero}, {Giglietto} et~al.}}]{fermi2009}
\bibinfo{author}{\bibfnamefont{D.}~\bibnamefont{{Grasso}}},
  \bibinfo{author}{\bibfnamefont{S.}~\bibnamefont{{Profumo}}},
  \bibinfo{author}{\bibfnamefont{A.~W.} \bibnamefont{{Strong}}},
  \bibinfo{author}{\bibfnamefont{L.}~\bibnamefont{{Baldini}}},
  \bibinfo{author}{\bibfnamefont{R.}~\bibnamefont{{Bellazzini}}},
  \bibinfo{author}{\bibfnamefont{E.~D.} \bibnamefont{{Bloom}}},
  \bibinfo{author}{\bibfnamefont{J.}~\bibnamefont{{Bregeon}}},
  \bibinfo{author}{\bibfnamefont{G.}~\bibnamefont{{Di Bernardo}}},
  \bibinfo{author}{\bibfnamefont{D.}~\bibnamefont{{Gaggero}}},
  \bibinfo{author}{\bibfnamefont{N.}~\bibnamefont{{Giglietto}}},
  \bibnamefont{et~al.}, \bibinfo{journal}{Astroparticle Physics}
  \textbf{\bibinfo{volume}{32}}, \bibinfo{pages}{140} (\bibinfo{year}{2009}),
  \eprint{0905.0636}.

\bibitem[{\citenamefont{{Serpico}}(2012)}]{serpico2012}
\bibinfo{author}{\bibfnamefont{P.~D.} \bibnamefont{{Serpico}}},
  \bibinfo{journal}{Astroparticle Physics} \textbf{\bibinfo{volume}{39}},
  \bibinfo{pages}{2} (\bibinfo{year}{2012}), \eprint{1108.4827}.

\bibitem[{\citenamefont{He}(2009)}]{He2009}
\bibinfo{author}{\bibfnamefont{X.-G.} \bibnamefont{He}}, \bibinfo{journal}{Mod.
  Phys. Lett.} \textbf{\bibinfo{volume}{A24}}, \bibinfo{pages}{2139}
  (\bibinfo{year}{2009}), \eprint{0908.2908}.

\bibitem[{\citenamefont{Abeysekara et~al.}(2017)}]{hawc2017}
\bibinfo{author}{\bibfnamefont{A.~U.} \bibnamefont{Abeysekara}}
  \bibnamefont{et~al.} (\bibinfo{collaboration}{HAWC}),
  \bibinfo{journal}{Science} \textbf{\bibinfo{volume}{358}},
  \bibinfo{pages}{911} (\bibinfo{year}{2017}), \eprint{1711.06223}.

\bibitem[{\citenamefont{{L{\'o}pez-Coto} and {Giacinti}}(2017)}]{giacinti2017}
\bibinfo{author}{\bibfnamefont{R.}~\bibnamefont{{L{\'o}pez-Coto}}}
  \bibnamefont{and}
  \bibinfo{author}{\bibfnamefont{G.}~\bibnamefont{{Giacinti}}},
  \bibinfo{journal}{ArXiv e-prints}  (\bibinfo{year}{2017}),
  \eprint{1712.04373}.

\bibitem[{\citenamefont{{Lipari}}(2017)}]{lipari2017}
\bibinfo{author}{\bibfnamefont{P.}~\bibnamefont{{Lipari}}},
  \bibinfo{journal}{Physical Review D} \textbf{\bibinfo{volume}{95}},
  \bibinfo{eid}{063009} (\bibinfo{year}{2017}), \eprint{1608.02018}.

\bibitem[{\citenamefont{Aharonian et~al.}(2008)}]{Aharonian:2008aa}
\bibinfo{author}{\bibfnamefont{F.}~\bibnamefont{Aharonian}}
  \bibnamefont{et~al.} (\bibinfo{collaboration}{H.E.S.S.}),
  \bibinfo{journal}{Phys. Rev. Lett.} \textbf{\bibinfo{volume}{101}},
  \bibinfo{pages}{261104} (\bibinfo{year}{2008}), \eprint{0811.3894}.

\bibitem[{\citenamefont{{DAMPE Collaboration} et~al.}(2017)\citenamefont{{DAMPE
  Collaboration}, {Ambrosi}, {An}, {Asfandiyarov}, {Azzarello}, {Bernardini},
  {Bertucci}, {Cai}, {Chang}, {Chen} et~al.}}]{dampe2017nature}
\bibinfo{author}{\bibnamefont{{DAMPE Collaboration}}},
  \bibinfo{author}{\bibfnamefont{G.}~\bibnamefont{{Ambrosi}}},
  \bibinfo{author}{\bibfnamefont{Q.}~\bibnamefont{{An}}},
  \bibinfo{author}{\bibfnamefont{R.}~\bibnamefont{{Asfandiyarov}}},
  \bibinfo{author}{\bibfnamefont{P.}~\bibnamefont{{Azzarello}}},
  \bibinfo{author}{\bibfnamefont{P.}~\bibnamefont{{Bernardini}}},
  \bibinfo{author}{\bibfnamefont{B.}~\bibnamefont{{Bertucci}}},
  \bibinfo{author}{\bibfnamefont{M.~S.} \bibnamefont{{Cai}}},
  \bibinfo{author}{\bibfnamefont{J.}~\bibnamefont{{Chang}}},
  \bibinfo{author}{\bibfnamefont{D.~Y.} \bibnamefont{{Chen}}},
  \bibnamefont{et~al.}, \bibinfo{journal}{Nature}
  \textbf{\bibinfo{volume}{552}}, \bibinfo{pages}{63} (\bibinfo{year}{2017}),
  \eprint{1711.10981}.

\bibitem[{\citenamefont{Ackermann et~al.}(2012)}]{Ackermann2012}
\bibinfo{author}{\bibfnamefont{M.}~\bibnamefont{Ackermann}}
  \bibnamefont{et~al.}, \bibinfo{journal}{\apj} \textbf{\bibinfo{volume}{750}},
  \bibinfo{eid}{3} (\bibinfo{year}{2012}), \eprint{1202.4039}.

\bibitem[{\citenamefont{Cholis et~al.}(2012)\citenamefont{Cholis, Tavakoli,
  Evoli, Maccione, and Ullio}}]{Cholis:2011un}
\bibinfo{author}{\bibfnamefont{I.}~\bibnamefont{Cholis}},
  \bibinfo{author}{\bibfnamefont{M.}~\bibnamefont{Tavakoli}},
  \bibinfo{author}{\bibfnamefont{C.}~\bibnamefont{Evoli}},
  \bibinfo{author}{\bibfnamefont{L.}~\bibnamefont{Maccione}}, \bibnamefont{and}
  \bibinfo{author}{\bibfnamefont{P.}~\bibnamefont{Ullio}},
  \bibinfo{journal}{JCAP} \textbf{\bibinfo{volume}{1205}}, \bibinfo{pages}{004}
  (\bibinfo{year}{2012}), \eprint{1106.5073}.

\bibitem[{\citenamefont{Pohl et~al.}(2008)\citenamefont{Pohl, Englmaier, and
  Bissantz}}]{Pohl:2007dz}
\bibinfo{author}{\bibfnamefont{M.}~\bibnamefont{Pohl}},
  \bibinfo{author}{\bibfnamefont{P.}~\bibnamefont{Englmaier}},
  \bibnamefont{and} \bibinfo{author}{\bibfnamefont{N.}~\bibnamefont{Bissantz}},
  \bibinfo{journal}{Astrophys. J.} \textbf{\bibinfo{volume}{677}},
  \bibinfo{pages}{283} (\bibinfo{year}{2008}), \eprint{0712.4264}.

\bibitem[{\citenamefont{Tavakoli}(2012)}]{Tavakoli:2012jx}
\bibinfo{author}{\bibfnamefont{M.}~\bibnamefont{Tavakoli}}
  (\bibinfo{year}{2012}), \eprint{1207.6150}.

\bibitem[{\citenamefont{Porter et~al.}(2006)\citenamefont{Porter, Moskalenko,
  and Strong}}]{Porter:2006tb}
\bibinfo{author}{\bibfnamefont{T.~A.} \bibnamefont{Porter}},
  \bibinfo{author}{\bibfnamefont{I.~V.} \bibnamefont{Moskalenko}},
  \bibnamefont{and} \bibinfo{author}{\bibfnamefont{A.~W.}
  \bibnamefont{Strong}}, \bibinfo{journal}{Astrophys. J.}
  \textbf{\bibinfo{volume}{648}}, \bibinfo{pages}{L29} (\bibinfo{year}{2006}),
  \eprint{astro-ph/0607344}.

\bibitem[{\citenamefont{Tavakoli et~al.}(2014)\citenamefont{Tavakoli, Cholis,
  Evoli, and Ullio}}]{Tavakoli:2013zva}
\bibinfo{author}{\bibfnamefont{M.}~\bibnamefont{Tavakoli}},
  \bibinfo{author}{\bibfnamefont{I.}~\bibnamefont{Cholis}},
  \bibinfo{author}{\bibfnamefont{C.}~\bibnamefont{Evoli}}, \bibnamefont{and}
  \bibinfo{author}{\bibfnamefont{P.}~\bibnamefont{Ullio}},
  \bibinfo{journal}{JCAP} \textbf{\bibinfo{volume}{1401}}, \bibinfo{pages}{017}
  (\bibinfo{year}{2014}), \eprint{1308.4135}.

\bibitem[{\citenamefont{{Erlykin} and {Wolfendale}}(2013)}]{2013APh_Erl_Wolf}
\bibinfo{author}{\bibfnamefont{A.~D.} \bibnamefont{{Erlykin}}}
  \bibnamefont{and} \bibinfo{author}{\bibfnamefont{A.~W.}
  \bibnamefont{{Wolfendale}}}, \bibinfo{journal}{Astroparticle Physics}
  \textbf{\bibinfo{volume}{42}}, \bibinfo{pages}{70} (\bibinfo{year}{2013}),
  \eprint{1212.2760}.

\bibitem[{\citenamefont{Gaggero
  et~al.}(2015{\natexlab{b}})\citenamefont{Gaggero, Grasso, Marinelli, Urbano,
  and Valli}}]{Gaggero:2015xza}
\bibinfo{author}{\bibfnamefont{D.}~\bibnamefont{Gaggero}},
  \bibinfo{author}{\bibfnamefont{D.}~\bibnamefont{Grasso}},
  \bibinfo{author}{\bibfnamefont{A.}~\bibnamefont{Marinelli}},
  \bibinfo{author}{\bibfnamefont{A.}~\bibnamefont{Urbano}}, \bibnamefont{and}
  \bibinfo{author}{\bibfnamefont{M.}~\bibnamefont{Valli}},
  \bibinfo{journal}{Astrophys. J.} \textbf{\bibinfo{volume}{815}},
  \bibinfo{pages}{L25} (\bibinfo{year}{2015}{\natexlab{b}}),
  \eprint{1504.00227}.

\bibitem[{\citenamefont{Acero et~al.}(2016)}]{Acero:2016qlg}
\bibinfo{author}{\bibfnamefont{F.}~\bibnamefont{Acero}} \bibnamefont{et~al.}
  (\bibinfo{collaboration}{Fermi-LAT}), \bibinfo{journal}{Astrophys. J. Suppl.}
  \textbf{\bibinfo{volume}{223}}, \bibinfo{pages}{26} (\bibinfo{year}{2016}),
  \eprint{1602.07246}.

\bibitem[{\citenamefont{Yang et~al.}(2016)\citenamefont{Yang, Aharonian, and
  Evoli}}]{Yang:2016jda}
\bibinfo{author}{\bibfnamefont{R.}~\bibnamefont{Yang}},
  \bibinfo{author}{\bibfnamefont{F.}~\bibnamefont{Aharonian}},
  \bibnamefont{and} \bibinfo{author}{\bibfnamefont{C.}~\bibnamefont{Evoli}},
  \bibinfo{journal}{Phys. Rev.} \textbf{\bibinfo{volume}{D93}},
  \bibinfo{pages}{123007} (\bibinfo{year}{2016}), \eprint{1602.04710}.

\bibitem[{\citenamefont{Nezri et~al.}(2012)\citenamefont{Nezri, Lavalle, and
  Teyssier}}]{Nezri:2012xu}
\bibinfo{author}{\bibfnamefont{E.}~\bibnamefont{Nezri}},
  \bibinfo{author}{\bibfnamefont{J.}~\bibnamefont{Lavalle}}, \bibnamefont{and}
  \bibinfo{author}{\bibfnamefont{R.}~\bibnamefont{Teyssier}},
  \bibinfo{journal}{Phys. Rev.} \textbf{\bibinfo{volume}{D86}},
  \bibinfo{pages}{063524} (\bibinfo{year}{2012}), \eprint{1204.4121}.

\bibitem[{\citenamefont{Charles et~al.}(2016)}]{Charles:2016pgz}
\bibinfo{author}{\bibfnamefont{E.}~\bibnamefont{Charles}} \bibnamefont{et~al.}
  (\bibinfo{collaboration}{Fermi-LAT}), \bibinfo{journal}{Phys. Rept.}
  \textbf{\bibinfo{volume}{636}}, \bibinfo{pages}{1} (\bibinfo{year}{2016}),
  \eprint{1605.02016}.

\bibitem[{\citenamefont{Conrad and Reimer}(2017)}]{Conrad:2017pms}
\bibinfo{author}{\bibfnamefont{J.}~\bibnamefont{Conrad}} \bibnamefont{and}
  \bibinfo{author}{\bibfnamefont{O.}~\bibnamefont{Reimer}},
  \bibinfo{journal}{Nature Phys.} \textbf{\bibinfo{volume}{13}},
  \bibinfo{pages}{224} (\bibinfo{year}{2017}), \eprint{1705.11165}.

\bibitem[{\citenamefont{Springel et~al.}(2008)\citenamefont{Springel, White,
  Frenk, Navarro, Jenkins, Vogelsberger, Wang, Ludlow, and
  Helmi}}]{Springel:2008by}
\bibinfo{author}{\bibfnamefont{V.}~\bibnamefont{Springel}},
  \bibinfo{author}{\bibfnamefont{S.~D.~M.} \bibnamefont{White}},
  \bibinfo{author}{\bibfnamefont{C.~S.} \bibnamefont{Frenk}},
  \bibinfo{author}{\bibfnamefont{J.~F.} \bibnamefont{Navarro}},
  \bibinfo{author}{\bibfnamefont{A.}~\bibnamefont{Jenkins}},
  \bibinfo{author}{\bibfnamefont{M.}~\bibnamefont{Vogelsberger}},
  \bibinfo{author}{\bibfnamefont{J.}~\bibnamefont{Wang}},
  \bibinfo{author}{\bibfnamefont{A.}~\bibnamefont{Ludlow}}, \bibnamefont{and}
  \bibinfo{author}{\bibfnamefont{A.}~\bibnamefont{Helmi}}
  (\bibinfo{year}{2008}), \eprint{0809.0894}.

\bibitem[{\citenamefont{Schaller et~al.}(2016)}]{Schaller:2015mua}
\bibinfo{author}{\bibfnamefont{M.}~\bibnamefont{Schaller}}
  \bibnamefont{et~al.}, \bibinfo{journal}{Mon. Not. Roy. Astron. Soc.}
  \textbf{\bibinfo{volume}{455}}, \bibinfo{pages}{4442} (\bibinfo{year}{2016}),
  \eprint{1509.02166}.

\bibitem[{\citenamefont{Taoso et~al.}(2008)\citenamefont{Taoso, Bertone, and
  Masiero}}]{Taoso:2007qk}
\bibinfo{author}{\bibfnamefont{M.}~\bibnamefont{Taoso}},
  \bibinfo{author}{\bibfnamefont{G.}~\bibnamefont{Bertone}}, \bibnamefont{and}
  \bibinfo{author}{\bibfnamefont{A.}~\bibnamefont{Masiero}},
  \bibinfo{journal}{JCAP} \textbf{\bibinfo{volume}{0803}}, \bibinfo{pages}{022}
  (\bibinfo{year}{2008}), \eprint{0711.4996}.

\bibitem[{\citenamefont{Bergstrom and Ullio}(1997)}]{Bergstrom:1997fh}
\bibinfo{author}{\bibfnamefont{L.}~\bibnamefont{Bergstrom}} \bibnamefont{and}
  \bibinfo{author}{\bibfnamefont{P.}~\bibnamefont{Ullio}},
  \bibinfo{journal}{Nucl. Phys.} \textbf{\bibinfo{volume}{B504}},
  \bibinfo{pages}{27} (\bibinfo{year}{1997}), \eprint{hep-ph/9706232}.

\bibitem[{\citenamefont{Ullio and Bergstrom}(1998)}]{Ullio:1997ke}
\bibinfo{author}{\bibfnamefont{P.}~\bibnamefont{Ullio}} \bibnamefont{and}
  \bibinfo{author}{\bibfnamefont{L.}~\bibnamefont{Bergstrom}},
  \bibinfo{journal}{Phys. Rev.} \textbf{\bibinfo{volume}{D57}},
  \bibinfo{pages}{1962} (\bibinfo{year}{1998}), \eprint{hep-ph/9707333}.

\bibitem[{\citenamefont{Bringmann et~al.}(2012)\citenamefont{Bringmann, Huang,
  Ibarra, Vogl, and Weniger}}]{Bringmann:2012vr}
\bibinfo{author}{\bibfnamefont{T.}~\bibnamefont{Bringmann}},
  \bibinfo{author}{\bibfnamefont{X.}~\bibnamefont{Huang}},
  \bibinfo{author}{\bibfnamefont{A.}~\bibnamefont{Ibarra}},
  \bibinfo{author}{\bibfnamefont{S.}~\bibnamefont{Vogl}}, \bibnamefont{and}
  \bibinfo{author}{\bibfnamefont{C.}~\bibnamefont{Weniger}},
  \bibinfo{journal}{JCAP} \textbf{\bibinfo{volume}{1207}}, \bibinfo{pages}{054}
  (\bibinfo{year}{2012}), \eprint{1203.1312}.

\bibitem[{\citenamefont{Weniger}(2012)}]{Weniger:2012tx}
\bibinfo{author}{\bibfnamefont{C.}~\bibnamefont{Weniger}},
  \bibinfo{journal}{JCAP} \textbf{\bibinfo{volume}{1208}}, \bibinfo{pages}{007}
  (\bibinfo{year}{2012}), \eprint{1204.2797}.

\bibitem[{\citenamefont{Finkbeiner et~al.}(2013)\citenamefont{Finkbeiner, Su,
  and Weniger}}]{Finkbeiner:2012ez}
\bibinfo{author}{\bibfnamefont{D.~P.} \bibnamefont{Finkbeiner}},
  \bibinfo{author}{\bibfnamefont{M.}~\bibnamefont{Su}}, \bibnamefont{and}
  \bibinfo{author}{\bibfnamefont{C.}~\bibnamefont{Weniger}},
  \bibinfo{journal}{JCAP} \textbf{\bibinfo{volume}{1301}}, \bibinfo{pages}{029}
  (\bibinfo{year}{2013}), \eprint{1209.4562}.

\bibitem[{\citenamefont{Weniger et~al.}(2013)\citenamefont{Weniger, Su,
  Finkbeiner, Bringmann, and Mirabal}}]{Weniger:2013tza}
\bibinfo{author}{\bibfnamefont{C.}~\bibnamefont{Weniger}},
  \bibinfo{author}{\bibfnamefont{M.}~\bibnamefont{Su}},
  \bibinfo{author}{\bibfnamefont{D.~P.} \bibnamefont{Finkbeiner}},
  \bibinfo{author}{\bibfnamefont{T.}~\bibnamefont{Bringmann}},
  \bibnamefont{and} \bibinfo{author}{\bibfnamefont{N.}~\bibnamefont{Mirabal}}
  (\bibinfo{year}{2013}), \eprint{1305.4710}.

\bibitem[{\citenamefont{Ackermann
  et~al.}(2015{\natexlab{a}})}]{Ackermann:2015lka}
\bibinfo{author}{\bibfnamefont{M.}~\bibnamefont{Ackermann}}
  \bibnamefont{et~al.} (\bibinfo{collaboration}{Fermi-LAT}),
  \bibinfo{journal}{Phys. Rev.} \textbf{\bibinfo{volume}{D91}},
  \bibinfo{pages}{122002} (\bibinfo{year}{2015}{\natexlab{a}}),
  \eprint{1506.00013}.

\bibitem[{\citenamefont{Abdalla et~al.}(2016)}]{Abdalla:2016olq}
\bibinfo{author}{\bibfnamefont{H.}~\bibnamefont{Abdalla}} \bibnamefont{et~al.}
  (\bibinfo{collaboration}{H.E.S.S.}), \bibinfo{journal}{Phys. Rev. Lett.}
  \textbf{\bibinfo{volume}{117}}, \bibinfo{pages}{151302}
  (\bibinfo{year}{2016}), \eprint{1609.08091}.

\bibitem[{\citenamefont{Gaggero
  et~al.}(2015{\natexlab{c}})\citenamefont{Gaggero, Taoso, Urbano, Valli, and
  Ullio}}]{Gaggero:2015nsa}
\bibinfo{author}{\bibfnamefont{D.}~\bibnamefont{Gaggero}},
  \bibinfo{author}{\bibfnamefont{M.}~\bibnamefont{Taoso}},
  \bibinfo{author}{\bibfnamefont{A.}~\bibnamefont{Urbano}},
  \bibinfo{author}{\bibfnamefont{M.}~\bibnamefont{Valli}}, \bibnamefont{and}
  \bibinfo{author}{\bibfnamefont{P.}~\bibnamefont{Ullio}},
  \bibinfo{journal}{JCAP} \textbf{\bibinfo{volume}{1512}}, \bibinfo{pages}{056}
  (\bibinfo{year}{2015}{\natexlab{c}}), \eprint{1507.06129}.

\bibitem[{\citenamefont{Carlson
  et~al.}(2016{\natexlab{a}})\citenamefont{Carlson, Linden, and
  Profumo}}]{Carlson:2015ona}
\bibinfo{author}{\bibfnamefont{E.}~\bibnamefont{Carlson}},
  \bibinfo{author}{\bibfnamefont{T.}~\bibnamefont{Linden}}, \bibnamefont{and}
  \bibinfo{author}{\bibfnamefont{S.}~\bibnamefont{Profumo}},
  \bibinfo{journal}{Phys. Rev. Lett.} \textbf{\bibinfo{volume}{117}},
  \bibinfo{pages}{111101} (\bibinfo{year}{2016}{\natexlab{a}}),
  \eprint{1510.04698}.

\bibitem[{\citenamefont{Carlson
  et~al.}(2016{\natexlab{b}})\citenamefont{Carlson, Linden, and
  Profumo}}]{Carlson:2016iis}
\bibinfo{author}{\bibfnamefont{E.}~\bibnamefont{Carlson}},
  \bibinfo{author}{\bibfnamefont{T.}~\bibnamefont{Linden}}, \bibnamefont{and}
  \bibinfo{author}{\bibfnamefont{S.}~\bibnamefont{Profumo}},
  \bibinfo{journal}{Phys. Rev.} \textbf{\bibinfo{volume}{D94}},
  \bibinfo{pages}{063504} (\bibinfo{year}{2016}{\natexlab{b}}),
  \eprint{1603.06584}.

\bibitem[{\citenamefont{{Vitale} et~al.}(2009)\citenamefont{{Vitale},
  {Morselli}, and {for the Fermi/LAT Collaboration}}}]{VitaleMorselli2009}
\bibinfo{author}{\bibfnamefont{V.}~\bibnamefont{{Vitale}}},
  \bibinfo{author}{\bibfnamefont{A.}~\bibnamefont{{Morselli}}},
  \bibnamefont{and} \bibinfo{author}{\bibnamefont{{for the Fermi/LAT
  Collaboration}}}, \bibinfo{journal}{ArXiv e-prints}  (\bibinfo{year}{2009}),
  \eprint{0912.3828}.

\bibitem[{\citenamefont{Goodenough and Hooper}(2009)}]{Goodenough:2009gk}
\bibinfo{author}{\bibfnamefont{L.}~\bibnamefont{Goodenough}} \bibnamefont{and}
  \bibinfo{author}{\bibfnamefont{D.}~\bibnamefont{Hooper}}
  (\bibinfo{year}{2009}), \eprint{0910.2998}.

\bibitem[{\citenamefont{Hooper and Goodenough}(2011)}]{Hooper:2010mq}
\bibinfo{author}{\bibfnamefont{D.}~\bibnamefont{Hooper}} \bibnamefont{and}
  \bibinfo{author}{\bibfnamefont{L.}~\bibnamefont{Goodenough}},
  \bibinfo{journal}{Phys. Lett.} \textbf{\bibinfo{volume}{B697}},
  \bibinfo{pages}{412} (\bibinfo{year}{2011}), \eprint{1010.2752}.

\bibitem[{\citenamefont{Boyarsky et~al.}(2011)\citenamefont{Boyarsky, Malyshev,
  and Ruchayskiy}}]{Boyarsky:2010dr}
\bibinfo{author}{\bibfnamefont{A.}~\bibnamefont{Boyarsky}},
  \bibinfo{author}{\bibfnamefont{D.}~\bibnamefont{Malyshev}}, \bibnamefont{and}
  \bibinfo{author}{\bibfnamefont{O.}~\bibnamefont{Ruchayskiy}},
  \bibinfo{journal}{Phys. Lett.} \textbf{\bibinfo{volume}{B705}},
  \bibinfo{pages}{165} (\bibinfo{year}{2011}), \eprint{1012.5839}.

\bibitem[{\citenamefont{Hooper and Linden}(2011)}]{Hooper:2011ti}
\bibinfo{author}{\bibfnamefont{D.}~\bibnamefont{Hooper}} \bibnamefont{and}
  \bibinfo{author}{\bibfnamefont{T.}~\bibnamefont{Linden}},
  \bibinfo{journal}{Phys. Rev.} \textbf{\bibinfo{volume}{D84}},
  \bibinfo{pages}{123005} (\bibinfo{year}{2011}), \eprint{1110.0006}.

\bibitem[{\citenamefont{Abazajian and Kaplinghat}(2012)}]{Abazajian:2012pn}
\bibinfo{author}{\bibfnamefont{K.~N.} \bibnamefont{Abazajian}}
  \bibnamefont{and}
  \bibinfo{author}{\bibfnamefont{M.}~\bibnamefont{Kaplinghat}},
  \bibinfo{journal}{Phys. Rev.} \textbf{\bibinfo{volume}{D86}},
  \bibinfo{pages}{083511} (\bibinfo{year}{2012}), \bibinfo{note}{[Erratum:
  Phys. Rev.D87,129902(2013)]}, \eprint{1207.6047}.

\bibitem[{\citenamefont{Hooper and Slatyer}(2013)}]{Hooper:2013rwa}
\bibinfo{author}{\bibfnamefont{D.}~\bibnamefont{Hooper}} \bibnamefont{and}
  \bibinfo{author}{\bibfnamefont{T.~R.} \bibnamefont{Slatyer}},
  \bibinfo{journal}{Phys. Dark Univ.} \textbf{\bibinfo{volume}{2}},
  \bibinfo{pages}{118} (\bibinfo{year}{2013}), \eprint{1302.6589}.

\bibitem[{\citenamefont{Gordon and Macias}(2013)}]{Gordon:2013vta}
\bibinfo{author}{\bibfnamefont{C.}~\bibnamefont{Gordon}} \bibnamefont{and}
  \bibinfo{author}{\bibfnamefont{O.}~\bibnamefont{Macias}},
  \bibinfo{journal}{Phys. Rev.} \textbf{\bibinfo{volume}{D88}},
  \bibinfo{pages}{083521} (\bibinfo{year}{2013}), \bibinfo{note}{[Erratum:
  Phys. Rev.D89,no.4,049901(2014)]}, \eprint{1306.5725}.

\bibitem[{\citenamefont{Huang et~al.}(2013)\citenamefont{Huang, Urbano, and
  Xue}}]{Huang:2013pda}
\bibinfo{author}{\bibfnamefont{W.-C.} \bibnamefont{Huang}},
  \bibinfo{author}{\bibfnamefont{A.}~\bibnamefont{Urbano}}, \bibnamefont{and}
  \bibinfo{author}{\bibfnamefont{W.}~\bibnamefont{Xue}} (\bibinfo{year}{2013}),
  \eprint{1307.6862}.

\bibitem[{\citenamefont{Macias and Gordon}(2014)}]{Macias:2013vya}
\bibinfo{author}{\bibfnamefont{O.}~\bibnamefont{Macias}} \bibnamefont{and}
  \bibinfo{author}{\bibfnamefont{C.}~\bibnamefont{Gordon}},
  \bibinfo{journal}{Phys. Rev.} \textbf{\bibinfo{volume}{D89}},
  \bibinfo{pages}{063515} (\bibinfo{year}{2014}), \eprint{1312.6671}.

\bibitem[{\citenamefont{Abazajian et~al.}(2014)\citenamefont{Abazajian, Canac,
  Horiuchi, and Kaplinghat}}]{Abazajian:2014fta}
\bibinfo{author}{\bibfnamefont{K.~N.} \bibnamefont{Abazajian}},
  \bibinfo{author}{\bibfnamefont{N.}~\bibnamefont{Canac}},
  \bibinfo{author}{\bibfnamefont{S.}~\bibnamefont{Horiuchi}}, \bibnamefont{and}
  \bibinfo{author}{\bibfnamefont{M.}~\bibnamefont{Kaplinghat}},
  \bibinfo{journal}{Phys. Rev.} \textbf{\bibinfo{volume}{D90}},
  \bibinfo{pages}{023526} (\bibinfo{year}{2014}), \eprint{1402.4090}.

\bibitem[{\citenamefont{Daylan et~al.}(2016)\citenamefont{Daylan, Finkbeiner,
  Hooper, Linden, Portillo, Rodd, and Slatyer}}]{Daylan:2014rsa}
\bibinfo{author}{\bibfnamefont{T.}~\bibnamefont{Daylan}},
  \bibinfo{author}{\bibfnamefont{D.~P.} \bibnamefont{Finkbeiner}},
  \bibinfo{author}{\bibfnamefont{D.}~\bibnamefont{Hooper}},
  \bibinfo{author}{\bibfnamefont{T.}~\bibnamefont{Linden}},
  \bibinfo{author}{\bibfnamefont{S.~K.~N.} \bibnamefont{Portillo}},
  \bibinfo{author}{\bibfnamefont{N.~L.} \bibnamefont{Rodd}}, \bibnamefont{and}
  \bibinfo{author}{\bibfnamefont{T.~R.} \bibnamefont{Slatyer}},
  \bibinfo{journal}{Phys. Dark Univ.} \textbf{\bibinfo{volume}{12}},
  \bibinfo{pages}{1} (\bibinfo{year}{2016}), \eprint{1402.6703}.

\bibitem[{\citenamefont{Zhou et~al.}(2015)\citenamefont{Zhou, Liang, Huang, Li,
  Fan, Feng, and Chang}}]{Zhou:2014lva}
\bibinfo{author}{\bibfnamefont{B.}~\bibnamefont{Zhou}},
  \bibinfo{author}{\bibfnamefont{Y.-F.} \bibnamefont{Liang}},
  \bibinfo{author}{\bibfnamefont{X.}~\bibnamefont{Huang}},
  \bibinfo{author}{\bibfnamefont{X.}~\bibnamefont{Li}},
  \bibinfo{author}{\bibfnamefont{Y.-Z.} \bibnamefont{Fan}},
  \bibinfo{author}{\bibfnamefont{L.}~\bibnamefont{Feng}}, \bibnamefont{and}
  \bibinfo{author}{\bibfnamefont{J.}~\bibnamefont{Chang}},
  \bibinfo{journal}{Phys. Rev.} \textbf{\bibinfo{volume}{D91}},
  \bibinfo{pages}{123010} (\bibinfo{year}{2015}), \eprint{1406.6948}.

\bibitem[{\citenamefont{Calore et~al.}(2015{\natexlab{a}})\citenamefont{Calore,
  Cholis, and Weniger}}]{Calore:2014xka}
\bibinfo{author}{\bibfnamefont{F.}~\bibnamefont{Calore}},
  \bibinfo{author}{\bibfnamefont{I.}~\bibnamefont{Cholis}}, \bibnamefont{and}
  \bibinfo{author}{\bibfnamefont{C.}~\bibnamefont{Weniger}},
  \bibinfo{journal}{JCAP} \textbf{\bibinfo{volume}{1503}}, \bibinfo{pages}{038}
  (\bibinfo{year}{2015}{\natexlab{a}}), \eprint{1409.0042}.

\bibitem[{\citenamefont{Horiuchi et~al.}(2016)\citenamefont{Horiuchi,
  Kaplinghat, and Kwa}}]{Horiuchi:2016zwu}
\bibinfo{author}{\bibfnamefont{S.}~\bibnamefont{Horiuchi}},
  \bibinfo{author}{\bibfnamefont{M.}~\bibnamefont{Kaplinghat}},
  \bibnamefont{and} \bibinfo{author}{\bibfnamefont{A.}~\bibnamefont{Kwa}},
  \bibinfo{journal}{JCAP} \textbf{\bibinfo{volume}{1611}}, \bibinfo{pages}{053}
  (\bibinfo{year}{2016}), \eprint{1604.01402}.

\bibitem[{\citenamefont{Ackermann
  et~al.}(2017{\natexlab{a}})}]{TheFermi-LAT:2017vmf}
\bibinfo{author}{\bibfnamefont{M.}~\bibnamefont{Ackermann}}
  \bibnamefont{et~al.} (\bibinfo{collaboration}{Fermi-LAT}),
  \bibinfo{journal}{Astrophys. J.} \textbf{\bibinfo{volume}{840}},
  \bibinfo{pages}{43} (\bibinfo{year}{2017}{\natexlab{a}}),
  \eprint{1704.03910}.

\bibitem[{\citenamefont{de~Boer et~al.}(2017)\citenamefont{de~Boer, Bosse,
  Gebauer, Neumann, and Biermann}}]{deBoer:2017sxb}
\bibinfo{author}{\bibfnamefont{W.}~\bibnamefont{de~Boer}},
  \bibinfo{author}{\bibfnamefont{L.}~\bibnamefont{Bosse}},
  \bibinfo{author}{\bibfnamefont{I.}~\bibnamefont{Gebauer}},
  \bibinfo{author}{\bibfnamefont{A.}~\bibnamefont{Neumann}}, \bibnamefont{and}
  \bibinfo{author}{\bibfnamefont{P.~L.} \bibnamefont{Biermann}},
  \bibinfo{journal}{Phys. Rev.} \textbf{\bibinfo{volume}{D96}},
  \bibinfo{pages}{043012} (\bibinfo{year}{2017}), \eprint{1707.08653}.

\bibitem[{\citenamefont{{Su} et~al.}(2010)\citenamefont{{Su}, {Slatyer}, and
  {Finkbeiner}}}]{2010ApJ...724.1044S}
\bibinfo{author}{\bibfnamefont{M.}~\bibnamefont{{Su}}},
  \bibinfo{author}{\bibfnamefont{T.~R.} \bibnamefont{{Slatyer}}},
  \bibnamefont{and} \bibinfo{author}{\bibfnamefont{D.~P.}
  \bibnamefont{{Finkbeiner}}}, \bibinfo{journal}{\apj}
  \textbf{\bibinfo{volume}{724}}, \bibinfo{pages}{1044} (\bibinfo{year}{2010}),
  \eprint{1005.5480}.

\bibitem[{\citenamefont{Huang et~al.}(2014)\citenamefont{Huang, Urbano, and
  Xue}}]{Huang:2013apa}
\bibinfo{author}{\bibfnamefont{W.-C.} \bibnamefont{Huang}},
  \bibinfo{author}{\bibfnamefont{A.}~\bibnamefont{Urbano}}, \bibnamefont{and}
  \bibinfo{author}{\bibfnamefont{W.}~\bibnamefont{Xue}},
  \bibinfo{journal}{JCAP} \textbf{\bibinfo{volume}{1404}}, \bibinfo{pages}{020}
  (\bibinfo{year}{2014}), \eprint{1310.7609}.

\bibitem[{\citenamefont{Alves et~al.}(2014)\citenamefont{Alves, Profumo,
  Queiroz, and Shepherd}}]{Alves:2014yha}
\bibinfo{author}{\bibfnamefont{A.}~\bibnamefont{Alves}},
  \bibinfo{author}{\bibfnamefont{S.}~\bibnamefont{Profumo}},
  \bibinfo{author}{\bibfnamefont{F.~S.} \bibnamefont{Queiroz}},
  \bibnamefont{and} \bibinfo{author}{\bibfnamefont{W.}~\bibnamefont{Shepherd}},
  \bibinfo{journal}{Phys. Rev.} \textbf{\bibinfo{volume}{D90}},
  \bibinfo{pages}{115003} (\bibinfo{year}{2014}), \eprint{1403.5027}.

\bibitem[{\citenamefont{Berlin et~al.}(2014)\citenamefont{Berlin, Hooper, and
  McDermott}}]{Berlin:2014tja}
\bibinfo{author}{\bibfnamefont{A.}~\bibnamefont{Berlin}},
  \bibinfo{author}{\bibfnamefont{D.}~\bibnamefont{Hooper}}, \bibnamefont{and}
  \bibinfo{author}{\bibfnamefont{S.~D.} \bibnamefont{McDermott}},
  \bibinfo{journal}{Phys. Rev.} \textbf{\bibinfo{volume}{D89}},
  \bibinfo{pages}{115022} (\bibinfo{year}{2014}), \eprint{1404.0022}.

\bibitem[{\citenamefont{Abdullah et~al.}(2014)\citenamefont{Abdullah, DiFranzo,
  Rajaraman, Tait, Tanedo, and Wijangco}}]{Abdullah:2014lla}
\bibinfo{author}{\bibfnamefont{M.}~\bibnamefont{Abdullah}},
  \bibinfo{author}{\bibfnamefont{A.}~\bibnamefont{DiFranzo}},
  \bibinfo{author}{\bibfnamefont{A.}~\bibnamefont{Rajaraman}},
  \bibinfo{author}{\bibfnamefont{T.~M.~P.} \bibnamefont{Tait}},
  \bibinfo{author}{\bibfnamefont{P.}~\bibnamefont{Tanedo}}, \bibnamefont{and}
  \bibinfo{author}{\bibfnamefont{A.~M.} \bibnamefont{Wijangco}},
  \bibinfo{journal}{Phys. Rev.} \textbf{\bibinfo{volume}{D90}},
  \bibinfo{pages}{035004} (\bibinfo{year}{2014}), \eprint{1404.6528}.

\bibitem[{\citenamefont{Agrawal et~al.}(2015)\citenamefont{Agrawal, Batell,
  Fox, and Harnik}}]{Agrawal:2014oha}
\bibinfo{author}{\bibfnamefont{P.}~\bibnamefont{Agrawal}},
  \bibinfo{author}{\bibfnamefont{B.}~\bibnamefont{Batell}},
  \bibinfo{author}{\bibfnamefont{P.~J.} \bibnamefont{Fox}}, \bibnamefont{and}
  \bibinfo{author}{\bibfnamefont{R.}~\bibnamefont{Harnik}},
  \bibinfo{journal}{JCAP} \textbf{\bibinfo{volume}{1505}}, \bibinfo{pages}{011}
  (\bibinfo{year}{2015}), \eprint{1411.2592}.

\bibitem[{\citenamefont{Calore et~al.}(2015{\natexlab{b}})\citenamefont{Calore,
  Cholis, McCabe, and Weniger}}]{Calore:2014nla}
\bibinfo{author}{\bibfnamefont{F.}~\bibnamefont{Calore}},
  \bibinfo{author}{\bibfnamefont{I.}~\bibnamefont{Cholis}},
  \bibinfo{author}{\bibfnamefont{C.}~\bibnamefont{McCabe}}, \bibnamefont{and}
  \bibinfo{author}{\bibfnamefont{C.}~\bibnamefont{Weniger}},
  \bibinfo{journal}{Phys. Rev.} \textbf{\bibinfo{volume}{D91}},
  \bibinfo{pages}{063003} (\bibinfo{year}{2015}{\natexlab{b}}),
  \eprint{1411.4647}.

\bibitem[{\citenamefont{Kaplinghat et~al.}(2015)\citenamefont{Kaplinghat,
  Linden, and Yu}}]{Kaplinghat:2015gha}
\bibinfo{author}{\bibfnamefont{M.}~\bibnamefont{Kaplinghat}},
  \bibinfo{author}{\bibfnamefont{T.}~\bibnamefont{Linden}}, \bibnamefont{and}
  \bibinfo{author}{\bibfnamefont{H.-B.} \bibnamefont{Yu}},
  \bibinfo{journal}{Phys. Rev. Lett.} \textbf{\bibinfo{volume}{114}},
  \bibinfo{pages}{211303} (\bibinfo{year}{2015}), \eprint{1501.03507}.

\bibitem[{\citenamefont{Karwin et~al.}(2017)\citenamefont{Karwin, Murgia, Tait,
  Porter, and Tanedo}}]{Karwin:2016tsw}
\bibinfo{author}{\bibfnamefont{C.}~\bibnamefont{Karwin}},
  \bibinfo{author}{\bibfnamefont{S.}~\bibnamefont{Murgia}},
  \bibinfo{author}{\bibfnamefont{T.~M.~P.} \bibnamefont{Tait}},
  \bibinfo{author}{\bibfnamefont{T.~A.} \bibnamefont{Porter}},
  \bibnamefont{and} \bibinfo{author}{\bibfnamefont{P.}~\bibnamefont{Tanedo}},
  \bibinfo{journal}{Phys. Rev.} \textbf{\bibinfo{volume}{D95}},
  \bibinfo{pages}{103005} (\bibinfo{year}{2017}), \eprint{1612.05687}.

\bibitem[{\citenamefont{Petrović et~al.}(2015)\citenamefont{Petrović,
  Serpico, and Zaharijas}}]{Petrovic:2014xra}
\bibinfo{author}{\bibfnamefont{J.}~\bibnamefont{Petrović}},
  \bibinfo{author}{\bibfnamefont{P.~D.} \bibnamefont{Serpico}},
  \bibnamefont{and}
  \bibinfo{author}{\bibfnamefont{G.}~\bibnamefont{Zaharijas}},
  \bibinfo{journal}{JCAP} \textbf{\bibinfo{volume}{1502}}, \bibinfo{pages}{023}
  (\bibinfo{year}{2015}), \eprint{1411.2980}.

\bibitem[{\citenamefont{Abazajian}(2011)}]{Abazajian:2010zy}
\bibinfo{author}{\bibfnamefont{K.~N.} \bibnamefont{Abazajian}},
  \bibinfo{journal}{JCAP} \textbf{\bibinfo{volume}{1103}}, \bibinfo{pages}{010}
  (\bibinfo{year}{2011}), \eprint{1011.4275}.

\bibitem[{\citenamefont{Yuan and Zhang}(2014)}]{Yuan:2014rca}
\bibinfo{author}{\bibfnamefont{Q.}~\bibnamefont{Yuan}} \bibnamefont{and}
  \bibinfo{author}{\bibfnamefont{B.}~\bibnamefont{Zhang}},
  \bibinfo{journal}{JHEAp} \textbf{\bibinfo{volume}{3-4}}, \bibinfo{pages}{1}
  (\bibinfo{year}{2014}), \eprint{1404.2318}.

\bibitem[{\citenamefont{Hooper et~al.}(2013)\citenamefont{Hooper, Cholis,
  Linden, Siegal-Gaskins, and Slatyer}}]{Hooper:2013nhl}
\bibinfo{author}{\bibfnamefont{D.}~\bibnamefont{Hooper}},
  \bibinfo{author}{\bibfnamefont{I.}~\bibnamefont{Cholis}},
  \bibinfo{author}{\bibfnamefont{T.}~\bibnamefont{Linden}},
  \bibinfo{author}{\bibfnamefont{J.}~\bibnamefont{Siegal-Gaskins}},
  \bibnamefont{and} \bibinfo{author}{\bibfnamefont{T.}~\bibnamefont{Slatyer}},
  \bibinfo{journal}{Phys. Rev.} \textbf{\bibinfo{volume}{D88}},
  \bibinfo{pages}{083009} (\bibinfo{year}{2013}), \eprint{1305.0830}.

\bibitem[{\citenamefont{Cholis et~al.}(2015{\natexlab{a}})\citenamefont{Cholis,
  Hooper, and Linden}}]{Cholis:2014lta}
\bibinfo{author}{\bibfnamefont{I.}~\bibnamefont{Cholis}},
  \bibinfo{author}{\bibfnamefont{D.}~\bibnamefont{Hooper}}, \bibnamefont{and}
  \bibinfo{author}{\bibfnamefont{T.}~\bibnamefont{Linden}},
  \bibinfo{journal}{JCAP} \textbf{\bibinfo{volume}{1506}}, \bibinfo{pages}{043}
  (\bibinfo{year}{2015}{\natexlab{a}}), \eprint{1407.5625}.

\bibitem[{\citenamefont{Carlson and Profumo}(2014)}]{Carlson:2014cwa}
\bibinfo{author}{\bibfnamefont{E.}~\bibnamefont{Carlson}} \bibnamefont{and}
  \bibinfo{author}{\bibfnamefont{S.}~\bibnamefont{Profumo}},
  \bibinfo{journal}{Phys. Rev.} \textbf{\bibinfo{volume}{D90}},
  \bibinfo{pages}{023015} (\bibinfo{year}{2014}), \eprint{1405.7685}.

\bibitem[{\citenamefont{Petrović et~al.}(2014)\citenamefont{Petrović,
  Serpico, and Zaharijaš}}]{Petrovic:2014uda}
\bibinfo{author}{\bibfnamefont{J.}~\bibnamefont{Petrović}},
  \bibinfo{author}{\bibfnamefont{P.~D.} \bibnamefont{Serpico}},
  \bibnamefont{and}
  \bibinfo{author}{\bibfnamefont{G.}~\bibnamefont{Zaharijaš}},
  \bibinfo{journal}{JCAP} \textbf{\bibinfo{volume}{1410}}, \bibinfo{pages}{052}
  (\bibinfo{year}{2014}), \eprint{1405.7928}.

\bibitem[{\citenamefont{Cholis et~al.}(2015{\natexlab{b}})\citenamefont{Cholis,
  Evoli, Calore, Linden, Weniger, and Hooper}}]{Cholis:2015dea}
\bibinfo{author}{\bibfnamefont{I.}~\bibnamefont{Cholis}},
  \bibinfo{author}{\bibfnamefont{C.}~\bibnamefont{Evoli}},
  \bibinfo{author}{\bibfnamefont{F.}~\bibnamefont{Calore}},
  \bibinfo{author}{\bibfnamefont{T.}~\bibnamefont{Linden}},
  \bibinfo{author}{\bibfnamefont{C.}~\bibnamefont{Weniger}}, \bibnamefont{and}
  \bibinfo{author}{\bibfnamefont{D.}~\bibnamefont{Hooper}},
  \bibinfo{journal}{JCAP} \textbf{\bibinfo{volume}{1512}}, \bibinfo{pages}{005}
  (\bibinfo{year}{2015}{\natexlab{b}}), \eprint{1506.05119}.

\bibitem[{\citenamefont{Figer et~al.}(2004)\citenamefont{Figer, Rich, Kim,
  Morris, and Serabyn}}]{Figer:2003tu}
\bibinfo{author}{\bibfnamefont{D.~F.} \bibnamefont{Figer}},
  \bibinfo{author}{\bibfnamefont{R.~M.} \bibnamefont{Rich}},
  \bibinfo{author}{\bibfnamefont{S.~S.} \bibnamefont{Kim}},
  \bibinfo{author}{\bibfnamefont{M.}~\bibnamefont{Morris}}, \bibnamefont{and}
  \bibinfo{author}{\bibfnamefont{E.}~\bibnamefont{Serabyn}},
  \bibinfo{journal}{Astrophys. J.} \textbf{\bibinfo{volume}{601}},
  \bibinfo{pages}{319} (\bibinfo{year}{2004}), \eprint{astro-ph/0309757}.

\bibitem[{\citenamefont{Crocker et~al.}(2011)\citenamefont{Crocker, Jones,
  Aharonian, Law, Melia, Oka, and Ott}}]{Crocker:2010qn}
\bibinfo{author}{\bibfnamefont{R.~M.} \bibnamefont{Crocker}},
  \bibinfo{author}{\bibfnamefont{D.~I.} \bibnamefont{Jones}},
  \bibinfo{author}{\bibfnamefont{F.}~\bibnamefont{Aharonian}},
  \bibinfo{author}{\bibfnamefont{C.~J.} \bibnamefont{Law}},
  \bibinfo{author}{\bibfnamefont{F.}~\bibnamefont{Melia}},
  \bibinfo{author}{\bibfnamefont{T.}~\bibnamefont{Oka}}, \bibnamefont{and}
  \bibinfo{author}{\bibfnamefont{J.}~\bibnamefont{Ott}}, \bibinfo{journal}{Mon.
  Not. Roy. Astron. Soc.} \textbf{\bibinfo{volume}{413}}, \bibinfo{pages}{763}
  (\bibinfo{year}{2011}), \eprint{1011.0206}.

\bibitem[{\citenamefont{Ferriere et~al.}(2007)\citenamefont{Ferriere, Gillard,
  and Jean}}]{Ferriere:2007yq}
\bibinfo{author}{\bibfnamefont{K.}~\bibnamefont{Ferriere}},
  \bibinfo{author}{\bibfnamefont{W.}~\bibnamefont{Gillard}}, \bibnamefont{and}
  \bibinfo{author}{\bibfnamefont{P.}~\bibnamefont{Jean}},
  \bibinfo{journal}{Astron. Astrophys.} \textbf{\bibinfo{volume}{467}},
  \bibinfo{pages}{611} (\bibinfo{year}{2007}), \eprint{astro-ph/0702532}.

\bibitem[{\citenamefont{Figer et~al.}(2002)}]{Figer:2002bi}
\bibinfo{author}{\bibfnamefont{D.~F.} \bibnamefont{Figer}}
  \bibnamefont{et~al.}, \bibinfo{journal}{Astrophys. J.}
  \textbf{\bibinfo{volume}{581}}, \bibinfo{pages}{258} (\bibinfo{year}{2002}),
  \eprint{astro-ph/0208145}.

\bibitem[{\citenamefont{{Montmerle}}(1979)}]{1979ApJ...231...95M}
\bibinfo{author}{\bibfnamefont{T.}~\bibnamefont{{Montmerle}}},
  \bibinfo{journal}{\apj} \textbf{\bibinfo{volume}{231}}, \bibinfo{pages}{95}
  (\bibinfo{year}{1979}).

\bibitem[{\citenamefont{{Schmidt}}(1959)}]{1959ApJ...129..243S}
\bibinfo{author}{\bibfnamefont{M.}~\bibnamefont{{Schmidt}}},
  \bibinfo{journal}{\apj} \textbf{\bibinfo{volume}{129}}, \bibinfo{pages}{243}
  (\bibinfo{year}{1959}).

\bibitem[{\citenamefont{Ajello et~al.}(2016)}]{TheFermi-LAT:2015kwa}
\bibinfo{author}{\bibfnamefont{M.}~\bibnamefont{Ajello}} \bibnamefont{et~al.}
  (\bibinfo{collaboration}{Fermi-LAT}), \bibinfo{journal}{Astrophys. J.}
  \textbf{\bibinfo{volume}{819}}, \bibinfo{pages}{44} (\bibinfo{year}{2016}),
  \eprint{1511.02938}.

\bibitem[{\citenamefont{Bartels et~al.}(2016)\citenamefont{Bartels,
  Krishnamurthy, and Weniger}}]{Bartels:2015aea}
\bibinfo{author}{\bibfnamefont{R.}~\bibnamefont{Bartels}},
  \bibinfo{author}{\bibfnamefont{S.}~\bibnamefont{Krishnamurthy}},
  \bibnamefont{and} \bibinfo{author}{\bibfnamefont{C.}~\bibnamefont{Weniger}},
  \bibinfo{journal}{Phys. Rev. Lett.} \textbf{\bibinfo{volume}{116}},
  \bibinfo{pages}{051102} (\bibinfo{year}{2016}), \eprint{1506.05104}.

\bibitem[{\citenamefont{Lee et~al.}(2016)\citenamefont{Lee, Lisanti, Safdi,
  Slatyer, and Xue}}]{Lee:2015fea}
\bibinfo{author}{\bibfnamefont{S.~K.} \bibnamefont{Lee}},
  \bibinfo{author}{\bibfnamefont{M.}~\bibnamefont{Lisanti}},
  \bibinfo{author}{\bibfnamefont{B.~R.} \bibnamefont{Safdi}},
  \bibinfo{author}{\bibfnamefont{T.~R.} \bibnamefont{Slatyer}},
  \bibnamefont{and} \bibinfo{author}{\bibfnamefont{W.}~\bibnamefont{Xue}},
  \bibinfo{journal}{Phys. Rev. Lett.} \textbf{\bibinfo{volume}{116}},
  \bibinfo{pages}{051103} (\bibinfo{year}{2016}), \eprint{1506.05124}.

\bibitem[{\citenamefont{Hooper and Mohlabeng}(2016)}]{Hooper:2015jlu}
\bibinfo{author}{\bibfnamefont{D.}~\bibnamefont{Hooper}} \bibnamefont{and}
  \bibinfo{author}{\bibfnamefont{G.}~\bibnamefont{Mohlabeng}},
  \bibinfo{journal}{JCAP} \textbf{\bibinfo{volume}{1603}}, \bibinfo{pages}{049}
  (\bibinfo{year}{2016}), \eprint{1512.04966}.

\bibitem[{\citenamefont{Ploeg et~al.}(2017)\citenamefont{Ploeg, Gordon,
  Crocker, and Macias}}]{Ploeg:2017vai}
\bibinfo{author}{\bibfnamefont{H.}~\bibnamefont{Ploeg}},
  \bibinfo{author}{\bibfnamefont{C.}~\bibnamefont{Gordon}},
  \bibinfo{author}{\bibfnamefont{R.}~\bibnamefont{Crocker}}, \bibnamefont{and}
  \bibinfo{author}{\bibfnamefont{O.}~\bibnamefont{Macias}},
  \bibinfo{journal}{JCAP} \textbf{\bibinfo{volume}{1708}}, \bibinfo{pages}{015}
  (\bibinfo{year}{2017}), \eprint{1705.00806}.

\bibitem[{\citenamefont{Ajello et~al.}(2017)}]{Fermi-LAT:2017yoi}
\bibinfo{author}{\bibfnamefont{M.}~\bibnamefont{Ajello}} \bibnamefont{et~al.}
  (\bibinfo{collaboration}{Fermi-LAT}), \bibinfo{journal}{Submitted to:
  Astrophys. J.}  (\bibinfo{year}{2017}), \eprint{1705.00009}.

\bibitem[{\citenamefont{Storm et~al.}(2017)\citenamefont{Storm, Weniger, and
  Calore}}]{Storm:2017arh}
\bibinfo{author}{\bibfnamefont{E.}~\bibnamefont{Storm}},
  \bibinfo{author}{\bibfnamefont{C.}~\bibnamefont{Weniger}}, \bibnamefont{and}
  \bibinfo{author}{\bibfnamefont{F.}~\bibnamefont{Calore}},
  \bibinfo{journal}{JCAP} \textbf{\bibinfo{volume}{1708}}, \bibinfo{pages}{022}
  (\bibinfo{year}{2017}), \eprint{1705.04065}.

\bibitem[{\citenamefont{Bartels et~al.}(2017)\citenamefont{Bartels, Storm,
  Weniger, and Calore}}]{Bartels:2017vsx}
\bibinfo{author}{\bibfnamefont{R.}~\bibnamefont{Bartels}},
  \bibinfo{author}{\bibfnamefont{E.}~\bibnamefont{Storm}},
  \bibinfo{author}{\bibfnamefont{C.}~\bibnamefont{Weniger}}, \bibnamefont{and}
  \bibinfo{author}{\bibfnamefont{F.}~\bibnamefont{Calore}}
  (\bibinfo{year}{2017}), \eprint{1711.04778}.

\bibitem[{\citenamefont{Calore et~al.}(2016)\citenamefont{Calore, Di~Mauro,
  Donato, Hessels, and Weniger}}]{Calore:2015bsx}
\bibinfo{author}{\bibfnamefont{F.}~\bibnamefont{Calore}},
  \bibinfo{author}{\bibfnamefont{M.}~\bibnamefont{Di~Mauro}},
  \bibinfo{author}{\bibfnamefont{F.}~\bibnamefont{Donato}},
  \bibinfo{author}{\bibfnamefont{J.~W.~T.} \bibnamefont{Hessels}},
  \bibnamefont{and} \bibinfo{author}{\bibfnamefont{C.}~\bibnamefont{Weniger}},
  \bibinfo{journal}{Astrophys. J.} \textbf{\bibinfo{volume}{827}},
  \bibinfo{pages}{143} (\bibinfo{year}{2016}), \eprint{1512.06825}.

\bibitem[{\citenamefont{Ackermann
  et~al.}(2017{\natexlab{b}})}]{Ackermann:2017nya}
\bibinfo{author}{\bibfnamefont{M.}~\bibnamefont{Ackermann}}
  \bibnamefont{et~al.} (\bibinfo{collaboration}{Fermi-LAT}),
  \bibinfo{journal}{Astrophys. J.} \textbf{\bibinfo{volume}{836}},
  \bibinfo{pages}{208} (\bibinfo{year}{2017}{\natexlab{b}}),
  \eprint{1702.08602}.

\bibitem[{\citenamefont{Eckner et~al.}(2017)}]{Eckner:2017oul}
\bibinfo{author}{\bibfnamefont{C.}~\bibnamefont{Eckner}} \bibnamefont{et~al.}
  (\bibinfo{year}{2017}), \eprint{1711.05127}.

\bibitem[{\citenamefont{Lake}(1990)}]{Lake:1990du}
\bibinfo{author}{\bibfnamefont{G.}~\bibnamefont{Lake}},
  \bibinfo{journal}{Nature} \textbf{\bibinfo{volume}{346}}, \bibinfo{pages}{39}
  (\bibinfo{year}{1990}).

\bibitem[{\citenamefont{Evans et~al.}(2004)\citenamefont{Evans, Ferrer, and
  Sarkar}}]{Evans:2003sc}
\bibinfo{author}{\bibfnamefont{N.~W.} \bibnamefont{Evans}},
  \bibinfo{author}{\bibfnamefont{F.}~\bibnamefont{Ferrer}}, \bibnamefont{and}
  \bibinfo{author}{\bibfnamefont{S.}~\bibnamefont{Sarkar}},
  \bibinfo{journal}{Phys. Rev.} \textbf{\bibinfo{volume}{D69}},
  \bibinfo{pages}{123501} (\bibinfo{year}{2004}), \eprint{astro-ph/0311145}.

\bibitem[{\citenamefont{Strigari et~al.}(2007)\citenamefont{Strigari,
  Koushiappas, Bullock, and Kaplinghat}}]{Strigari:2006rd}
\bibinfo{author}{\bibfnamefont{L.~E.} \bibnamefont{Strigari}},
  \bibinfo{author}{\bibfnamefont{S.~M.} \bibnamefont{Koushiappas}},
  \bibinfo{author}{\bibfnamefont{J.~S.} \bibnamefont{Bullock}},
  \bibnamefont{and}
  \bibinfo{author}{\bibfnamefont{M.}~\bibnamefont{Kaplinghat}},
  \bibinfo{journal}{Phys. Rev.} \textbf{\bibinfo{volume}{D75}},
  \bibinfo{pages}{083526} (\bibinfo{year}{2007}), \eprint{astro-ph/0611925}.

\bibitem[{\citenamefont{Strigari et~al.}(2008)\citenamefont{Strigari,
  Koushiappas, Bullock, Kaplinghat, Simon, Geha, and
  Willman}}]{Strigari:2007at}
\bibinfo{author}{\bibfnamefont{L.~E.} \bibnamefont{Strigari}},
  \bibinfo{author}{\bibfnamefont{S.~M.} \bibnamefont{Koushiappas}},
  \bibinfo{author}{\bibfnamefont{J.~S.} \bibnamefont{Bullock}},
  \bibinfo{author}{\bibfnamefont{M.}~\bibnamefont{Kaplinghat}},
  \bibinfo{author}{\bibfnamefont{J.~D.} \bibnamefont{Simon}},
  \bibinfo{author}{\bibfnamefont{M.}~\bibnamefont{Geha}}, \bibnamefont{and}
  \bibinfo{author}{\bibfnamefont{B.}~\bibnamefont{Willman}},
  \bibinfo{journal}{Astrophys. J.} \textbf{\bibinfo{volume}{678}},
  \bibinfo{pages}{614} (\bibinfo{year}{2008}), \eprint{0709.1510}.

\bibitem[{\citenamefont{Ackermann et~al.}(2011)}]{Ackermann:2011wa}
\bibinfo{author}{\bibfnamefont{M.}~\bibnamefont{Ackermann}}
  \bibnamefont{et~al.} (\bibinfo{collaboration}{Fermi-LAT}),
  \bibinfo{journal}{Phys. Rev. Lett.} \textbf{\bibinfo{volume}{107}},
  \bibinfo{pages}{241302} (\bibinfo{year}{2011}), \eprint{1108.3546}.

\bibitem[{\citenamefont{Ackermann
  et~al.}(2015{\natexlab{b}})}]{Ackermann:2015zua}
\bibinfo{author}{\bibfnamefont{M.}~\bibnamefont{Ackermann}}
  \bibnamefont{et~al.} (\bibinfo{collaboration}{Fermi-LAT}),
  \bibinfo{journal}{Phys. Rev. Lett.} \textbf{\bibinfo{volume}{115}},
  \bibinfo{pages}{231301} (\bibinfo{year}{2015}{\natexlab{b}}),
  \eprint{1503.02641}.

\bibitem[{\citenamefont{Ahnen et~al.}(2016)}]{Ahnen:2016qkx}
\bibinfo{author}{\bibfnamefont{M.~L.} \bibnamefont{Ahnen}} \bibnamefont{et~al.}
  (\bibinfo{collaboration}{Fermi-LAT, MAGIC}), \bibinfo{journal}{JCAP}
  \textbf{\bibinfo{volume}{1602}}, \bibinfo{pages}{039} (\bibinfo{year}{2016}),
  \eprint{1601.06590}.

\bibitem[{\citenamefont{Albert et~al.}(2017)}]{Fermi-LAT:2016uux}
\bibinfo{author}{\bibfnamefont{A.}~\bibnamefont{Albert}} \bibnamefont{et~al.}
  (\bibinfo{collaboration}{DES, Fermi-LAT}), \bibinfo{journal}{Astrophys. J.}
  \textbf{\bibinfo{volume}{834}}, \bibinfo{pages}{110} (\bibinfo{year}{2017}),
  \eprint{1611.03184}.

\bibitem[{\citenamefont{Archambault et~al.}(2017)}]{Archambault:2017wyh}
\bibinfo{author}{\bibfnamefont{S.}~\bibnamefont{Archambault}}
  \bibnamefont{et~al.} (\bibinfo{collaboration}{VERITAS}),
  \bibinfo{journal}{Phys. Rev.} \textbf{\bibinfo{volume}{D95}},
  \bibinfo{pages}{082001} (\bibinfo{year}{2017}), \eprint{1703.04937}.

\bibitem[{\citenamefont{Abazajian and Keeley}(2016)}]{Abazajian:2015raa}
\bibinfo{author}{\bibfnamefont{K.~N.} \bibnamefont{Abazajian}}
  \bibnamefont{and} \bibinfo{author}{\bibfnamefont{R.~E.}
  \bibnamefont{Keeley}}, \bibinfo{journal}{Phys. Rev.}
  \textbf{\bibinfo{volume}{D93}}, \bibinfo{pages}{083514}
  (\bibinfo{year}{2016}), \eprint{1510.06424}.

\bibitem[{\citenamefont{Keeley et~al.}(2017)\citenamefont{Keeley, Abazajian,
  Kwa, Rodd, and Safdi}}]{Keeley:2017fbz}
\bibinfo{author}{\bibfnamefont{R.}~\bibnamefont{Keeley}},
  \bibinfo{author}{\bibfnamefont{K.}~\bibnamefont{Abazajian}},
  \bibinfo{author}{\bibfnamefont{A.}~\bibnamefont{Kwa}},
  \bibinfo{author}{\bibfnamefont{N.}~\bibnamefont{Rodd}}, \bibnamefont{and}
  \bibinfo{author}{\bibfnamefont{B.}~\bibnamefont{Safdi}}
  (\bibinfo{year}{2017}), \eprint{1710.03215}.

\bibitem[{\citenamefont{Bonnivard
  et~al.}(2015{\natexlab{a}})\citenamefont{Bonnivard, Combet, Maurin, and
  Walker}}]{Bonnivard:2014kza}
\bibinfo{author}{\bibfnamefont{V.}~\bibnamefont{Bonnivard}},
  \bibinfo{author}{\bibfnamefont{C.}~\bibnamefont{Combet}},
  \bibinfo{author}{\bibfnamefont{D.}~\bibnamefont{Maurin}}, \bibnamefont{and}
  \bibinfo{author}{\bibfnamefont{M.~G.} \bibnamefont{Walker}},
  \bibinfo{journal}{Mon. Not. Roy. Astron. Soc.}
  \textbf{\bibinfo{volume}{446}}, \bibinfo{pages}{3002}
  (\bibinfo{year}{2015}{\natexlab{a}}), \eprint{1407.7822}.

\bibitem[{\citenamefont{Ullio and Valli}(2016)}]{Ullio:2016kvy}
\bibinfo{author}{\bibfnamefont{P.}~\bibnamefont{Ullio}} \bibnamefont{and}
  \bibinfo{author}{\bibfnamefont{M.}~\bibnamefont{Valli}},
  \bibinfo{journal}{JCAP} \textbf{\bibinfo{volume}{1607}}, \bibinfo{pages}{025}
  (\bibinfo{year}{2016}), \eprint{1603.07721}.

\bibitem[{\citenamefont{Hayashi et~al.}(2016)\citenamefont{Hayashi, Ichikawa,
  Matsumoto, Ibe, Ishigaki, and Sugai}}]{Hayashi:2016kcy}
\bibinfo{author}{\bibfnamefont{K.}~\bibnamefont{Hayashi}},
  \bibinfo{author}{\bibfnamefont{K.}~\bibnamefont{Ichikawa}},
  \bibinfo{author}{\bibfnamefont{S.}~\bibnamefont{Matsumoto}},
  \bibinfo{author}{\bibfnamefont{M.}~\bibnamefont{Ibe}},
  \bibinfo{author}{\bibfnamefont{M.~N.} \bibnamefont{Ishigaki}},
  \bibnamefont{and} \bibinfo{author}{\bibfnamefont{H.}~\bibnamefont{Sugai}},
  \bibinfo{journal}{Mon. Not. Roy. Astron. Soc.}
  \textbf{\bibinfo{volume}{461}}, \bibinfo{pages}{2914} (\bibinfo{year}{2016}),
  \eprint{1603.08046}.

\bibitem[{\citenamefont{Ichikawa et~al.}(2017)\citenamefont{Ichikawa, Ishigaki,
  Matsumoto, Ibe, Sugai, Hayashi, and Horigome}}]{Ichikawa:2016nbi}
\bibinfo{author}{\bibfnamefont{K.}~\bibnamefont{Ichikawa}},
  \bibinfo{author}{\bibfnamefont{M.~N.} \bibnamefont{Ishigaki}},
  \bibinfo{author}{\bibfnamefont{S.}~\bibnamefont{Matsumoto}},
  \bibinfo{author}{\bibfnamefont{M.}~\bibnamefont{Ibe}},
  \bibinfo{author}{\bibfnamefont{H.}~\bibnamefont{Sugai}},
  \bibinfo{author}{\bibfnamefont{K.}~\bibnamefont{Hayashi}}, \bibnamefont{and}
  \bibinfo{author}{\bibfnamefont{S.-i.} \bibnamefont{Horigome}},
  \bibinfo{journal}{Mon. Not. Roy. Astron. Soc.}
  \textbf{\bibinfo{volume}{468}}, \bibinfo{pages}{2884} (\bibinfo{year}{2017}),
  \eprint{1608.01749}.

\bibitem[{\citenamefont{Geringer-Sameth
  et~al.}(2015)\citenamefont{Geringer-Sameth, Walker, Koushiappas, Koposov,
  Belokurov, Torrealba, and Evans}}]{Geringer-Sameth:2015lua}
\bibinfo{author}{\bibfnamefont{A.}~\bibnamefont{Geringer-Sameth}},
  \bibinfo{author}{\bibfnamefont{M.~G.} \bibnamefont{Walker}},
  \bibinfo{author}{\bibfnamefont{S.~M.} \bibnamefont{Koushiappas}},
  \bibinfo{author}{\bibfnamefont{S.~E.} \bibnamefont{Koposov}},
  \bibinfo{author}{\bibfnamefont{V.}~\bibnamefont{Belokurov}},
  \bibinfo{author}{\bibfnamefont{G.}~\bibnamefont{Torrealba}},
  \bibnamefont{and} \bibinfo{author}{\bibfnamefont{N.~W.} \bibnamefont{Evans}},
  \bibinfo{journal}{Phys. Rev. Lett.} \textbf{\bibinfo{volume}{115}},
  \bibinfo{pages}{081101} (\bibinfo{year}{2015}), \eprint{1503.02320}.

\bibitem[{\citenamefont{Hooper and Linden}(2015)}]{Hooper:2015ula}
\bibinfo{author}{\bibfnamefont{D.}~\bibnamefont{Hooper}} \bibnamefont{and}
  \bibinfo{author}{\bibfnamefont{T.}~\bibnamefont{Linden}},
  \bibinfo{journal}{JCAP} \textbf{\bibinfo{volume}{1509}}, \bibinfo{pages}{016}
  (\bibinfo{year}{2015}), \eprint{1503.06209}.

\bibitem[{\citenamefont{Bonnivard
  et~al.}(2015{\natexlab{b}})\citenamefont{Bonnivard, Combet, Maurin,
  Geringer-Sameth, Koushiappas, Walker, Mateo, Olszewski, and
  Bailey~III}}]{Bonnivard:2015tta}
\bibinfo{author}{\bibfnamefont{V.}~\bibnamefont{Bonnivard}},
  \bibinfo{author}{\bibfnamefont{C.}~\bibnamefont{Combet}},
  \bibinfo{author}{\bibfnamefont{D.}~\bibnamefont{Maurin}},
  \bibinfo{author}{\bibfnamefont{A.}~\bibnamefont{Geringer-Sameth}},
  \bibinfo{author}{\bibfnamefont{S.~M.} \bibnamefont{Koushiappas}},
  \bibinfo{author}{\bibfnamefont{M.~G.} \bibnamefont{Walker}},
  \bibinfo{author}{\bibfnamefont{M.}~\bibnamefont{Mateo}},
  \bibinfo{author}{\bibfnamefont{E.~W.} \bibnamefont{Olszewski}},
  \bibnamefont{and} \bibinfo{author}{\bibfnamefont{J.~I.}
  \bibnamefont{Bailey~III}}, \bibinfo{journal}{Astrophys. J.}
  \textbf{\bibinfo{volume}{808}}, \bibinfo{pages}{L36}
  (\bibinfo{year}{2015}{\natexlab{b}}), \eprint{1504.03309}.

\bibitem[{\citenamefont{Regis et~al.}(2017)\citenamefont{Regis, Richter, and
  Colafrancesco}}]{Regis:2017oet}
\bibinfo{author}{\bibfnamefont{M.}~\bibnamefont{Regis}},
  \bibinfo{author}{\bibfnamefont{L.}~\bibnamefont{Richter}}, \bibnamefont{and}
  \bibinfo{author}{\bibfnamefont{S.}~\bibnamefont{Colafrancesco}},
  \bibinfo{journal}{JCAP} \textbf{\bibinfo{volume}{1707}}, \bibinfo{pages}{025}
  (\bibinfo{year}{2017}), \eprint{1703.09921}.

\bibitem[{\citenamefont{Colafrancesco et~al.}(2007)\citenamefont{Colafrancesco,
  Profumo, and Ullio}}]{Colafrancesco:2006he}
\bibinfo{author}{\bibfnamefont{S.}~\bibnamefont{Colafrancesco}},
  \bibinfo{author}{\bibfnamefont{S.}~\bibnamefont{Profumo}}, \bibnamefont{and}
  \bibinfo{author}{\bibfnamefont{P.}~\bibnamefont{Ullio}},
  \bibinfo{journal}{Phys. Rev.} \textbf{\bibinfo{volume}{D75}},
  \bibinfo{pages}{023513} (\bibinfo{year}{2007}), \eprint{astro-ph/0607073}.

\bibitem[{\citenamefont{Profumo and Ullio}(2010)}]{Profumo:2010ya}
\bibinfo{author}{\bibfnamefont{S.}~\bibnamefont{Profumo}} \bibnamefont{and}
  \bibinfo{author}{\bibfnamefont{P.}~\bibnamefont{Ullio}}
  (\bibinfo{year}{2010}), \eprint{1001.4086}.

\bibitem[{\citenamefont{Regis et~al.}(2014)\citenamefont{Regis, Colafrancesco,
  Profumo, de~Blok, Massardi, and Richter}}]{Regis:2014tga}
\bibinfo{author}{\bibfnamefont{M.}~\bibnamefont{Regis}},
  \bibinfo{author}{\bibfnamefont{S.}~\bibnamefont{Colafrancesco}},
  \bibinfo{author}{\bibfnamefont{S.}~\bibnamefont{Profumo}},
  \bibinfo{author}{\bibfnamefont{W.~J.~G.} \bibnamefont{de~Blok}},
  \bibinfo{author}{\bibfnamefont{M.}~\bibnamefont{Massardi}}, \bibnamefont{and}
  \bibinfo{author}{\bibfnamefont{L.}~\bibnamefont{Richter}},
  \bibinfo{journal}{JCAP} \textbf{\bibinfo{volume}{1410}}, \bibinfo{pages}{016}
  (\bibinfo{year}{2014}), \eprint{1407.4948}.

\bibitem[{\citenamefont{{De Angelis} et~al.}(2017)\citenamefont{{De Angelis},
  {Tatischeff}, {Grenier}, {McEnery}, {Mallamaci}, {Tavani}, {Oberlack},
  {Hanlon}, {Walter}, {Argan} et~al.}}]{astrogam2017}
\bibinfo{author}{\bibfnamefont{A.}~\bibnamefont{{De Angelis}}},
  \bibinfo{author}{\bibfnamefont{V.}~\bibnamefont{{Tatischeff}}},
  \bibinfo{author}{\bibfnamefont{I.~A.} \bibnamefont{{Grenier}}},
  \bibinfo{author}{\bibfnamefont{J.}~\bibnamefont{{McEnery}}},
  \bibinfo{author}{\bibfnamefont{M.}~\bibnamefont{{Mallamaci}}},
  \bibinfo{author}{\bibfnamefont{M.}~\bibnamefont{{Tavani}}},
  \bibinfo{author}{\bibfnamefont{U.}~\bibnamefont{{Oberlack}}},
  \bibinfo{author}{\bibfnamefont{L.}~\bibnamefont{{Hanlon}}},
  \bibinfo{author}{\bibfnamefont{R.}~\bibnamefont{{Walter}}},
  \bibinfo{author}{\bibfnamefont{A.}~\bibnamefont{{Argan}}},
  \bibnamefont{et~al.}, \bibinfo{journal}{ArXiv e-prints}
  (\bibinfo{year}{2017}), \eprint{1711.01265}.

\bibitem[{\citenamefont{Moiseev et~al.}(2015)}]{Moiseev:2015lva}
\bibinfo{author}{\bibfnamefont{A.~A.} \bibnamefont{Moiseev}}
  \bibnamefont{et~al.} (\bibinfo{year}{2015}), \eprint{1508.07349}.

\bibitem[{\citenamefont{Hunter et~al.}(2014)}]{Hunter:2013wla}
\bibinfo{author}{\bibfnamefont{S.~D.} \bibnamefont{Hunter}}
  \bibnamefont{et~al.}, \bibinfo{journal}{Astropart. Phys.}
  \textbf{\bibinfo{volume}{59}}, \bibinfo{pages}{18} (\bibinfo{year}{2014}),
  \eprint{1311.2059}.

\bibitem[{\citenamefont{Essig et~al.}(2013)\citenamefont{Essig, Kuflik,
  McDermott, Volansky, and Zurek}}]{Essig:2013goa}
\bibinfo{author}{\bibfnamefont{R.}~\bibnamefont{Essig}},
  \bibinfo{author}{\bibfnamefont{E.}~\bibnamefont{Kuflik}},
  \bibinfo{author}{\bibfnamefont{S.~D.} \bibnamefont{McDermott}},
  \bibinfo{author}{\bibfnamefont{T.}~\bibnamefont{Volansky}}, \bibnamefont{and}
  \bibinfo{author}{\bibfnamefont{K.~M.} \bibnamefont{Zurek}},
  \bibinfo{journal}{JHEP} \textbf{\bibinfo{volume}{11}}, \bibinfo{pages}{193}
  (\bibinfo{year}{2013}), \eprint{1309.4091}.

\bibitem[{\citenamefont{Boddy and Kumar}(2015)}]{Boddy:2015efa}
\bibinfo{author}{\bibfnamefont{K.~K.} \bibnamefont{Boddy}} \bibnamefont{and}
  \bibinfo{author}{\bibfnamefont{J.}~\bibnamefont{Kumar}},
  \bibinfo{journal}{Phys. Rev.} \textbf{\bibinfo{volume}{D92}},
  \bibinfo{pages}{023533} (\bibinfo{year}{2015}), \eprint{1504.04024}.

\bibitem[{\citenamefont{Strong et~al.}(1999)\citenamefont{Strong, Bloemen,
  Diehl, Hermsen, and Schoenfelder}}]{Strong:1998ck}
\bibinfo{author}{\bibfnamefont{A.~W.} \bibnamefont{Strong}},
  \bibinfo{author}{\bibfnamefont{H.}~\bibnamefont{Bloemen}},
  \bibinfo{author}{\bibfnamefont{R.}~\bibnamefont{Diehl}},
  \bibinfo{author}{\bibfnamefont{W.}~\bibnamefont{Hermsen}}, \bibnamefont{and}
  \bibinfo{author}{\bibfnamefont{V.}~\bibnamefont{Schoenfelder}},
  \bibinfo{journal}{Astrophys. Lett. Commun.} \textbf{\bibinfo{volume}{39}},
  \bibinfo{pages}{209} (\bibinfo{year}{1999}), \eprint{astro-ph/9811211}.

\bibitem[{\citenamefont{Hunter et~al.}(1997)}]{Hunter:1997we}
\bibinfo{author}{\bibfnamefont{S.~D.} \bibnamefont{Hunter}}
  \bibnamefont{et~al.}, \bibinfo{journal}{Astrophys. J.}
  \textbf{\bibinfo{volume}{481}}, \bibinfo{pages}{205} (\bibinfo{year}{1997}).

\bibitem[{\citenamefont{{Bartels} et~al.}(2017)\citenamefont{{Bartels},
  {Gaggero}, and {Weniger}}}]{bartels2017}
\bibinfo{author}{\bibfnamefont{R.}~\bibnamefont{{Bartels}}},
  \bibinfo{author}{\bibfnamefont{D.}~\bibnamefont{{Gaggero}}},
  \bibnamefont{and}
  \bibinfo{author}{\bibfnamefont{C.}~\bibnamefont{{Weniger}}},
  \bibinfo{journal}{JCAP} \textbf{\bibinfo{volume}{5}}, \bibinfo{eid}{001}
  (\bibinfo{year}{2017}), \eprint{1703.02546}.

\bibitem[{\citenamefont{{B{\oe}hm} and {Ascasibar}}(2004)}]{boehm2004PRD}
\bibinfo{author}{\bibfnamefont{C.}~\bibnamefont{{B{\oe}hm}}} \bibnamefont{and}
  \bibinfo{author}{\bibfnamefont{Y.}~\bibnamefont{{Ascasibar}}},
  \bibinfo{journal}{Physical review D} \textbf{\bibinfo{volume}{70}},
  \bibinfo{eid}{115013} (\bibinfo{year}{2004}), \eprint{hep-ph/0408213}.

\bibitem[{\citenamefont{{Boehm} et~al.}(2004)\citenamefont{{Boehm}, {Hooper},
  {Silk}, {Casse}, and {Paul}}}]{boehm2004PRL}
\bibinfo{author}{\bibfnamefont{C.}~\bibnamefont{{Boehm}}},
  \bibinfo{author}{\bibfnamefont{D.}~\bibnamefont{{Hooper}}},
  \bibinfo{author}{\bibfnamefont{J.}~\bibnamefont{{Silk}}},
  \bibinfo{author}{\bibfnamefont{M.}~\bibnamefont{{Casse}}}, \bibnamefont{and}
  \bibinfo{author}{\bibfnamefont{J.}~\bibnamefont{{Paul}}},
  \bibinfo{journal}{Physical Review Letters} \textbf{\bibinfo{volume}{92}},
  \bibinfo{eid}{101301} (\bibinfo{year}{2004}), \eprint{astro-ph/0309686}.

\bibitem[{\citenamefont{{Lorenz} and {The MAGIC
  Collaboration}}(2004)}]{Lorenz2004}
\bibinfo{author}{\bibfnamefont{E.}~\bibnamefont{{Lorenz}}} \bibnamefont{and}
  \bibinfo{author}{\bibnamefont{{The MAGIC Collaboration}}},
  \bibinfo{journal}{New astronomy reviews} \textbf{\bibinfo{volume}{48}},
  \bibinfo{pages}{339} (\bibinfo{year}{2004}).

\bibitem[{\citenamefont{{Abeysekara} et~al.}(2014)\citenamefont{{Abeysekara},
  {Alfaro}, {Alvarez}, {{\'A}lvarez}, {Arceo}, {Arteaga-Vel{\'a}zquez}, {Ayala
  Solares}, {Barber}, {Baughman}, {Bautista-Elivar} et~al.}}]{hawc2014}
\bibinfo{author}{\bibfnamefont{A.~U.} \bibnamefont{{Abeysekara}}},
  \bibinfo{author}{\bibfnamefont{R.}~\bibnamefont{{Alfaro}}},
  \bibinfo{author}{\bibfnamefont{C.}~\bibnamefont{{Alvarez}}},
  \bibinfo{author}{\bibfnamefont{J.~D.} \bibnamefont{{{\'A}lvarez}}},
  \bibinfo{author}{\bibfnamefont{R.}~\bibnamefont{{Arceo}}},
  \bibinfo{author}{\bibfnamefont{J.~C.} \bibnamefont{{Arteaga-Vel{\'a}zquez}}},
  \bibinfo{author}{\bibfnamefont{H.~A.} \bibnamefont{{Ayala Solares}}},
  \bibinfo{author}{\bibfnamefont{A.~S.} \bibnamefont{{Barber}}},
  \bibinfo{author}{\bibfnamefont{B.~M.} \bibnamefont{{Baughman}}},
  \bibinfo{author}{\bibfnamefont{N.}~\bibnamefont{{Bautista-Elivar}}},
  \bibnamefont{et~al.}, \bibinfo{journal}{Physical Review D}
  \textbf{\bibinfo{volume}{90}}, \bibinfo{eid}{122002} (\bibinfo{year}{2014}),
  \eprint{1405.1730}.

\bibitem[{\citenamefont{{Abramowski} et~al.}(2011)\citenamefont{{Abramowski},
  {Acero}, {Aharonian}, {Akhperjanian}, {Anton}, {Barnacka}, {Barres de
  Almeida}, {Bazer-Bachi}, {Becherini}, {Becker} et~al.}}]{hess2011DM}
\bibinfo{author}{\bibfnamefont{A.}~\bibnamefont{{Abramowski}}},
  \bibinfo{author}{\bibfnamefont{F.}~\bibnamefont{{Acero}}},
  \bibinfo{author}{\bibfnamefont{F.}~\bibnamefont{{Aharonian}}},
  \bibinfo{author}{\bibfnamefont{A.~G.} \bibnamefont{{Akhperjanian}}},
  \bibinfo{author}{\bibfnamefont{G.}~\bibnamefont{{Anton}}},
  \bibinfo{author}{\bibfnamefont{A.}~\bibnamefont{{Barnacka}}},
  \bibinfo{author}{\bibfnamefont{U.}~\bibnamefont{{Barres de Almeida}}},
  \bibinfo{author}{\bibfnamefont{A.~R.} \bibnamefont{{Bazer-Bachi}}},
  \bibinfo{author}{\bibfnamefont{Y.}~\bibnamefont{{Becherini}}},
  \bibinfo{author}{\bibfnamefont{J.}~\bibnamefont{{Becker}}},
  \bibnamefont{et~al.}, \bibinfo{journal}{Physical Review Letters}
  \textbf{\bibinfo{volume}{106}}, \bibinfo{eid}{161301} (\bibinfo{year}{2011}),
  \eprint{1103.3266}.

\bibitem[{\citenamefont{{Doro} et~al.}(2013)\citenamefont{{Doro}, {Conrad},
  {Emmanoulopoulos}, {S{\`a}nchez-Conde}, {Barrio}, {Birsin}, {Bolmont},
  {Brun}, {Colafrancesco}, {Connell} et~al.}}]{doro2013}
\bibinfo{author}{\bibfnamefont{M.}~\bibnamefont{{Doro}}},
  \bibinfo{author}{\bibfnamefont{J.}~\bibnamefont{{Conrad}}},
  \bibinfo{author}{\bibfnamefont{D.}~\bibnamefont{{Emmanoulopoulos}}},
  \bibinfo{author}{\bibfnamefont{M.~A.} \bibnamefont{{S{\`a}nchez-Conde}}},
  \bibinfo{author}{\bibfnamefont{J.~A.} \bibnamefont{{Barrio}}},
  \bibinfo{author}{\bibfnamefont{E.}~\bibnamefont{{Birsin}}},
  \bibinfo{author}{\bibfnamefont{J.}~\bibnamefont{{Bolmont}}},
  \bibinfo{author}{\bibfnamefont{P.}~\bibnamefont{{Brun}}},
  \bibinfo{author}{\bibfnamefont{S.}~\bibnamefont{{Colafrancesco}}},
  \bibinfo{author}{\bibfnamefont{S.~H.} \bibnamefont{{Connell}}},
  \bibnamefont{et~al.}, \bibinfo{journal}{Astroparticle Physics}
  \textbf{\bibinfo{volume}{43}}, \bibinfo{pages}{189} (\bibinfo{year}{2013}),
  \eprint{1208.5356}.

\bibitem[{\citenamefont{{Silverwood} et~al.}(2015)\citenamefont{{Silverwood},
  {Weniger}, {Scott}, and {Bertone}}}]{silverwood2015}
\bibinfo{author}{\bibfnamefont{H.}~\bibnamefont{{Silverwood}}},
  \bibinfo{author}{\bibfnamefont{C.}~\bibnamefont{{Weniger}}},
  \bibinfo{author}{\bibfnamefont{P.}~\bibnamefont{{Scott}}}, \bibnamefont{and}
  \bibinfo{author}{\bibfnamefont{G.}~\bibnamefont{{Bertone}}},
  \bibinfo{journal}{JCAP} \textbf{\bibinfo{volume}{3}}, \bibinfo{eid}{055}
  (\bibinfo{year}{2015}), \eprint{1408.4131}.

\bibitem[{\citenamefont{{Conrad}}(2017)}]{conrad2017}
\bibinfo{author}{\bibfnamefont{J.}~\bibnamefont{{Conrad}}}, in
  \emph{\bibinfo{booktitle}{6th International Symposium on High Energy
  Gamma-Ray Astronomy}} (\bibinfo{year}{2017}), vol. \bibinfo{volume}{1792} of
  \emph{\bibinfo{series}{American Institute of Physics Conference Series}}, p.
  \bibinfo{pages}{030002}, \eprint{1610.03258}.

\bibitem[{\citenamefont{{Morselli}}(2017)}]{morselli2017}
\bibinfo{author}{\bibfnamefont{A.}~\bibnamefont{{Morselli}}},
  \bibinfo{journal}{ArXiv e-prints}  (\bibinfo{year}{2017}),
  \eprint{1709.01483}.

\bibitem[{\citenamefont{{Aharonian} et~al.}(2006)\citenamefont{{Aharonian},
  {Akhperjanian}, {Bazer-Bachi}, {Beilicke}, {Benbow}, {Berge}, {Bernl{\"o}hr},
  {Boisson}, {Bolz}, {Borrel} et~al.}}]{hess2006}
\bibinfo{author}{\bibfnamefont{F.}~\bibnamefont{{Aharonian}}},
  \bibinfo{author}{\bibfnamefont{A.~G.} \bibnamefont{{Akhperjanian}}},
  \bibinfo{author}{\bibfnamefont{A.~R.} \bibnamefont{{Bazer-Bachi}}},
  \bibinfo{author}{\bibfnamefont{M.}~\bibnamefont{{Beilicke}}},
  \bibinfo{author}{\bibfnamefont{W.}~\bibnamefont{{Benbow}}},
  \bibinfo{author}{\bibfnamefont{D.}~\bibnamefont{{Berge}}},
  \bibinfo{author}{\bibfnamefont{K.}~\bibnamefont{{Bernl{\"o}hr}}},
  \bibinfo{author}{\bibfnamefont{C.}~\bibnamefont{{Boisson}}},
  \bibinfo{author}{\bibfnamefont{O.}~\bibnamefont{{Bolz}}},
  \bibinfo{author}{\bibfnamefont{V.}~\bibnamefont{{Borrel}}},
  \bibnamefont{et~al.}, \bibinfo{journal}{Nature}
  \textbf{\bibinfo{volume}{439}}, \bibinfo{pages}{695} (\bibinfo{year}{2006}),
  \eprint{astro-ph/0603021}.

\bibitem[{\citenamefont{{HESS Collaboration} et~al.}(2016)\citenamefont{{HESS
  Collaboration}, {Abramowski}, {Aharonian}, {Benkhali}, {Akhperjanian},
  {Ang{\"u}ner}, {Backes}, {Balzer}, {Becherini}, {Tjus} et~al.}}]{hess2016}
\bibinfo{author}{\bibnamefont{{HESS Collaboration}}},
  \bibinfo{author}{\bibfnamefont{A.}~\bibnamefont{{Abramowski}}},
  \bibinfo{author}{\bibfnamefont{F.}~\bibnamefont{{Aharonian}}},
  \bibinfo{author}{\bibfnamefont{F.~A.} \bibnamefont{{Benkhali}}},
  \bibinfo{author}{\bibfnamefont{A.~G.} \bibnamefont{{Akhperjanian}}},
  \bibinfo{author}{\bibfnamefont{E.~O.} \bibnamefont{{Ang{\"u}ner}}},
  \bibinfo{author}{\bibfnamefont{M.}~\bibnamefont{{Backes}}},
  \bibinfo{author}{\bibfnamefont{A.}~\bibnamefont{{Balzer}}},
  \bibinfo{author}{\bibfnamefont{Y.}~\bibnamefont{{Becherini}}},
  \bibinfo{author}{\bibfnamefont{J.~B.} \bibnamefont{{Tjus}}},
  \bibnamefont{et~al.}, \bibinfo{journal}{Nature}
  \textbf{\bibinfo{volume}{531}}, \bibinfo{pages}{476} (\bibinfo{year}{2016}),
  \eprint{1603.07730}.

\bibitem[{\citenamefont{{H.~E.~S.~S.~Collaboration}
  et~al.}(2017)\citenamefont{{H.~E.~S.~S.~Collaboration}, {:}, {Abdalla},
  {Abramowski}, {Aharonian}, {Ait Benkhali}, {Akhperjaniany}, {Andersson},
  {Ang{\"u}ner}, {Arakawa} et~al.}}]{hess2017}
\bibinfo{author}{\bibnamefont{{H.~E.~S.~S.~Collaboration}}},
  \bibinfo{author}{\bibnamefont{{:}}},
  \bibinfo{author}{\bibfnamefont{H.}~\bibnamefont{{Abdalla}}},
  \bibinfo{author}{\bibfnamefont{A.}~\bibnamefont{{Abramowski}}},
  \bibinfo{author}{\bibfnamefont{F.}~\bibnamefont{{Aharonian}}},
  \bibinfo{author}{\bibfnamefont{F.}~\bibnamefont{{Ait Benkhali}}},
  \bibinfo{author}{\bibfnamefont{A.~G.} \bibnamefont{{Akhperjaniany}}},
  \bibinfo{author}{\bibfnamefont{T.}~\bibnamefont{{Andersson}}},
  \bibinfo{author}{\bibfnamefont{E.~O.} \bibnamefont{{Ang{\"u}ner}}},
  \bibinfo{author}{\bibfnamefont{M.}~\bibnamefont{{Arakawa}}},
  \bibnamefont{et~al.}, \bibinfo{journal}{ArXiv e-prints}
  (\bibinfo{year}{2017}), \eprint{1706.04535}.

\bibitem[{\citenamefont{Hütten et~al.}(2016)\citenamefont{Hütten, Combet,
  Maier, and Maurin}}]{Hutten:2016jko}
\bibinfo{author}{\bibfnamefont{M.}~\bibnamefont{Hütten}},
  \bibinfo{author}{\bibfnamefont{C.}~\bibnamefont{Combet}},
  \bibinfo{author}{\bibfnamefont{G.}~\bibnamefont{Maier}}, \bibnamefont{and}
  \bibinfo{author}{\bibfnamefont{D.}~\bibnamefont{Maurin}},
  \bibinfo{journal}{JCAP} \textbf{\bibinfo{volume}{1609}}, \bibinfo{pages}{047}
  (\bibinfo{year}{2016}), \eprint{1606.04898}.

\bibitem[{\citenamefont{Lefranc
  et~al.}(2016{\natexlab{a}})\citenamefont{Lefranc, Mamon, and
  Panci}}]{Lefranc:2016dgx}
\bibinfo{author}{\bibfnamefont{V.}~\bibnamefont{Lefranc}},
  \bibinfo{author}{\bibfnamefont{G.~A.} \bibnamefont{Mamon}}, \bibnamefont{and}
  \bibinfo{author}{\bibfnamefont{P.}~\bibnamefont{Panci}},
  \bibinfo{journal}{JCAP} \textbf{\bibinfo{volume}{1609}}, \bibinfo{pages}{021}
  (\bibinfo{year}{2016}{\natexlab{a}}), \eprint{1605.02793}.

\bibitem[{\citenamefont{Lefranc
  et~al.}(2016{\natexlab{b}})\citenamefont{Lefranc, Moulin, Panci, Sala, and
  Silk}}]{Lefranc:2016fgn}
\bibinfo{author}{\bibfnamefont{V.}~\bibnamefont{Lefranc}},
  \bibinfo{author}{\bibfnamefont{E.}~\bibnamefont{Moulin}},
  \bibinfo{author}{\bibfnamefont{P.}~\bibnamefont{Panci}},
  \bibinfo{author}{\bibfnamefont{F.}~\bibnamefont{Sala}}, \bibnamefont{and}
  \bibinfo{author}{\bibfnamefont{J.}~\bibnamefont{Silk}},
  \bibinfo{journal}{JCAP} \textbf{\bibinfo{volume}{1609}}, \bibinfo{pages}{043}
  (\bibinfo{year}{2016}{\natexlab{b}}), \eprint{1608.00786}.

\bibitem[{\citenamefont{{Chang} et~al.}(2017)\citenamefont{{Chang}, {Ambrosi},
  {An}, {Asfandiyarov}, {Azzarello}, {Bernardini}, {Bertucci}, {Cai},
  {Caragiulo}, {Chen} et~al.}}]{dampe2017}
\bibinfo{author}{\bibfnamefont{J.}~\bibnamefont{{Chang}}},
  \bibinfo{author}{\bibfnamefont{G.}~\bibnamefont{{Ambrosi}}},
  \bibinfo{author}{\bibfnamefont{Q.}~\bibnamefont{{An}}},
  \bibinfo{author}{\bibfnamefont{R.}~\bibnamefont{{Asfandiyarov}}},
  \bibinfo{author}{\bibfnamefont{P.}~\bibnamefont{{Azzarello}}},
  \bibinfo{author}{\bibfnamefont{P.}~\bibnamefont{{Bernardini}}},
  \bibinfo{author}{\bibfnamefont{B.}~\bibnamefont{{Bertucci}}},
  \bibinfo{author}{\bibfnamefont{M.~S.} \bibnamefont{{Cai}}},
  \bibinfo{author}{\bibfnamefont{M.}~\bibnamefont{{Caragiulo}}},
  \bibinfo{author}{\bibfnamefont{D.~Y.} \bibnamefont{{Chen}}},
  \bibnamefont{et~al.}, \bibinfo{journal}{Astroparticle Physics}
  \textbf{\bibinfo{volume}{95}}, \bibinfo{pages}{6} (\bibinfo{year}{2017}),
  \eprint{1706.08453}.

\bibitem[{\citenamefont{{Marrocchesi}}(2015)}]{calet2015}
\bibinfo{author}{\bibfnamefont{P.~S.} \bibnamefont{{Marrocchesi}}},
  \bibinfo{journal}{ArXiv e-prints}  (\bibinfo{year}{2015}),
  \eprint{1512.08059}.

\bibitem[{\citenamefont{{Ong} et~al.}(2017)\citenamefont{{Ong}, {Aramaki},
  {Bird}, {Boezio}, {Boggs}, {Carr}, {Craig}, {von Doetinchem}, {Fabris},
  {Gahbauer} et~al.}}]{gaps2017}
\bibinfo{author}{\bibfnamefont{R.~A.} \bibnamefont{{Ong}}},
  \bibinfo{author}{\bibfnamefont{T.}~\bibnamefont{{Aramaki}}},
  \bibinfo{author}{\bibfnamefont{R.}~\bibnamefont{{Bird}}},
  \bibinfo{author}{\bibfnamefont{M.}~\bibnamefont{{Boezio}}},
  \bibinfo{author}{\bibfnamefont{S.~E.} \bibnamefont{{Boggs}}},
  \bibinfo{author}{\bibfnamefont{R.}~\bibnamefont{{Carr}}},
  \bibinfo{author}{\bibfnamefont{W.~W.} \bibnamefont{{Craig}}},
  \bibinfo{author}{\bibfnamefont{P.}~\bibnamefont{{von Doetinchem}}},
  \bibinfo{author}{\bibfnamefont{L.}~\bibnamefont{{Fabris}}},
  \bibinfo{author}{\bibfnamefont{F.}~\bibnamefont{{Gahbauer}}},
  \bibnamefont{et~al.}, \bibinfo{journal}{ArXiv e-prints}
  (\bibinfo{year}{2017}), \eprint{1710.00452}.

\end{thebibliography}

\end{document}